\documentclass[twocolumn]{aastex631}

\begin{document}

\title{Sleeping Giants Arise: Monitoring the Return of Three Changing-Look Quasars to Their High States
\footnote{Based on observations obtained with the Hobby-Eberly Telescope (HET), which is a joint project of the University of Texas at Austin, the Pennsylvania State University, Ludwig-Maximillians-Universitaet Muenchen, and Georg-August Universitaet Goettingen. The HET is named in honor of its principal benefactors, William P. Hobby and Robert E. Eberly.}}

\author[0000-0003-1752-679X]{Laura Duffy}
\affiliation{Department of Astronomy and Astrophysics and Institute for Gravitation and the Cosmos, The Pennsylvania State University, 525 Davey Lab, University Park, PA 16803, USA}
\email{lrd48@psu.edu}

\author[0000-0002-3719-940X]{Michael Eracleous}
\affil{Department of Astronomy and Astrophysics and Institute for Gravitation and the Cosmos, The Pennsylvania State University, 525 Davey Lab, University Park, PA 16803, USA}

\author[0000-0001-8665-5523]{John J. Ruan}
\affil{Department of Physics and Astronomy, Bishop’s University, 2600 College St., Sherbrooke, QC J1M 1Z7, Canada}

\author[0000-0002-6893-3742]{Qian Yang}
\affil{Center for Astrophysics, Harvard \& Smithsonian, 60 Garden Street, Cambridge, MA 02138, USA}

\author[0000-0001-8557-2822]{Jessie C. Runnoe}
\affil{Department of Physics and Astronomy, Vanderbilt University, Nashville, TN 37235, USA}




\begin{abstract}
Changing-look quasars challenge many models of the quasar central engine. Their extreme variability in both the continuum and broad emission-line fluxes on timescales on the order of years is difficult to explain. To investigate the cause of the observed transitions, we present new contemporaneous optical and X-ray observations of three faded changing-look quasars as they return to a state of high optical luminosity. Two of these three remained in a quiescent state for more than ten years before returning to a new high state. We find that before, during, and after transition, the spectral energy distributions of all three follow predictions for quasars based on X-ray binary outbursts, suggesting that the underlying mechanism is likely a changing accretion rate causing changes in the accretion flow structure. In two of the three cases, the transition between the initial high and low state and the transition between the low and new high state took nearly identical amounts of time, on the order of hundreds of days. This transition timescale is a useful constraint on models of the accretion state changes. The behavior of the broad line profiles suggests that the BLR structure is changing during the transition.

\end{abstract}



\section{Introduction} \label{sec:intro}
Changing-look quasars \citep[CLQs; e.g.,][]{lamassa15} are characterized by dramatic changes in their broad emission-line fluxes and non-stellar continuum levels on short (month-to-year) timescales \citep[see][for a review]{ricci23}. As a result of this variation, they provide a unique opportunity to probe accretion flow variations. In the high state, light from the quasar dominates, and both broad and narrow emission lines are observable in addition to a strong non-stellar continuum. In the low state, the broad emission lines and non-stellar continuum weaken dramatically, and the light from the host galaxy may dominate over light from the quasar. This sort of transition was first detected in the optical spectra of lower-luminosity Seyfert galaxies \citep[e.g.,][]{cohen86, goodrich89, goodrich90, goodrich95, denney14}. Quasars, however, are significantly more luminous than Seyfert galaxies, making changes of the observed magnitude surprising \citep[see][for relationship between AGN luminosity and variability]{vandenberk04, macleod10}. Nonetheless, the explanation for the observed behavior may be shared between CLQs and changing-look Seyfert galaxies. The term `changing-look AGN' has also been used to describe changes in the X-ray spectra of AGN \citep[e.g.,][]{matt03, piconcelli03, marchese12, ricci16}, but here we are specifically referring to behavior in the ultraviolet (UV), optical, and infrared (IR).

CLQs are frequently identified through spectroscopic monitoring of quasars \citep[e.g.][]{lamassa15, macleod16, runnoe16, ruan16, yang18, zeltyn22}, often as a part of large, time-domain surveys such as the Sloan Digital Sky Survey \citep[SDSS; see][]{macleod19, green22, zeltyn22} or the Dark Energy Spectroscopic Instrument \cite[DESI; see][]{guo24, guo24_2}. Both `turn-on' and `turn-off' CLQs and changing-look AGN have also been identified through large changes in their optical or infrared light curves, and later confirmed through repeat spectroscopy \citep[e.g.][]{macleod16, gezari17, yang18, macleod19, sheng20, ln22, wang24, yang24}. A few studies have also found previously-identified CLQs that return to their initial states, be it through fading back to a low state or returning to a high state \citep[e.g.][]{zeltyn22, yang24, lyu24, wang24b}. \cite{yang24} found two previously-identified turn-off CLQs that have recently turned on, as well as a large sample of turn-on CLQs that have indications of recent nuclear activity, through the presence of strong narrow [O~III] emission. Some studies have also investigated in detail changing-look Seyfert galaxies that have exhibited multiple transitions  \citep[e.g.,][]{denney14, veronese24, palit25}.

Many mechanisms have been suggested to explain the observed transitions, but the short transition timescale \citep[see][]{trakhtenbrot19} and the fact that the transitions sometimes reccur \citep[see][]{zeltyn22, yang24, lyu24} remain a challenge. Nuclear supernovae and tidal disruption events \citep[TDEs; e.g.][]{merloni15, blanchard17} have been proposed as explanations in the past, but are generally considered unlikely in most cases \citep[see][for discussion]{yang19}, and especially when the CLQ transitions between high and low states multiple times. TDEs do sometimes show multiple flares \citep[e.g.,][]{somalwar23,sun24,veres24,sun25} but such repeated flaring occurs in $<30$\% of cases \citep[][]{somalwar23}, and is usually from repeating partial TDEs, which appear periodic, unlike CLQs that go through multiple state transitions. Importantly, CLQ bright states have much longer durations than typical TDE flares \citep[][and references to repeated TDEs above]{runnoe16, macleod19}, disfavoring the TDE explanation for the majority of CLQ events. In a small number of cases variable dust extinction of the broad-line region (BLR) and continuum source are able to explain the observed changes \citep[e.g.][]{maiolino10, markowitz14, zeltyn22}. However, in the majority of cases dust extinction alone is insufficient to explain the observed transition \citep[e.g.][]{lamassa15, macleod16, zeltyn24, ricci23, duffy25}. 

The widely favored explanation for the observed behavior is variable accretion rates or changing accretion flow structures \citep[e.g.][]{lamassa15, macleod19, ruan19, trakhtenbrot19, DexterBegelman19, jin21, liska22, liska23}. Some work draws analogies between CLQs and X-ray binaries \citep[XRBs;][Gilbert et al. in prep]{ruan19, jin21, yang23}, which have been observed to undergo similarly dramatic changes on short timescales. This analogy is further motivated both by the observation of a fundamental plane in radio and X-ray luminosity and black hole mass, which links weakly-accreting black holes across many orders of magnitude in mass \citep{merloni03, falcke04, gultekin22} and by observations linking characteristic timescales in X-ray variations with the black hole mass, again across many decades in mass \citep{mchardy06}. Furthermore, both XRBs and AGN show similar correlations between X-ray photon index and the Eddington ratio \citep[e.g.][]{yang15}. Galactic XRBs have been observed to show transitions in their accretion states as a function of the Eddington ratio, going between a low Eddington ratio, hard-spectrum state and a high Eddington ratio, soft-spectrum state \citep[see][for a review]{done07} and transitioning at an Eddington ratio $\sim10^{-2}$. The cause of the transition has often been attributed to a changing accretion flow structure, from a geometrically thin, optically thick disk \citep[i.e.][]{shakura73} in the high, soft state to a radiatively inefficient, advection dominated accretion flow \citep[ADAF; see][]{narayan94, shapiro76} in the inner regions of a truncated thin disk in the low, hard state. This mechanism has been previously invoked to explain the transition seen in CLQs \citep[][Gilbert et al. in prep]{ruan19, jin21, yang23, yang24, veronese24}. In XRBs, the changing disk structure is probed using the X-ray photon index, $\Gamma$. In models that compare CLQs to XRBs, the UV-optical spectral index, $\alpha_{ox}$ is used instead, because the accretion flows are generally much cooler.

ADAFs are also commonly used to explain the spectral energy distributions (SEDs) of low-luminosity AGNs (LLAGN), which lack the `big blue bump' usually seen in quasar SEDs \citep[e.g.,][]{ho99, eracleous10}. LLAGN SEDs have been modelled as the combination of an ADAF, a truncated thin disk, and some jet contribution \citep[e.g.,][]{nemmen14}. Intriguingly, prior studies of CLQs have found that the SEDs of low-state CLQs also lack the big blue bump seen in most quasar SEDs \citep[e.g.,][]{duffy25}.

It is still uncertain whether, in a manner similar to XRBs, CLQs represent the transition between accretion flow states. Discerning if CLQ transitions can generally be attributed to a change between an ADAF \citep[with or without a luminous jet; see][]{markoff01} and a geometrically thin, optically thick disk would be useful -- it may allow us to use our understanding of XRBs to explain AGN. This hypothesis faces challenges, however. The commonly-invoked timescale for large changes in accretion disk luminosity as a result of changes in the accretion rate is the viscous time \citep{shakura73, frank02}, which is multiple orders of magnitude longer than the observed transition times, which are on the order of months to years \citep[see][]{ricci23}. Other models suggest that the changes seen in the accretion flow should reflect the time required for the accretion disk to heat or cool (the thermal timescale), or the time required for a heating or cooling front to propagate through the accretion disk, which is slightly longer than the thermal time \citep{stern18, ross18}. Those models suggest that the source of the changing-look phenomenon could be due to heating and cooling fronts propagating outwards from the innermost stable circular orbit through a geometrically thin, optically thick disk. Recent models have also identified other mechanisms that act on a much shorter time -- modifications to the standard thin disk model and models based on galactic X-ray binaries harboring black holes both significantly shorten the relevant timescale \citep[][]{DexterBegelman19, noda18, sniegowska20}. Other models and simulations that include the effects of magneto-hydrodynamics also find shorter timescales for the evaporation of a geometrically thin disk into a truncated thin disk and two-phase, thick corona \cite[][]{liska22}, again on shorter timescales than the viscous time. \cite{liska23} suggest that the cause of changing-look phenomena in AGN may be due to warps and tears in geometrically thin accretion disks that are misaligned with the black-hole spin axis. These torn disks can lead to temporary increases in the accretion rate on timescales much shorter than the viscous time.

To further explore the hypothesis that the transitions of CLQs are associated with changes in the structure of their accretion flows \citep[e.g.][]{ruan19}, we monitored three previously known turn-off CLQs as they returned to the high state. Our aim was to trace the spectroscopic transformations and check if the shapes of the spectral energy distributions in the high state resembled those of luminous quasars. Previous observations of two of these CLQs in their low states showed them to have spectral energy distributions resembling those of ADAFs \citep{duffy25}.  In this work, we present new contemporaneous X-ray, UV, and optical observations of each CLQ and combine them with multi-epoch X-ray, UV and optical observations of the previous high and low states. We trace the transition between the high and low states and back again, characterize the emission line profiles over this time, and compare our findings to predictions for quasars based on analogies with XRBs.

We list our three targets and their basic properties in Table~\ref{table:props} along with the truncated identifier we use to refer to them hereafter. The return of one of the three CLQs, J1011, to a bright state has been previously reported by \cite{yang24}, \cite{wang24b} and \cite{lyu24}. All three studies note that the new high state of J1011 is dimmer than the previous high state, and all three identified J1011 as a returning CLQ on the basis of its optical and IR light curves.

We describe our target selection and observations in Section \ref{sec:obs}. In Section \ref{sec:meas}, we explain our data processing and measurement methods. We perform analysis of the multi-epoch measurements in Section \ref{sec:analysis}, and further explore those measurements and their implications in Section \ref{sec:disc}. Finally, we summarize our work in Section \ref{sec:conc}. Throughout, we adopt a $\Lambda$CDM cosmology with $H_0 = 70$ km~s$^{-1}$, $\Omega_{m,0}=0.30$, and $\Omega_\Lambda=0.7$.

\section{Targets \& Observations}\label{sec:obs}
\subsection{Target Selection}

We identified the targets serendipitously (see properties in Table~\ref{table:props}). Two of the three targets, J1011 and J2336, attracted our attention because they showed some continuum re-brightening in spectra taken for different reasons and presented in \cite{duffy25} The other target, J2333, came to our attention because it displayed an increase in flux in its ZTF g-band flux by a factor of $\sim4$ since the last low-state observation in 2018 \citep[presented in][]{jin21}. We elaborate further below.

\begin{figure*}
    \centering
    \includegraphics[width=0.45\textwidth]{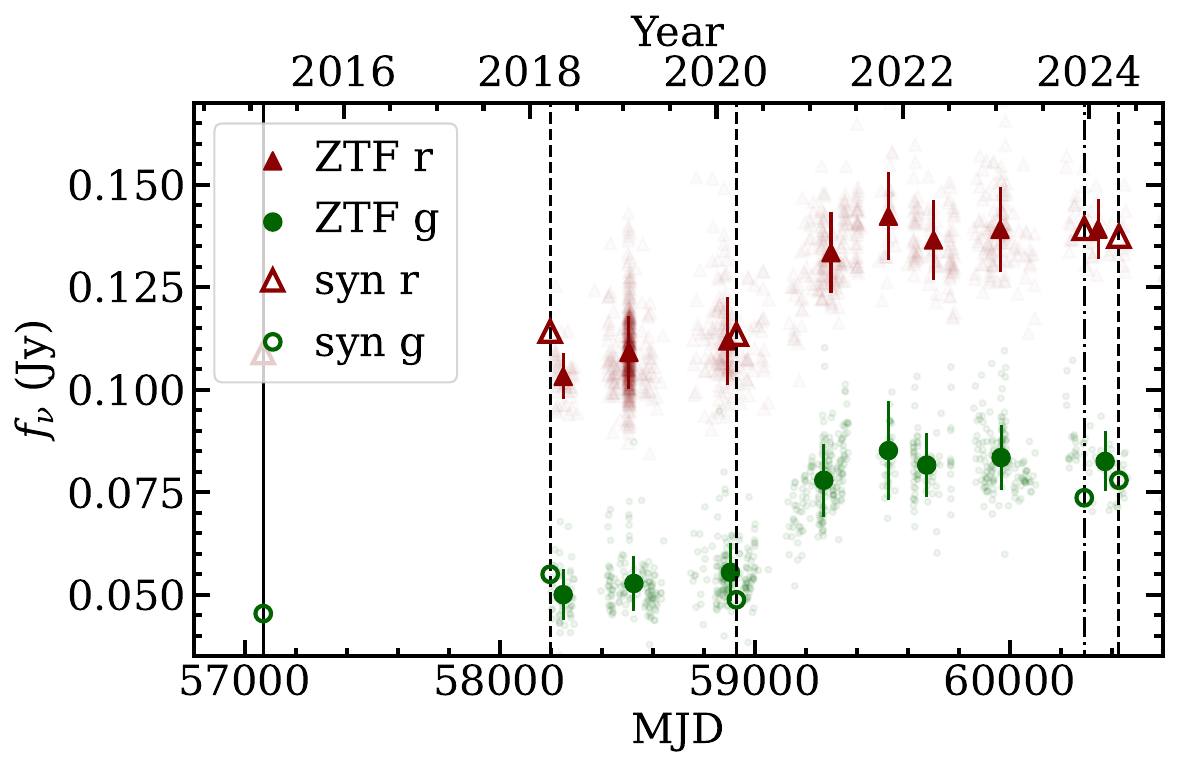}
    \hskip 1cm
    \includegraphics[width=0.45\textwidth]{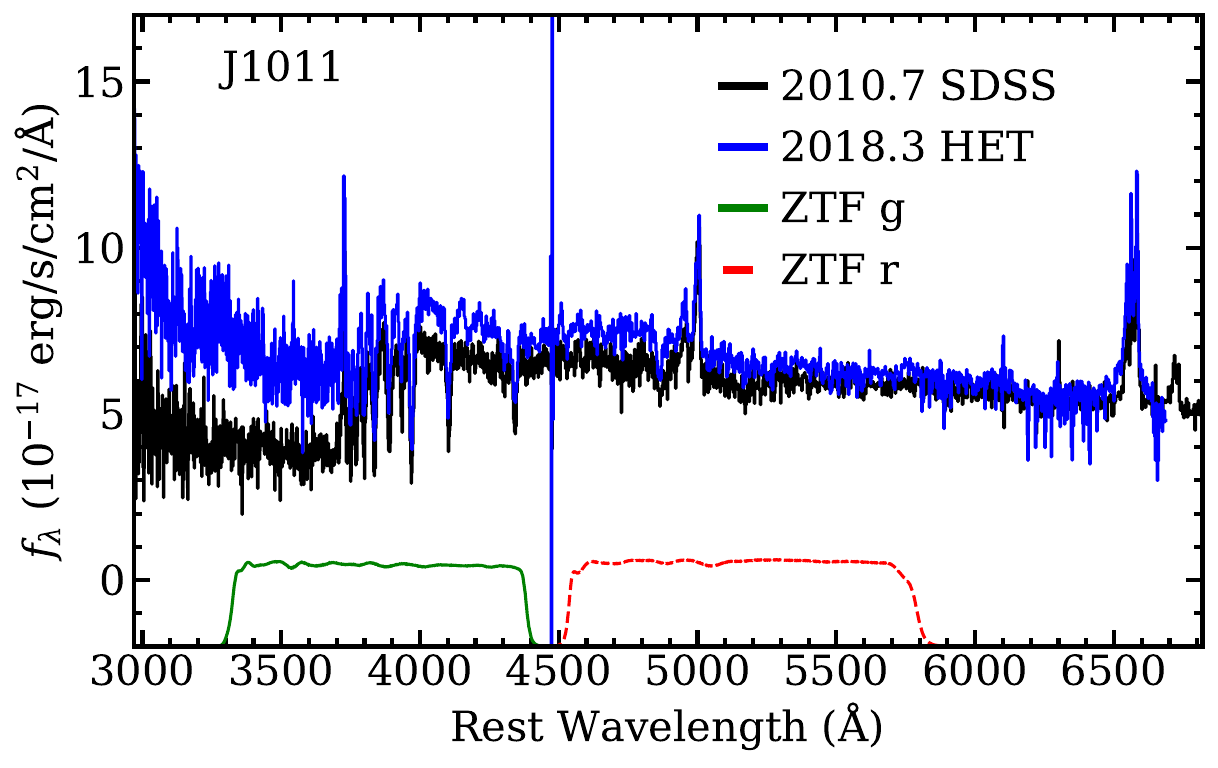}
    \includegraphics[width=0.45\textwidth]{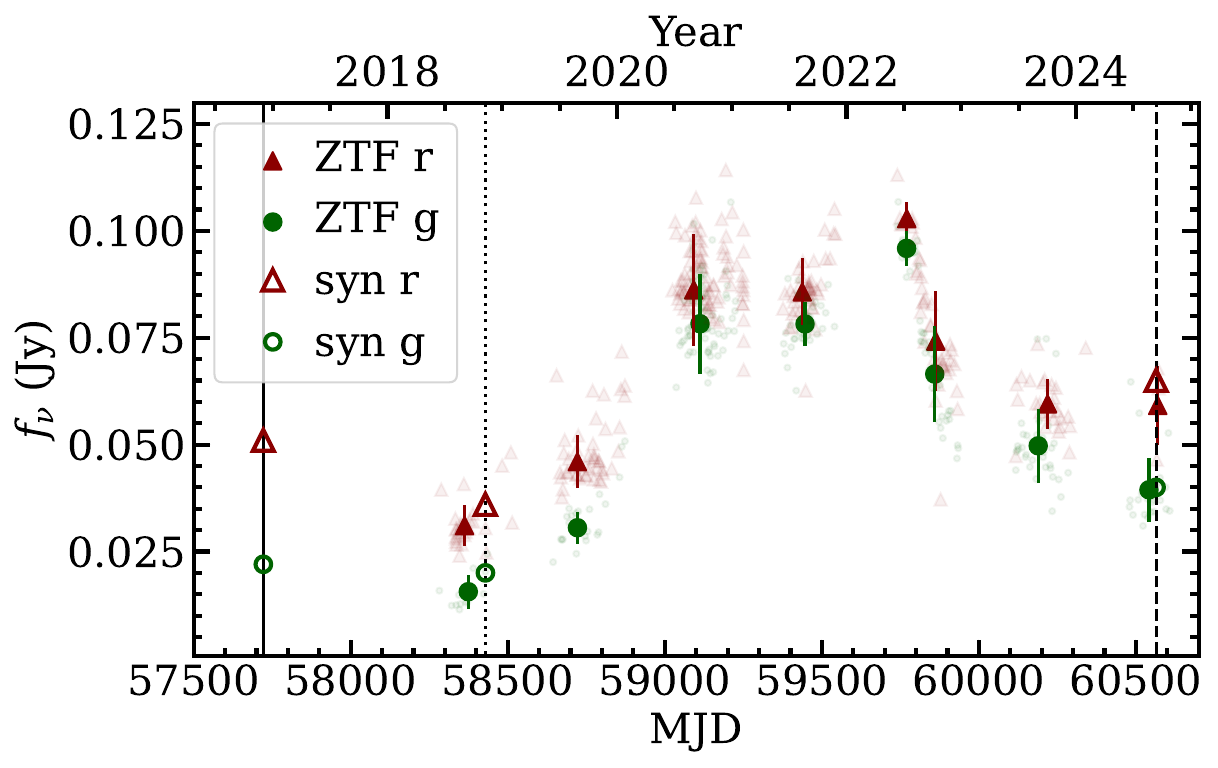}
    \hskip 1cm
    \includegraphics[width=0.45\textwidth]{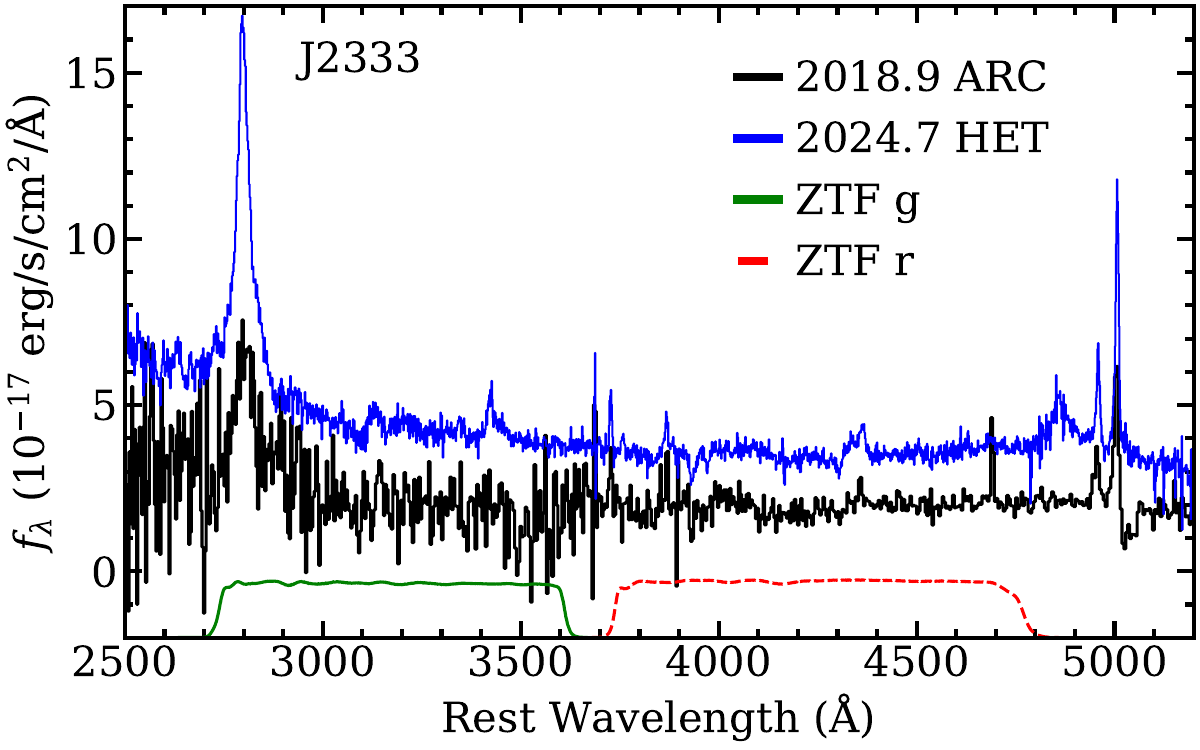}
    \includegraphics[width=0.45\textwidth]{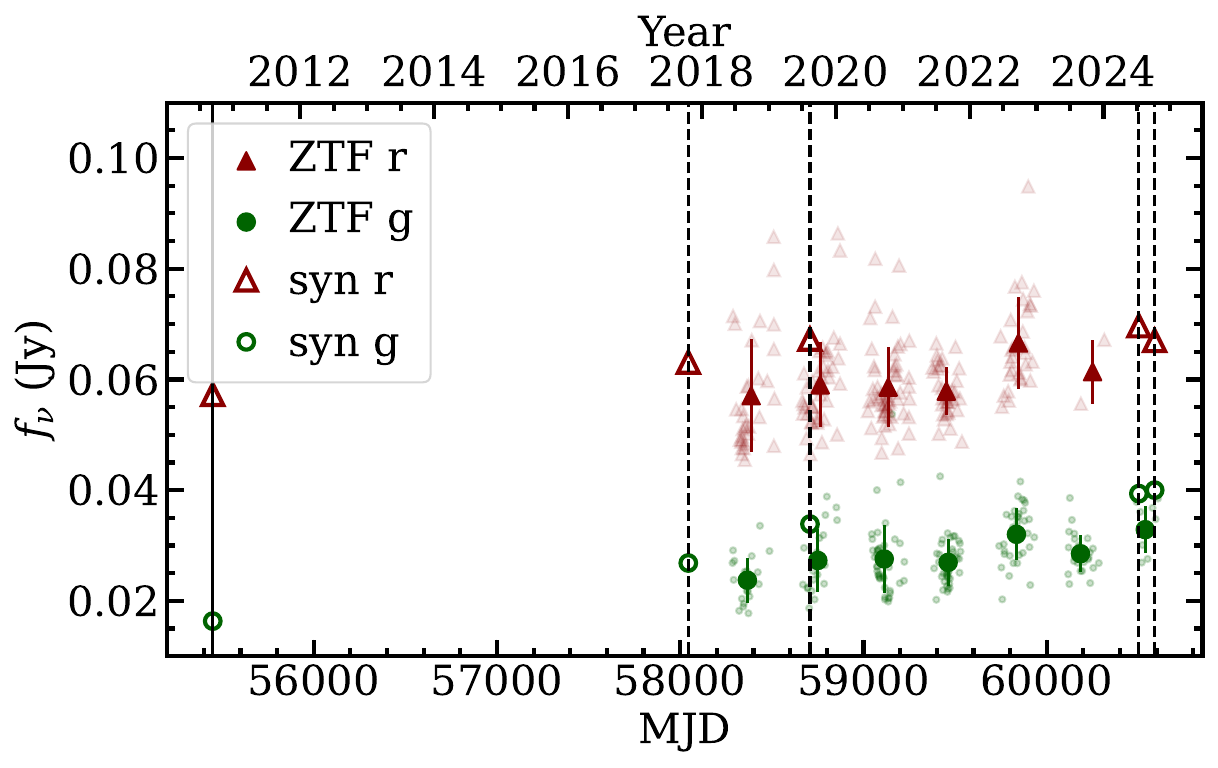}
    \hskip 1cm
    \includegraphics[width=0.45\textwidth]{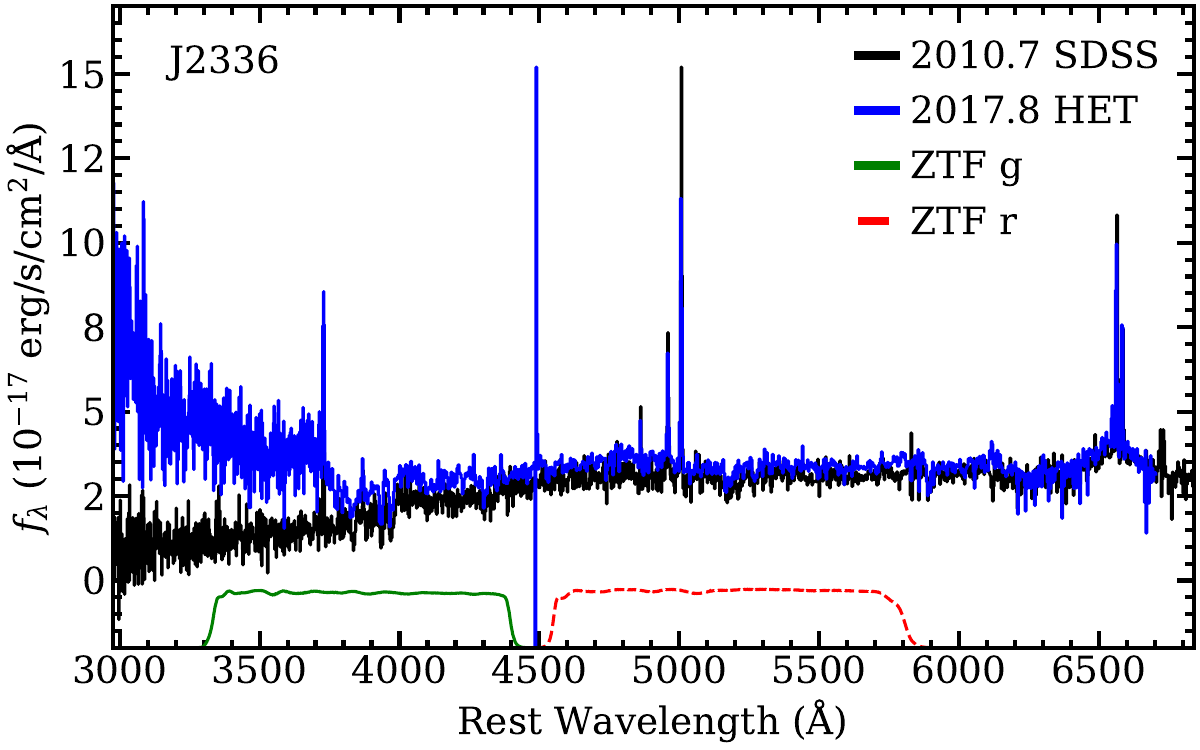}
    \caption{\textit{Left}: Zwicky Transient facility light curves of J1011 (top), J2333 (middle), and J2336 (bottom). In the light red triangles and light green circles, we plot all available ZTF r- and g-band photometry. We show r- and g-band data binned by observation epoch in the filled red triangles and green circles respectively. The error bars represent the standard deviation in the measurements incorporated in each bin. Finally, we show synthetic ZTF r- and g-band photometry for our spectra in the hollow red triangles and hollow green circles. We mark the date of the SDSS low-state spectra with the solid black line, the HET spectra with dashed lines, and spectra from MMT (J1011) and ARC (J2333) with the dot-dashed line. Synthetic ZTF photometry from optical spectra is consistent with observed photometry. \textit{Right}: Spectra of J1011 (top), J2333 (middle), and J2336 (bottom) in their low states (black) and in first follow-up observation where we noticed continuum or spectral evolution (blue). In solid green and dashed red, respectively, we plot the ZTF g- and r-bandpasses. For all three CLQs, changes in the blue continuum are caught by the ZTF g-band. The sharp feature at an observed wavelength of 5577~\AA{}, which corresponds to a rest-frame wavelength around 4500~\AA{} for J1011 and J2336 and 3700~\AA{} for J2333, is a telluric [O~I] emission line, which was not perfectly removed from the HET spectra.}
    \label{fig:lightcurve}
\end{figure*}

The first target, J1011 at $z=0.246$, had shown some continuum evolution between a 2010 low-state spectrum and a 2018 Hobby-Eberly Telescope \citep[see Figure~\ref{fig:lightcurve}, top right and][]{duffy25}. We verified the re-brightening by examining Zwicky Transient Facility \citep[ZTF;][]{bellm19, graham19, masci19} r- and g-band light curves (see Figure~\ref{fig:lightcurve}, top left). J1011 has archival optical spectra dating back to 2003 from the SDSS. We supplement the two SDSS observations with further optical spectra taken by the HET in 2018, 2020, and 2024 \citep[see][and Gilbert et al., in prep for presentation of the 2018 and 2020 spectra, respectively]{duffy25} and by the MMT in late 2023 \citep[see][for presentation of the 2023 spectrum]{yang24}.

We noticed and decided to follow up the second target, J2333 at $z=0.513$, while visually inspecting the ZTF light curves of known CLQs. The middle left panel of Figure~\ref{fig:lightcurve} shows the light curve evolution of J2333, and the middle right panel shows two representaive optical spectra.

The third target, J2336 at redshift $z=0.243$, was selected for further monitoring in a similar manner to J1011. A follow-up low-state optical spectrum from 2017 showed brightening in the near-UV continuum, although little change elsewhere. The ZTF light curve of the object showed small increases in g-band flux, although nowhere near the level of increase generally used to identify CLQ candidates in large statistical samples. Figure~\ref{fig:lightcurve} (bottom left and right panels) displays the ZTF light curve and spectral evolution used for candidate identification. This CLQ has archival spectra from 2001 and 2010 from the SDSS. We supplement SDSS observations with follow-up optical spectra taken by the HET in 2017, 2019, and 2024 \citep[see][and Gilbert et al., in prep for presentation of the 2017 and 2019 spectra, respectively]{duffy25}.

All of the spectra that we used in this work can be seen in Figure \ref{fig:low_spec}. For all CLQs, we also make use of published results from Chandra X-ray observations \citep{ruan19, jin21} and newly obtained Swift/XRT and Swift/UVOT observations. The Chandra observations and the new Swift/XRT and Swift/UVOT observations were taken contemporaneously with the three most recent spectra for J1011 and J2336, and the two most recent spectra for J2333. Thus, we are able to trace out the behavior of all CLQs across a broad range of wavelengths over time.

\begin{deluxetable*}{ccccccc}
\tablecaption{Intrinsic CLQ Properties}
\tablewidth{0pt}
\label{table:props}
\setlength{\tabcolsep}{6pt}
\tablehead{
{SDSS} &{Truncated} & {} & {Luminosity}  & {} \\
{Identifier} &{Identifier} & {\textit{z}}& {Distance (Mpc)} & {log$\left(M_{\rm BH}/{\rm M_\odot}\right)$}\\
{(1)} & (2) & (3) & (4) & (5)}
\startdata
{J$101152.98+544206.4$} & {J1011} & {0.246} & {1240} & {7.6$\pm$0.1}\\
{J$233317.38-002303.5$} & {J2333} & {0.513} & {2260} & {8.5$\pm$0.5}\\
{J$233602.98+001728.7$} & {J2336} & {0.243} & {1220} & {8.1$\pm$0.5}\\
\enddata
\tablecomments{Column 1: Full SDSS object name, constructed from J2000 coordinates, Column 2: Truncated name used in this paper, Column 3: Redshift, Column 4: Luminosity Distance (Mpc) calculated using the cosmology specified in Section \ref{sec:intro}, Column 5: Values for black hole mass are taken from \cite{runnoe16}, \cite{jin21}, and \cite{ruan16} respectively, and were measured with single-epoch scaling relations using the widths of the Balmer emission lines as measured in the high state.} 
\end{deluxetable*}

\subsection{Sloan Digital Sky Survey Observations}\label{ssec:SDSS}
All SDSS spectra were taken with the SDSS~2.5~m telescope at Apache Point Observatory \citep{gunn06}. Early spectra were taken with the SDSS spectrograph, whereas later spectra were taken with the Bayon Osciallation Spectroscopic Survey \citep[BOSS;][]{dawson13, smee13, dawson16} spectrograph. The SDSS spectrograph employed $3''$ fibers and produced spectra that cover a range 3800--9200~\AA{} with a resolving power, $R\sim2000$ \citep{york00, smee13}. Later spectra were taken with the upgraded BOSS spectrograph, which uses $2''$ diameter fibers and covers a wavelength range 3600--10400~\AA{} with $R\sim2000$.

\begin{deluxetable*}{cccccccc}
\tablecaption{CLQ Observation Log}
\tablewidth{0pt}
\label{table:dates}
\setlength{\tabcolsep}{10pt}
\tablehead{
{Object} & {State}  & {Telescope} & {MJD} & {Decimal Year} & {Exposure Time (s)}\\
{(1)} & (2) & (3) & (4) & (5) & (6)}
\startdata
{J1011} & {\textit{High}} & {SDSS} & {52652} & {2003.03} & {4800}\\
{} & {\textit{Low}} & {SDSS} & {57073} & {2015.14} & {4500}\\
{} & {\textit{Low}} & {HET} & {58231} & {2018.31} & {2400}\\
{} & {\textit{Transition}} & {HET} & {58928} & {2020.22} & {1500}\\
{} & {\textit{High}} & {MMT} & {60291} & {2023.95} & {1600\tablenotemark{a}}\\
{} & {\textit{High}} & {HET} & {60428} & {2024.32} & {2400}\\
{} & {\textit{High}} & {Swift} & {60455} & {2024.40} & {1700}\\
{} & {\textit{High}} & {HET} & {60621} & {2024.85} & {2400}\\
{} & {} & {} & {}\\
{J2333} & {\textit{Low}} & {SDSS} & {52199} & {2001.79} & {3000}\\
{} & {\textit{High}} & {SDSS} & {55447} & {2010.68} & {4500}\\
{} & {\textit{Transition}} & {SDSS} & {57722} & {2016.91} & {4500}\\
{} & {\textit{Low}} & {ARC} & {58429} & {2018.85} & {900}\\
{} & {\textit{High}} & {HET} & {60564} & {2024.69} & {2640}\\
{} & {\textit{High}} & {Swift} & {60589} & {2024.76} & {1800}\\
{} & {} & {} & {}\\
{J2336} & {\textit{High}} & {SDSS} & {52096}\tablenotemark{b} & {2001.51} & {3800}\\
{} & {\textit{Low}} & {SDSS} & {55449} & {2010.69} & {5400}\\
{} & {\textit{Low}} & {HET} & {58044} & {2017.79} & {3000}\\
{} & {\textit{Transition}} & {HET} & {58708} & {2019.61} & {3200}\\
{} & {\textit{High}} & {HET} & {60503} & {2024.53} & {3000}\\
{} & {\textit{High}} & {Swift} & {60523} & {2024.58} & {1600}\\
{} & {\textit{High}} & {HET} & {60583} & {2024.75} & {3000}\\
\enddata
\tablecomments{Column 1: Truncated SDSS object name, Column 2: State at the time of observations. ``High" refers to the high state, where broad H$\beta$ is obvious, ``transition" refers to a transition state, where H$\beta$ is declining or growing in strength, and ``low" refers to the low state, where broad H$\beta$ has diminished, Column 3: Telescope used for the obsercation. ``SDSS" refers to the Sloan Digital Sky Survey, ``HET" is the Hobby-Eberly Telescope, ``MMT" refers to MMT at the MMT Observatory, ``ARC" is the Astrophysical Research Consortium 3.5 m telescope, and ``Swift" is the Niel Gehrels Swift Observatory, Column 4: MJD of observation, Column 5: Decimal year of observation, Column 6: Exposure time of observation in seconds. Swift UVOT and XRT observations were taken simultaneously and thus had identical exposure times.}
\tablenotetext{a}{This observation was taken in two subexposures, each of 800 s.}
\tablenotetext{b}{This spectrum is the average of four exposures taken in the same high state over the course of two years. See details in \citet{ruan16,ruan19}.}

\end{deluxetable*}

All three CLQs were first observed in SDSS-I/II and were later re-observed. Following the discussion in Section 3.1 of \cite{ruan16}, we co-add the four early spectra of J2336 in order to achieve a higher signal-to-noise ratio, and take the mean date as the date observed \citep[for individual MJDs, see][]{ruan16}. All four of the early spectra were taken over a period of two years and represent the same high spectral state. We list the modified Julian dates (MJDs) of all observations in Table~\ref{table:dates}.

\subsection{Hobby-Eberly Telescope Observations}\label{ssec:HET}
We obtained new high-state optical spectra of the three CLQs using the HET \citep{ramsey98, hill21}. These spectra were obtained with the blue and red arms of the second generation low-resolution spectrograph \citep[LRS2-B and LRS2-R;][]{chonis16}. LRS2 observes with a bundle of hexagonal fibers, each of diameter $\sim0\farcs6$, that cover a large area on the sky ($\sim 12'' \times 6''$). The bundle feeds either LRS2-B or LRS2-R, both of which are double spectrographs. The `uv' and `orange' arms of LRS2-B together cover a wavelength range 3640--7000~\AA{} at $R\sim1140$--1910. The `red' and `farred' arms of LRS2-R together cover a wavelength range 6430--1056~\AA{} at $R\sim1760$--1920. We used the `uv' and `orange' channels of LRS2-B and the `red' channel of LRS2-R to produce spectra that cover an observed wavelength range 3640--8450~\AA{}. 

\subsection{Neil Gehrels Swift Observatory Observations}\label{ssec:swift}
On the basis of the HET spectra, we obtained contemporaneous target-of-opportunity X-ray observations and UV photometry from the Neil Gehrels Swift X-ray Telescope (XRT) and the UV/Optical Telescope (UVOT). Once we confirmed that the CLQs had returned to a high state, regardless of predicted X-ray brightness, we requested one spacecraft orbit of observations in the UV and X-ray to maximize scheduling flexibility. XRT is an X-ray CCD imaging spectrometer that operates between $0.2-10$ keV and has a field of view of $23\farcs6\times23\farcs6$ \citep{burrows05}. The XRT point-spread function (PSF) has an $18''$ half power diameter at 1.5 keV, which encompasses the entire CLQ and host galaxy in all cases. For our observations, the XRT operated in photon counting mode. 

UVOT is a 30-cm UV/optical telescope that is co-aligned with the XRT. It has a $17\arcmin\times17\arcmin$ field of view, and an effective plate scale of $1^{\prime\prime}$~pixel$^{-1}$ \citep{roming05}. Because the UVOT detector is a CCD operating in a photon counting mode, sometimes bright sources suffer from coincidence loss \citep{poole08, breeveld10}. Our sources are too faint for this to be a problem in our observations. We used the uvw1 filter, which has a nominal wavelength of 2600~\AA{} and a PSF FWHM of $2\farcs37$ \citep{breeveld10}.

\section{Data Processing \& Measurements}\label{sec:meas}
\subsection{HET Data Processing}
The raw LRS2 data were initially processed with \texttt{Panacea}\footnote{\url{https://github.com/grzeimann/Panacea}}, which carries out bias
subtraction, dark subtraction, fiber tracing, fiber wavelength evaluation, fiber extraction, fiber-to-fiber normalization, source detection, source
extraction, and flux calibration for each channel. The absolute flux calibration comes from response curves constructed from observations of standard stars and measures of the mirror illumination,
as well as the exposure throughput from guider images.

Some observations had multiple subexposures -- in those cases, we combined the subexposures, weighted by their inverse square errors such that the subexposure with the highest signal-to-noise ratio had the most weight. Spectra from the two arms of both the LRS2-B and the LRS2-R spectrographs also overlap in a wavelength region. For each object, we re-normalized all spectra from every spectrograph to the level of the `orange' spectrograph, using the overlapping regions, and then combined them into a single spectrum. We then corrected for telluric absorption bands from O$_2$ and H$_2$O using standard star observations and continuous atmospheric absorption. Finally, we corrected for Galactic extinction using a \cite{fitzpatrick99} extinction law and assuming $R_\mathrm{V} = 3.1$. We also corrected SDSS spectra for Milky Way extinction in the same manner.

\begin{figure*}
    \centering
    \includegraphics[width=0.75\textwidth]{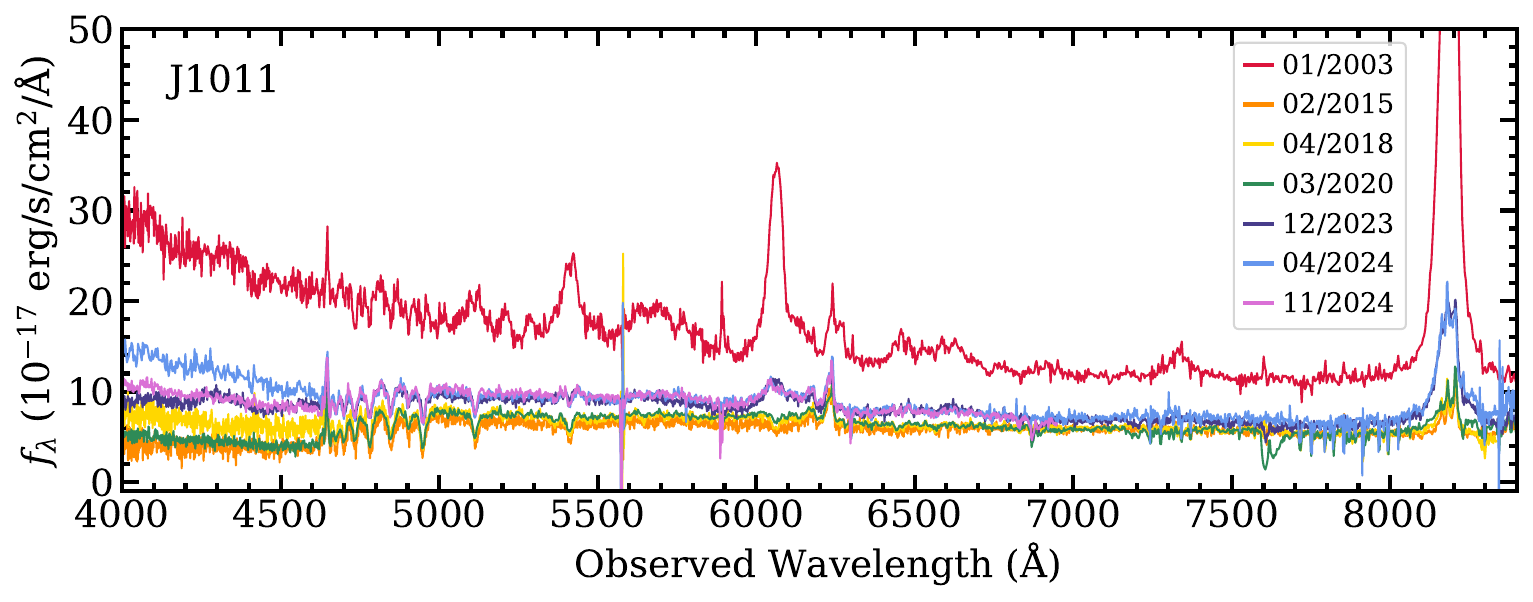}
    \includegraphics[width=0.75\textwidth]{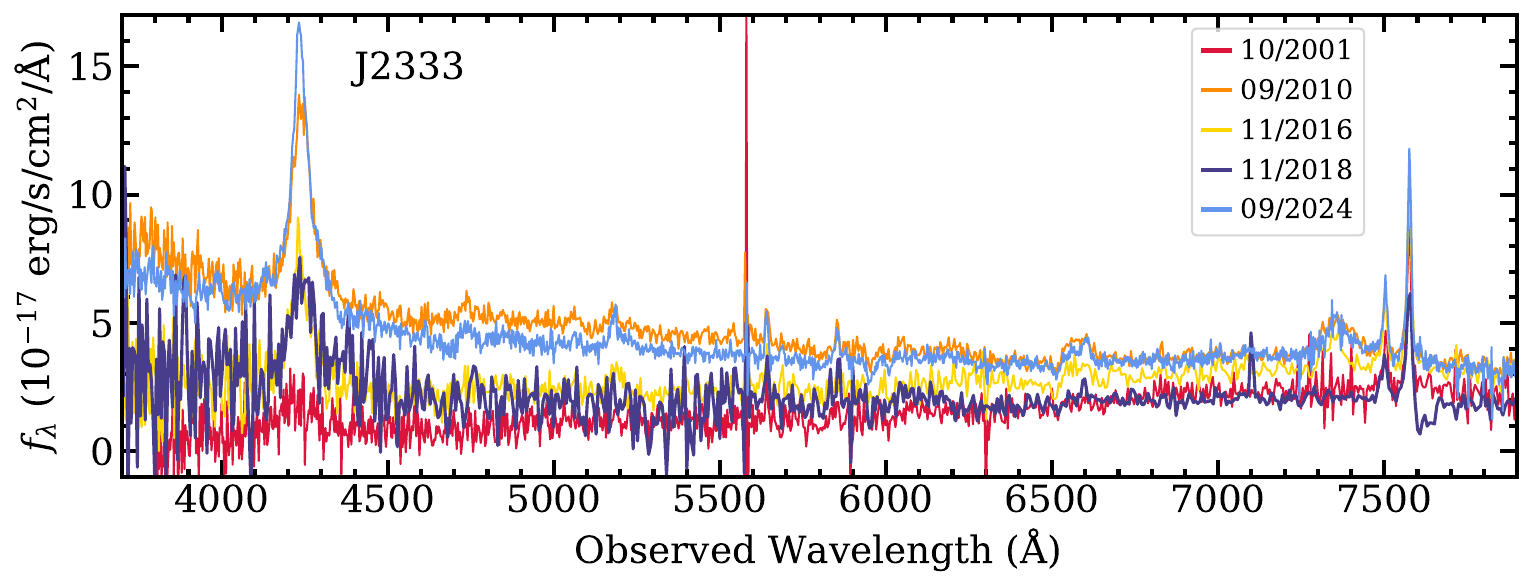}
    \includegraphics[width=0.75\textwidth]{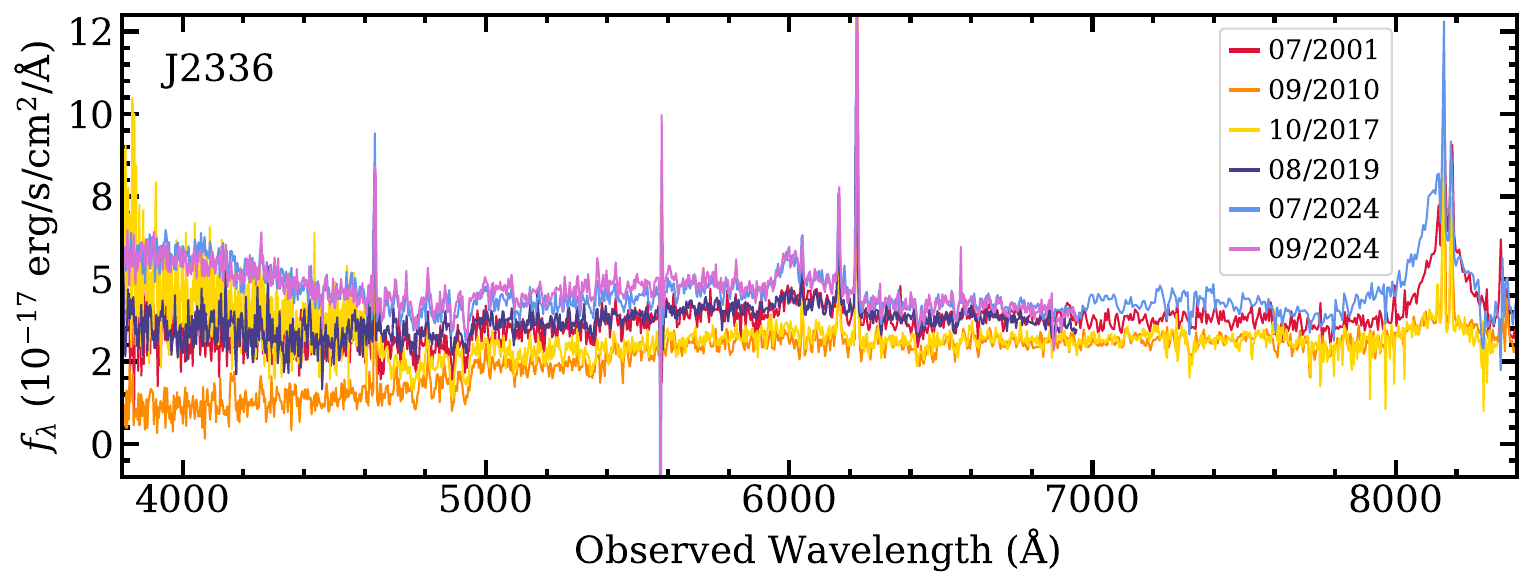}
    \caption{All observed states of J1011 (top), J2333 (middle) and J2336 (bottom). In J1011, the new high state has significant broad H$\beta$ but is notably dimmer than the previous high state. J1011 has maintained the new high state for at least the time between December 2023 to November 2024. In J2333, the new high state is nearly identical to the SDSS high state. In J2336, the new high state is higher than the original high state, and has been maintained at least between July and September of 2024. The sharp feature at an observed wavelength of 5577~\AA{} is a telluric [O~I] emission line, which was not perfectly removed from the HET spectra.}
    \label{fig:low_spec}
\end{figure*}

In order to maintain consistent flux comparisons between the high and low-states, we scaled low-state HET spectra to archival low-state SDSS spectra by measuring the [O~III] doublet flux in both the SDSS low-state and HET low-state spectra. We then re-scaled the HET spectra until the [O~III] flux matched the level from the most recent low-state SDSS observation. This step accounts for differences in the flux calibration between the HET and the SDSS spectrographs, and for the difference in the aperture sizes of the two instruments because the narrow-line region for quasars of this luminosity has typical sizes of $\sim2''$ \citep{bennert02}, and thus falls entirely within both the SDSS and the HET LRS2 apertures. We show the new HET spectra in Figure~\ref{fig:low_spec}, plotted with the most recent high and low states.

\subsection{Swift/XRT \& Swift/UVOT Data Processing}
For the XRT, we accessed level 2 data products from the UK Swift Science Data Centre. The CLQ observations have Swift target IDs 16640, 16834, 16733 for J1011, J2333, and J2336 respectively. While we detected J2333 in the X-ray, neither J1011 nor J2336 had any detected X-ray counts during the one orbit, target of opportunity exposure. Thus, we used the background counts to determine upper limits on the X-ray count rate following \cite{kraft91}. We converted the observed 2--10~keV count rates and upper limits to rest-frame 2~keV monochromatic luminosities or meaningful upper limits, $L_{2\;\mathrm{keV}}$, using the WebPIMMs tool\footnote{\url{http://cxc.harvard.edu/toolkit/pimms.jsp}} and assuming a photon index of $\Gamma=1.8$, a value commonly used for low-luminosity AGN \citep{constantin09, gu09, younes11}. Following discussion in \cite{ruan19}, we test a range of $\Gamma$ between 1.6 and 2, and confirm that variations in $\Gamma$ produce systematic uncertainties of $<5\%$. Thus, we list the line of sight column density, the count rates, and the corresponding 2 keV luminosities in Table~\ref{table:swift_obs}.

\begin{deluxetable*}{cccccc}
\tablecaption{Swift/XRT and UVOT properties of CLQs}
\tablewidth{0pt}
\label{table:swift_obs}
\setlength{\tabcolsep}{5pt}
\tablehead{
{}& {} & {} &  {X-ray}& {} & {}\\
{}& {N$_\mathrm{H}$} & {A$_\mathrm{V}$} & {Count Rate}& {$L_{2\mathrm{keV}}$} & {$L_{uvw1}$}\\
{Object} & {(10$^{19}$ cm$^{-2}$)} & {(mag)} & {(s$^{-1}$)} & {($10^{42}$ erg s$^{-1}$)} & {($10^{42}$ erg s$^{-1}$)}\\
{(1)} & (2) & (3) & (4) & (5) & (6)}
\startdata
{J1011} & {8.7} & {0.024} & {$<1.4\times10^{-3}$} & {$<2.1$} & {$34\pm2$}\\
{J2333} & {37.4} & {0.095} & {$8.7\times10^{-3}$} & {$31\pm 8$} & {$96\pm4$}\\
{J2336} & {34.7} & {0.084} & {$<1.5\times10^{-3}$} & {$<2.4$} & {$9\pm1$}\\
\enddata
\tablecomments{Column 1: Truncated SDSS object name, Column 2: H I column density reported by \cite{hi4pi}, Column 3: Milky Way line of sight A$_\mathrm{V}$ reported by \cite{schlafly11}, Column 4: Swift/XRT count rate or upper limit, calculated following \cite{kraft91}, Column 5: Rest-frame 2 keV luminosity or upper limit, Column 6: uvw1 luminosity.} 
\end{deluxetable*}

\begin{figure*}
    \centering
    \includegraphics[width=0.44\textwidth]{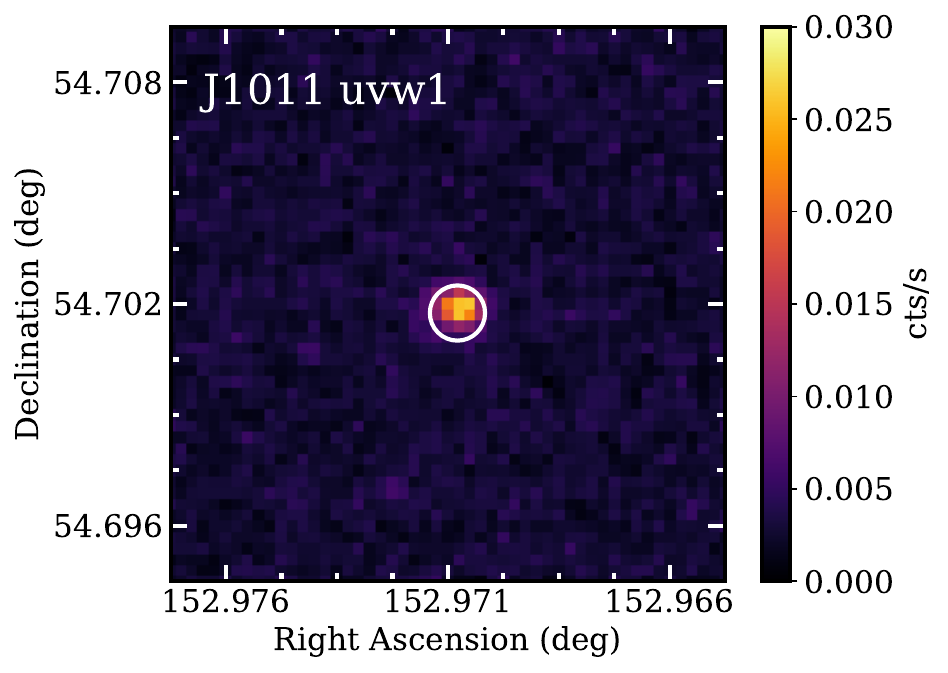}
    \includegraphics[width=0.44\textwidth]{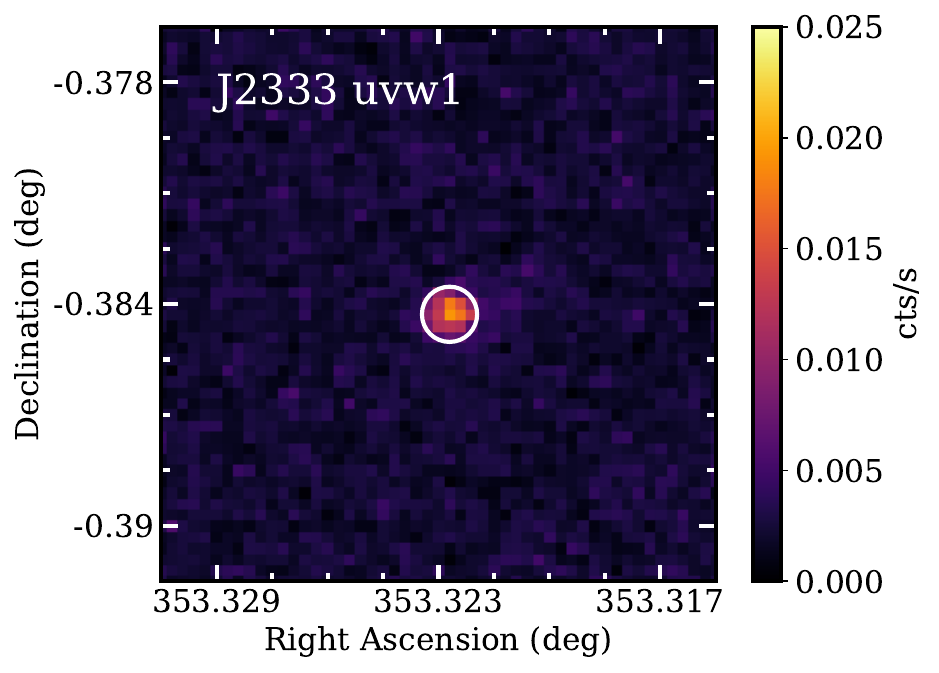}
    \includegraphics[width=0.44\textwidth]{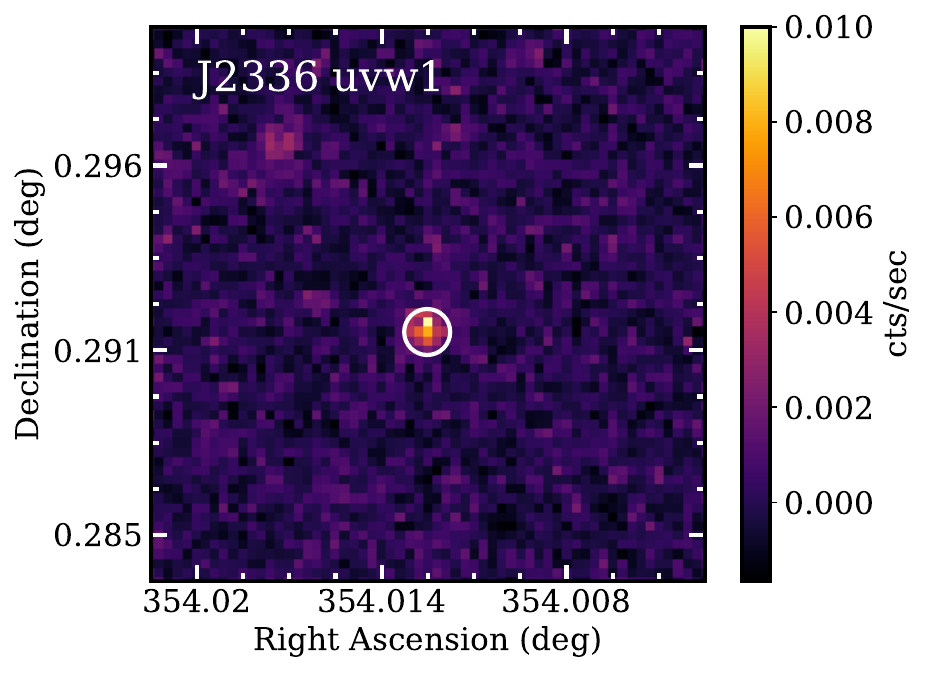}
    \caption{Swift UVOT uvw1 images of J1011 (top left), J2333 (top right) and J2336 (bottom). Overplotted in white are apertures on the nucleus of each CLQ with radii of $2\farcs37$, the size of one uvw1 resolution element.}
    \label{fig:uvw1_phot}
\end{figure*}

Contemporaneously with the XRT observations, we also exposed in the uvw1 filter with UVOT. For the UVOT data, we divided counts by the exposure map to create count rate images. We subtracted the background from the UVOT images by placing five apertures in the field, each with a radius of 10 pixels, averaged the count rate per pixel within each aperture, and then took the median of those averages and subtracted it from the image. We then corrected the images for Milky Way line of sight extinction using the dust extinction law from \cite{fitzpatrick99}. We report the values we used for V-band extinction in Table~\ref{table:swift_obs}. Because we are interested only in the flux from the unresolved point source at the center of the extended galaxy, we then placed an aperture with a radius equal to one Swift resolution element ($2\farcs37$ for the uvw1) on the center of each CLQ. This ensures that we maximize light from the unresolved point source while minimizing any flux from the extended galaxy. We show uvw1 images with these apertures in Figure~\ref{fig:uvw1_phot}. Following the conversions in \cite{poole08}, we translated the count rate into the flux, and then  flux into the luminosity. We report the uvw1 luminosity in Table~\ref{table:swift_obs}.

\subsection{Photometric Measurements}
We accessed photometric data from the Catalina Sky Survey's \citep[CSS;][]{css} second public data release by retrieving observations within 6\arcsec\ of the coordinates of our three targets. We similarly accessed IR lightcurves from the Near-Earth Object Wide-field Infrared Survey \citep[NEOWISE;][]{neowise}, which uses the WISE spacecraft \citep{wright10}\footnote{NEOWISE data can be accessed via doi:\dataset[10.26131/IRSA124]{} at \url{https://doi.org/10.26131/IRSA124}}. We selected WISE counterparts based on photometry at the location of the SDSS sources. Finally, we accessed lightcurves from ZTF data release 23 \citep{bellm19, graham19, masci19} within 1\arcsec\ of the coordinates of the CLQs\footnote{ZTF data can be accessed via doi:\dataset[10.26131/IRSA598]{} at {https://doi.org/10.26131/IRSA598}}. In all cases, we binned the resulting lightcurves by observation epoch, and took the standard deviation of measurements within the bin as the error bars on the binned measurement. For ZTF, the bins were, on average, 300 days in width. The NEOWISE bins were 100 days in width, and the CSS bins were, on average, 300 days in width.

\subsection{Optical Spectral Decomposition}
We decomposed all spectra using a modified version of the Python implementation of \texttt{pPXF} \citep{cappellari23}. Here, we describe some of the additions necessary for our purposes, but for a more detailed discussion of the modifications, see Section 3.4 of \cite{duffy25}. In order to fit spectra with significant quasar contributions well, we added templates to represent quasar power-law continuua for a range of indices. We also added templates to model the UV and optical broad Fe~II emission, higher-order Balmer lines, and the Balmer continuum. We excluded the regions around broad emission lines in our fit, and then used our implementation of \texttt{pPXF} to fit all templates simultaneously.

We first fit the low-state spectra, and removed starlight. We then used the host-subtracted spectra to make monochromatic continuum measurements before we removed any remaining contributions from the quasar, such as the non-stellar continuum and broad Fe~II. We then fit each emission line with a combination of Gaussian components, and treated the narrow and broad emission lines separately. We do not attach any physical meaning to the individual Gaussian components, but treat the combination of Gaussians corresponding to each line as a whole. In the low-state spectra, broad H$\beta$ was sometimes undetectable -- in those cases, we report upper limits on the broad-line fluxes assuming that broad H$\beta$ had the same width as broad H$\alpha$, which was detectable. For one of the spectra of J1011, broad H$\beta$ is barely detectable. In that case, we report the flux, but no other associated parameters. Once we fit the spectra, we bootstrap resampled each spectrum 1000 times by assuming the flux at each pixel of the spectrum followed a normal distribution with an amplitude equivalent to the observed flux at that pixel and a standard deviation set by the observed error at each pixel. We then resampled each pixel to create 1000 realizations of the spectrum, and performed each step of the emission line fitting again to obtain uncertainties.

\begin{figure*}
    \centering
    \includegraphics[width=0.75\textwidth]{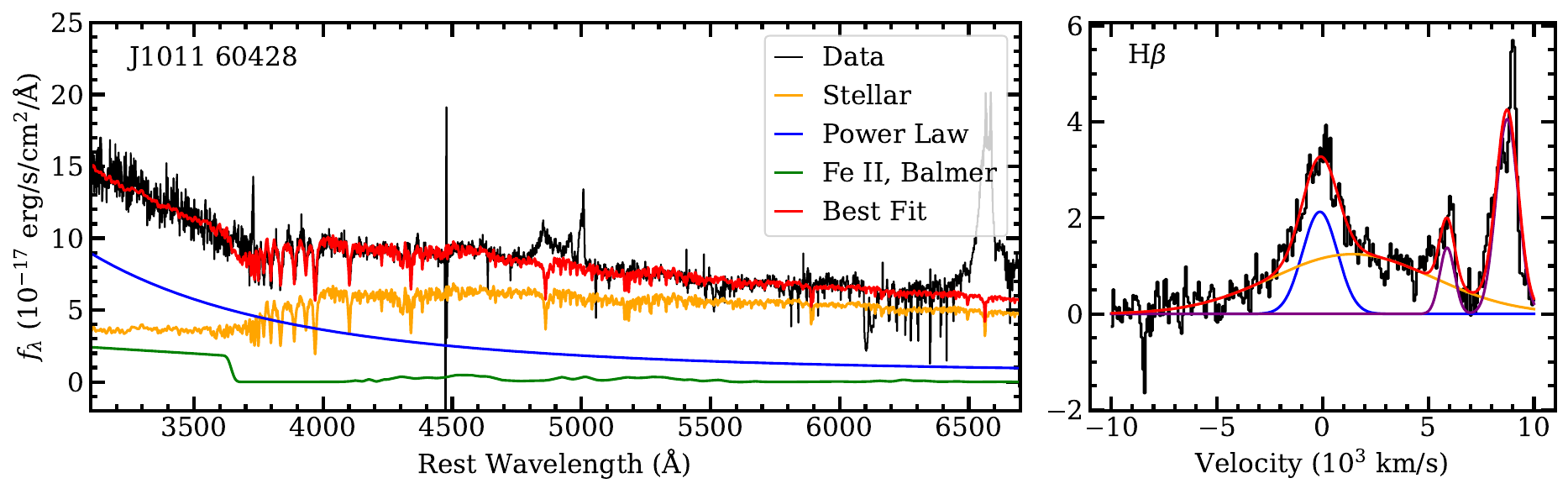}
    \includegraphics[width=0.75\textwidth]{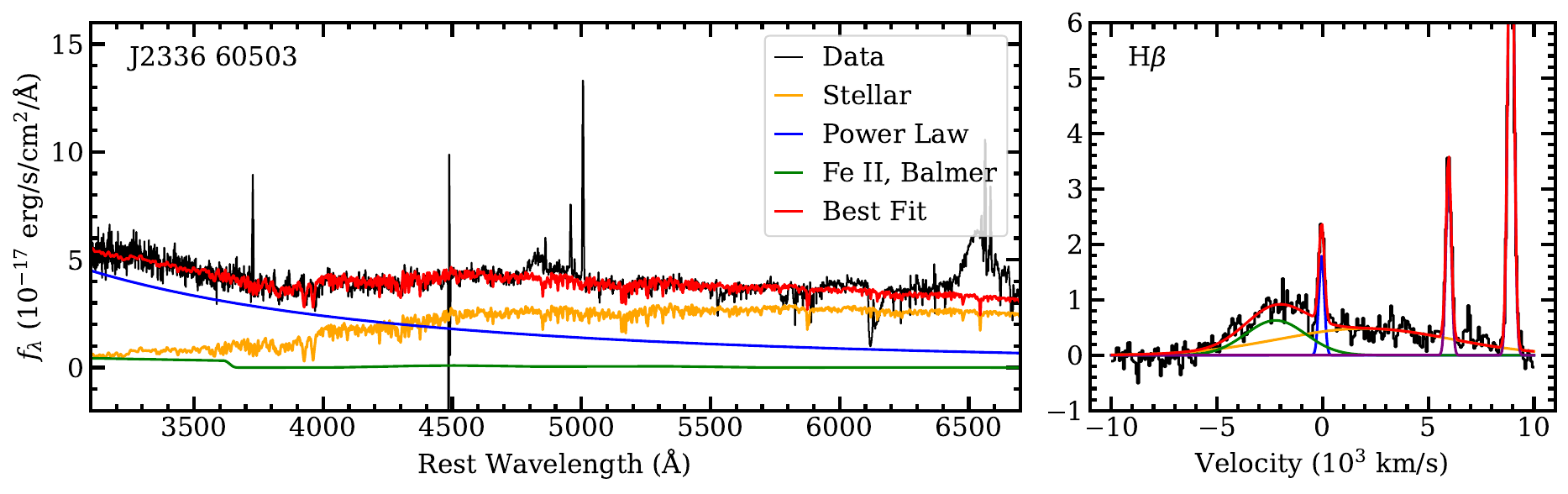}
    \includegraphics[width=\textwidth]{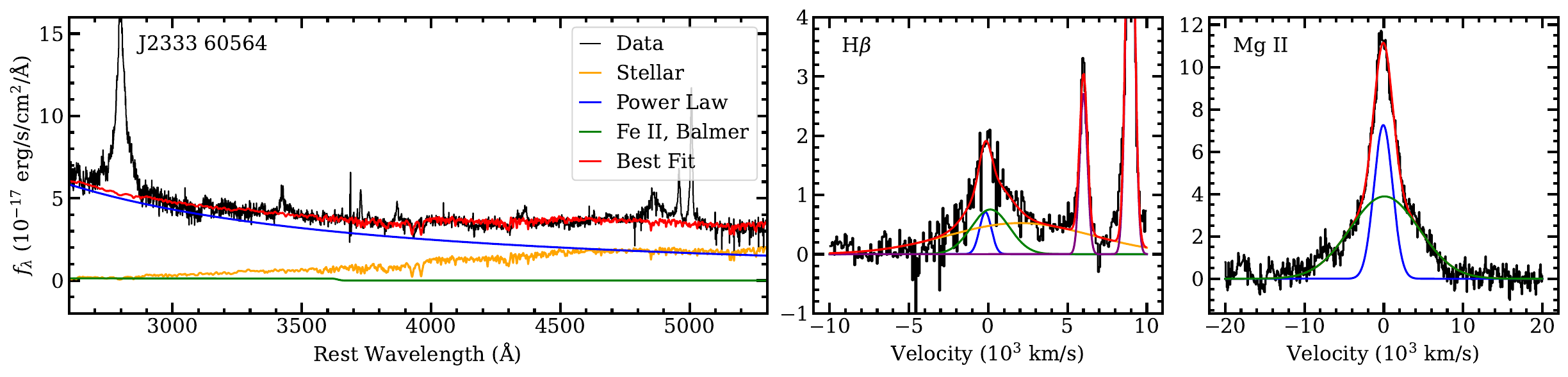}
    \caption{Example spectral decomposition for the new high states of J1011, J2336, and J2333 (left panels) -- the MJD of the example fit is given in the upper left. The observed spectrum is shown in black, the best \texttt{pPXF} fit is shown in red, the stellar population fit is shown in orange, the non-stellar continuum is shown in blue, and other quasar components such as Fe II and Balmer continuum emission are shown in green. H$\beta$ (and Mg~II, for J2333) emission line fits are shown on the right panels. J1011 is shown on the top, J2336 in the middle, and J2333 is shown on the bottom. H$\beta$ and Mg~II are fit with a combination of broad and narrow Gaussians. The total best emission line complex fits are shown in red, and individual components are shown in various colors. The purple Gaussians represent narrow [O~III].} 
    \label{fig:j1011specdecomp}
\end{figure*}

We fit high-state spectra with the same procedure. Although in many studies, the high-state stellar contributions are fixed to the low-state stellar populations, we did not impose this constraint due to the differences in aperture sizes of the SDSS fibers between observations, and the SDSS fibers with the HET. However, we compared the resulting stellar populations between spectra, and found very similar stellar populations in the high and low states. We show examples of the best-fit spectral decomposition for a representative spectrum of the new high states of each CLQ in Figure~\ref{fig:j1011specdecomp}.

To describe how the profile of H$\beta$ changes with time, we measured a number of quantities related to the emission line profile at each epoch. Following \cite{eracleous12} and \cite{runnoe15}, we measured the first four moments \citep[$\mu_1$ -- $\mu_4$; see][for the adopted definition of the moments]{eracleous12} of the emission line profile and used them to obtain the velocity dispersion, the skewness coefficient of the profile, the Pearson skewness coefficient, and the kurtosis. If the line profile is perfectly symmetric, the skewness coefficient has a value of zero, and negative values of the skewness coefficient indicate that the line profile leans red. We define kurtosis as $\mu_4$/$\mu_2^2$. The kurtosis indicates how sharply peaked an emission line profile is; a perfectly Gaussian distribution has a value of kurtosis of three, and values less than three indicate boxier profiles, whereas values greater than three indicate a more sharply peaked profile (see Section \ref{sec:blr} for discussion of physical implications). We also report the H$\beta$ emission line centroid and the full width at half maximum (FWHM) of each emission line profile. We made all measurements using the best fit broad Gaussian components to the emission line at each epoch, and used bootstrap resampling to determine the error bars. The measured values can be found in Table~\ref{table:emlines}.

For J2333, we were also able to study the broad Mg~II emission line. We calculated all the same quantities as above for Mg~II, and compared them to the values we find in H$\beta$ in that quasar.

\begin{deluxetable*}{ccccccccccc}
\tablecaption{Measured Broad Emission Line Properties}
\tablewidth{0pt}
\label{table:emlines}
\setlength{\tabcolsep}{2pt}
\tablehead{
{}& {} & {}& {Integrated} & {} & {Centroid} & {Velocity} & {} & {Pearson} & {} & {}\\
{}& {Emission}& {} & {Flux} & {FWHM} & {Shift} & {Dispersion} & {Skewness} & {Skewness} & {}\\
{Object} &{Line} & {MJD} & {(erg s$^{-1}$ cm$^{-2}$)} & {(km s$^{-1}$)} & {(km s$^{-1}$)} & {(km s$^{-1}$)} & {Coefficient} & {Coefficient} & {Kurtosis}\\
{(1)} & (2) & (3) & (4) & (5) & (6) & (7) & (8) & (9) & (10)}
\startdata
{J1011} & {H$\beta$} & {52652} & {991 $\pm$ 14} & {2280 $\pm$ 40} & {10 $\pm$ 40} & {1310 $\pm$ 70} & {$-0.15$ $\pm$ 0.14} & {$-0.08$ $\pm$ 0.03} & {4.1 $\pm$ 0.6}\\
{} & {} & {57073} & {$<14$} & {\dots} & {\dots} & {\dots} & {\dots} & {\dots} & {\dots}\\
{} & {} & {58231} & {25 $\pm$ 4} & {\dots} & {\dots} & {\dots} & {\dots} & {\dots} & {\dots}\\
{} & {} & {58928} & {115 $\pm$ 22} & {8290 $\pm$ 2350} & {$-270$ $\pm$ 750} & {3550 $\pm$ 1000} & {0.12 $\pm$ 0.32} & {0.02 $\pm$ 0.10} & {3.0 $\pm$ 0.6}\\
{} & {} & {60292} & {281 $\pm$ 7} & {2420 $\pm$ 100} & {420 $\pm$ 100} & {2080 $\pm$ 130} & {0.19 $\pm$ 0.05} & {0.04 $\pm$ 0.03} & {3.8 $\pm$ 0.1}\\
{} & {} & {60428} & {284 $\pm$ 15} & {2490 $\pm$ 180} & {850 $\pm$ 240} & {3440 $\pm$ 318} & {0.20 $\pm$ 0.05} & {0.10 $\pm$ 0.03} & {3.6 $\pm$ 0.1}\\
{} & {} & {60621} & {273 $\pm$ 7} & {2900 $\pm$ 100} & {310 $\pm$ 120} & {2430 $\pm$ 130} & {0.16 $\pm$ 0.06} & {0.04 $\pm$ 0.02} & {3.8 $\pm$ 0.1}\\
{} & {} & {} & {} & {} & {} & {} & {} & {} & {} & {}\\
{J2333} & {Mg II} & {52199} & {121 $\pm$ 22} & {10720 $\pm$ 1890} & {$-750$ $\pm$ 900} & {3180 $\pm$ 650} & {0.00 $\pm$ 0.33} & {-0.01 $\pm$ 0.10} & {3.2 $\pm$ 0.1}\\
{} & {} & {55447} & {341 $\pm$ 8} & {4290 $\pm$ 140} & {540 $\pm$ 110} & {1370 $\pm$ 50} & {0.31 $\pm$ 0.15} & {0.04 $\pm$ 0.03} & {4.0 $\pm$ 0.1}\\
{} & {} & {57722} & {305 $\pm$ 20} & {3800 $\pm$ 450} & {$-110$ $\pm$ 210} & {1910 $\pm$ 260} & {0.06 $\pm$ 0.11} & {0.01 $\pm$ 0.05} & {3.8 $\pm$ 0.4}\\
{} & {} & {58429} & {445 $\pm$ 22} & {5810 $\pm$ 400} & {2140 $\pm$ 210} & {2980 $\pm$ 210} & {0.68 $\pm$ 0.07} & {0.23 $\pm$ 0.01} & {4.1 $\pm$ 0.2}\\
{} & {} & {60564} & {583 $\pm$ 17} & {3660 $\pm$ 200} & {0 $\pm$ 110} & {2010 $\pm$ 130} & {0.04 $\pm$ 0.05} & {0.00 $\pm$ 0.03} & {4.2 $\pm$ 0.2}\\
{} & {} & {} & {}\\
{J2333} & {H$\beta$} & {52199} & {$<5$} & {\dots} & {\dots} & {\dots} & {\dots} & {\dots} & {\dots}\\
{} & {} & {55447} & {141 $\pm$ 10} & {4080 $\pm$ 300} & {800 $\pm$ 190} & {3260 $\pm$ 550} & {0.66 $\pm$ 0.12} & {0.10 $\pm$ 0.02} & {5.0 $\pm$ 0.8}\\
{} & {} & {57722} & {85 $\pm$ 15} & {2140 $\pm$ 410} & {740 $\pm$ 370} & {3890 $\pm$ 1730} & {0.66 $\pm$ 0.36} & {0.16 $\pm$ 0.06} & {5.3 $\pm$ 3.1}\\
{} & {} & {58429} & {$<15$} & {\dots} & {\dots} & {\dots} & {\dots} & {\dots} & {\dots}\\
{} & {} & {60564} & {139 $\pm$ 5} & {2560 $\pm$ 240} & {1300 $\pm$ 60} & {3570 $\pm$ 240} & {0.43 $\pm$ 0.01} & {0.18 $\pm$ 0.01} & {3.8 $\pm$ 0.1}\\
{} & {} & {} & {} & {} & {} & {} & {} & {} & {} & {}\\
{J2336} & {H$\beta$} & {52096} & {69 $\pm$ 6} & {6360 $\pm$ 580} & {370 $\pm$ 250} & {2460 $\pm$ 740} & {0.11 $\pm$ 0.23} & {$-0.03$ $\pm$ 0.04} & {2.6 $\pm$ 2.3}\\
{} & {} & {55449} & {$<29$} & {\dots} & {\dots} & {\dots} & {\dots} & {\dots} & {\dots}\\
{} & {} & {58044} & {$<21$} & {\dots} & {\dots} & {\dots} & {\dots} & {\dots} & {\dots}\\
{} & {} & {58708} & {62 $\pm$ 6} & {9480 $\pm$ 1440} & {990 $\pm$ 250} & {4530 $\pm$ 740} & {0.34 $\pm$ 0.11} & {0.11 $\pm$ 0.04} & {3.0 $\pm$ 0.4}\\
{} & {} & {60503} & {111 $\pm$ 5} & {6400 $\pm$ 710} & {$-60$ $\pm$ 250} & {3400 $\pm$ 260} & {0.6 $\pm$ 0.06} & {0.17 $\pm$ 0.02} & {2.9 $\pm$ 0.1}\\
{} & {} & {60583} & {117 $\pm$ 4} & {4890 $\pm$ 430} & {$-500$ $\pm$ 190} & {2530 $\pm$ 170} & {0.56 $\pm$ 0.11} & {0.13 $\pm$ 0.03} & { 3.0 $\pm$ 0.2}\\
\enddata
\tablecomments{Column 1: Truncated SDSS object name, Column 2: Emission line for which the measurements were made, Column 3: MJD of observation, Column 4: Broad emission line flux, Column 5: Broad emission line FWHM, Column 6: Centroid wavelength of the broad emission line, Column 7: Velocity dispersion of the broad emission line, Column 8: Skewness coefficient of the broad emission line, Column 9: Pearson skewness coefficient of the broad emission line, Column 10: Kurtosis, which characterizes the boxiness, of the broad emission line. Values $>3$ are cuspier than a Gaussian distributions, and values $<3$ are boxier than a Gaussian distribution, Columns 5-10: If there was no significant flux detection, we do not measure any further parameters for the line.} 
\end{deluxetable*}

\section{Analysis}\label{sec:analysis}
\subsection{Dust Extinction}
Following \cite{duffy25}, we conducted a series of tests to ensure that the variability we see does not result from dust extinction alone. We measured the flux in broad H$\beta$ in both states and calculated the V-band extinction, $A_\mathrm{V}$, required to go from the original high-state value to the original low-state value, and to go from the new high-state to the most recent low-state. We did the same calculation using the non-stellar continuum level at $5100$~\AA{} and $3300$~\AA{} and, in J2333, with Mg~II. For these calculations, we assumed a \cite{cardelli89} extinction law with $R_\mathrm{V}=3.1$. Although various attenuation and extinction laws differ substantially in the UV, they are very similar in the optical \citep[see][for review]{salim20}. Thus, the choice of extinction law has a negligible effect on our calculated values and does not impact our results. If variable dust obscuration was causing the observed behavior, the $A_\mathrm{V}$ values calculated for each CLQ should be self-consistent. In this case, however, we find that dust extinction alone is not sufficient to explain the spectral evolution we see in any of the three CLQs.

\cite{duffy25} find that dust extinction alone is unable to explain the initial drop from the high state to the low state for J1011. For later epochs of observations of J1011, we find that the $A_\mathrm{V}$ necessary to redden the 2024 high state to the level of the 2018 low state, when calculated using broad H$\beta$, is $2.3\pm0.2$ whereas if the same value is calculated using the continuum at $3300$~\AA{}, an $A_\mathrm{V}$ of $0.5\pm0.1$ is required -- these values are not consistent and thus a change in dust attenuation alone is not sufficient to explain the return of J1011.

To redden broad H$\beta$ in the 2024 high state of J2333 to the level of the 2017 low state, an $A_\mathrm{V}$ value $>2.1$ is required whereas, if we repeat the calculation for the flux at 4200~\AA{}, a value of $0.2 \pm 0.1$ is all that is required. This is also not self-consistent, and thus a change in dust attenuation cannot explain the return of J2333.

We reach the same conclusion for J2336. \cite{duffy25} find that dust extinction is also unable to explain the transition from a high state to a low state, and we find that a change in extinction alone is also insufficient to explain the transition to a new high state. When values are measured using broad H$\beta$ in the 2024 high state and the 2018 low state, a value of $A_\mathrm{V} > 1.6$ is required, whereas for the change in the continuum at 4200~\AA{}, the value required is $0.4\pm0.2$. Thus, a change in dust attenuation is also unlikely to explain the behavior of J2336.

\subsection{Continuum Light Curves}

\begin{figure*}
    \centering
    \includegraphics[width=0.44\textwidth]{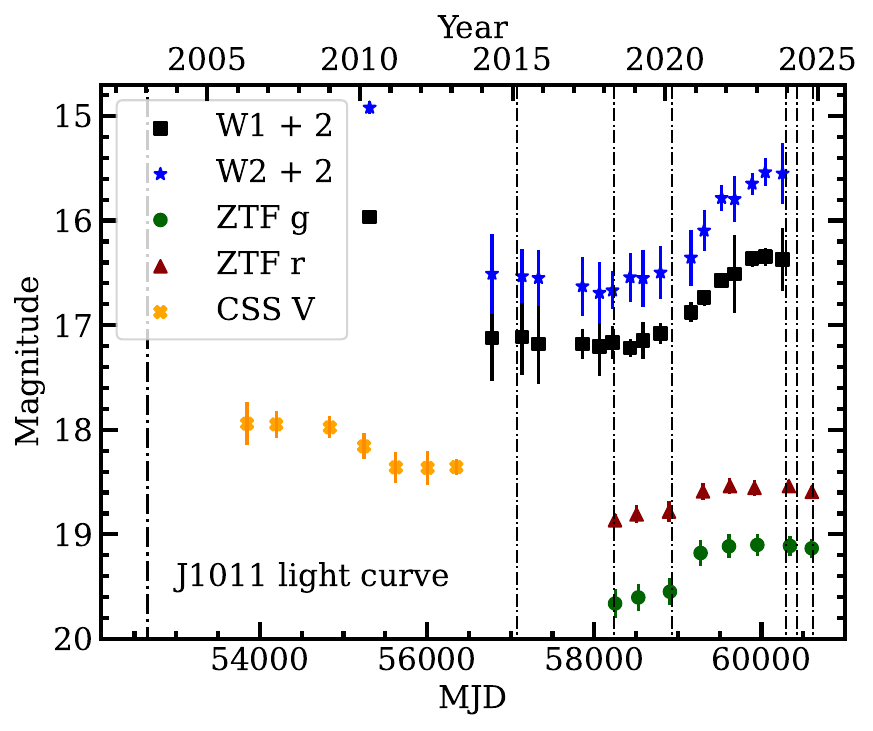}
    \includegraphics[width=0.44\textwidth]{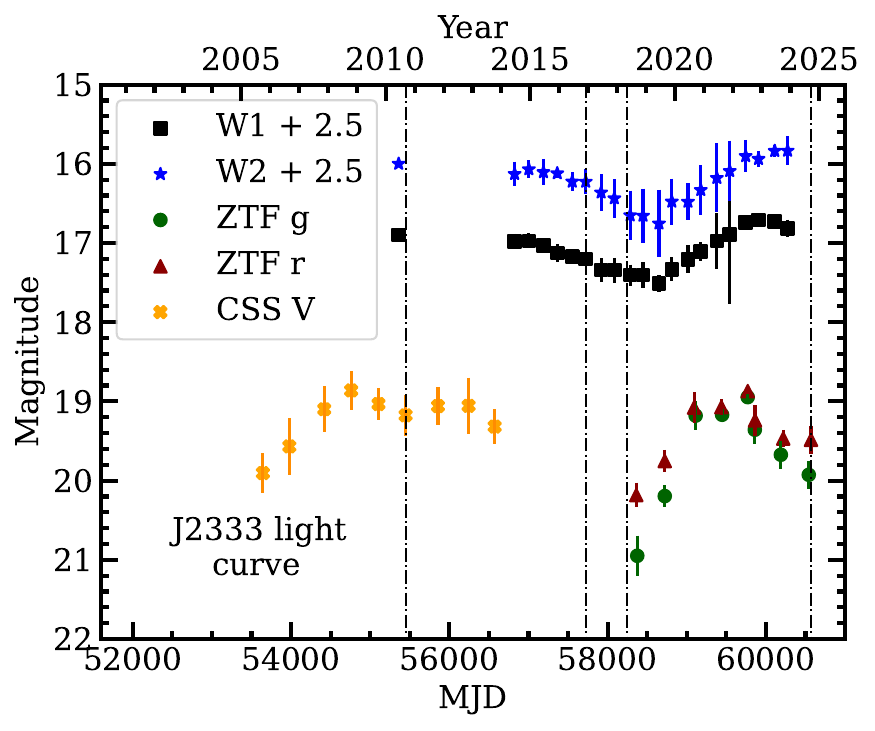}
    \includegraphics[width=0.44\textwidth]{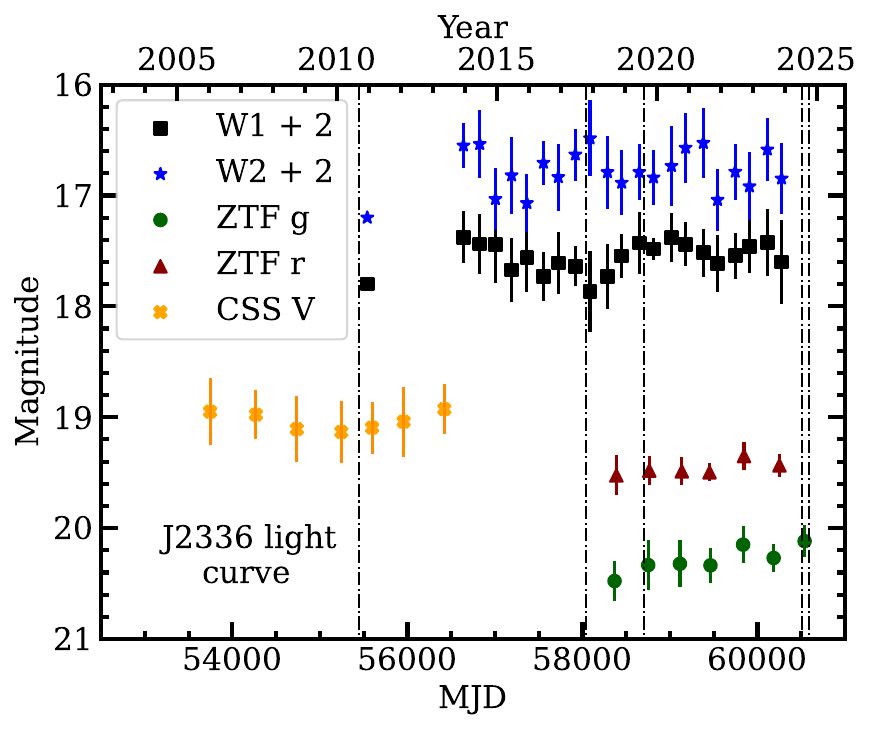}
    \caption{Light curves for J1011 (top left), J2333 (top right), and J2336 (bottom) featuring data from CSS (orange xs), WISE or NEOWISE W1 (blue stars) and W2 (black squares), and ZTF g (green circles) and r (red triangles) bands. We mark the dates of spectra within the range of dates shown with vertical dot dashed lines. All light curves are binned by observation epoch, and the error bars represent the standard deviation within the bin. Both J1011 and J2333 display coordinated behavior between the IR data from NEOWISE and the continuum traced by ZTF, whereas the changes in J2336 are more complex.}
    \label{fig:j1011_lc}
\end{figure*}
We compared the optical and IR lightcurves of each of the three CLQs over the times of observation to both constrain the duration of the low and high states and to characterize the transition time scale. 

\begin{figure*}
    \centering
    \includegraphics[width=\columnwidth]{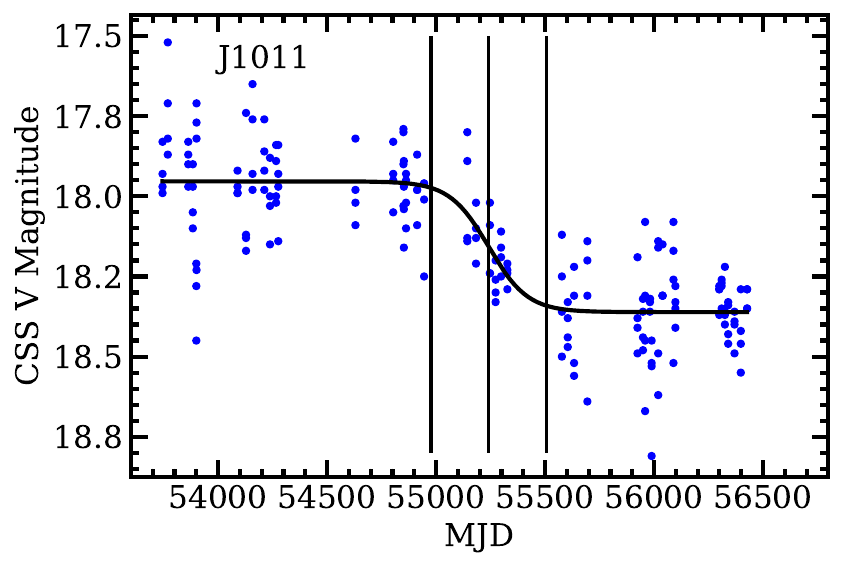}
    \includegraphics[width=\columnwidth]{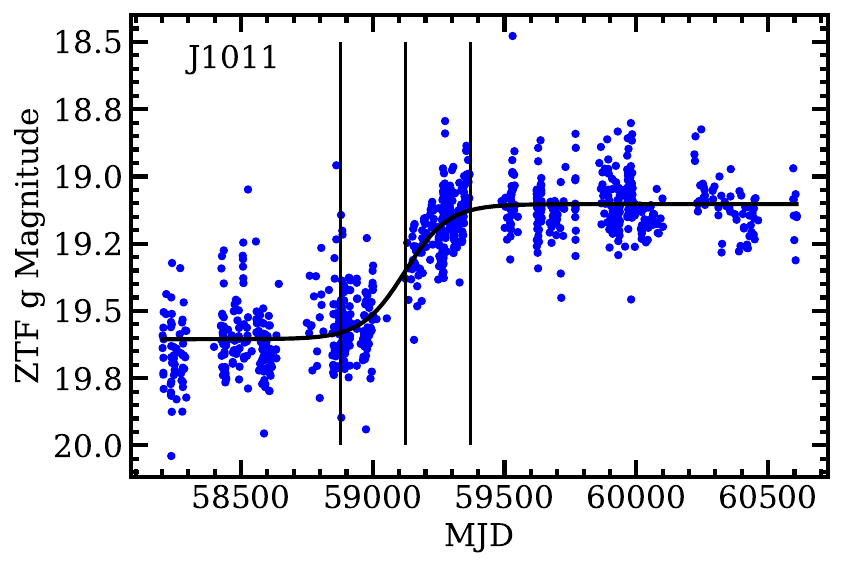}
    \includegraphics[width=\columnwidth]{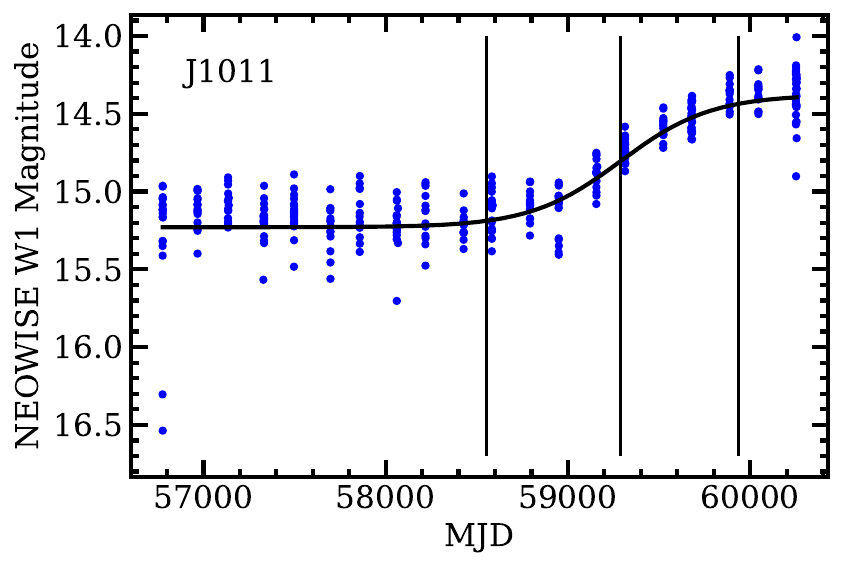}
    \includegraphics[width=\columnwidth]{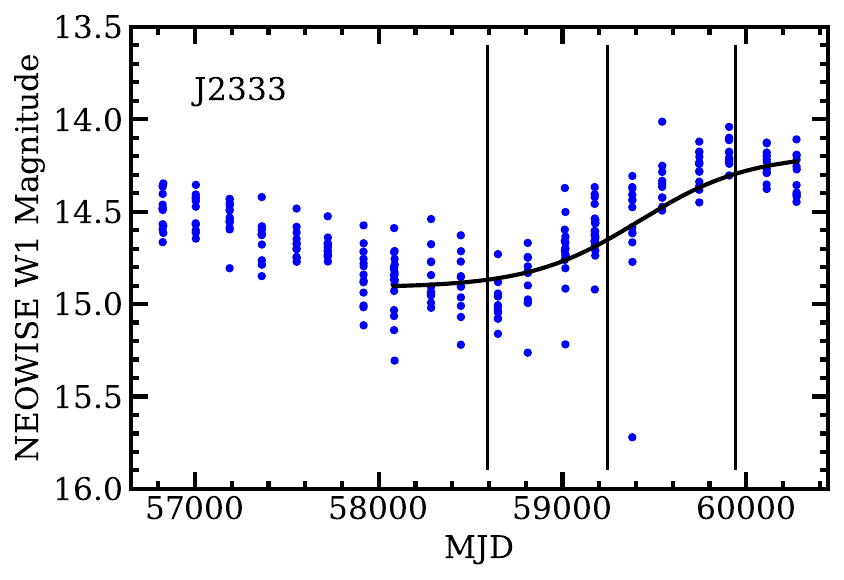}

    \caption{J1011 light curves (blue points) from CSS (top left), ZTF (top right) and NEOWISE (bottom left) fit with best fit sigmoid functions (shown as the black curve). We show the J2333 NEOWISE light curve (bottom right), also with the best fit sigmoid function. We also place vertical black lines to mark the center of the sigmoid, the place where the transition has either increased or decreased the flux by 10\% and the place where the flux has reached 90\% of its final value. For J1011, we note that the transition to a faint state traced by the CSS light curve takes almost exactly the same amount of time (528 days) as the transition to another high state, traced by ZTF (493 days). In comparison, the transition traced by the NEOWISE W1 is much longer (1380 days).}
    \label{fig:sigmoids}
\end{figure*}

J1011 appears to make a transition in flux between MJD $\sim 59000$ and MJD $\sim 59400$ (see Figure~\ref{fig:j1011_lc}). We also observe a transition in the NEOWISE IR lightcurve, which may lag behind the optical transition \citep[see also][]{lyu22, lyu24}. We fit the CSS, ZTF, and NEOWISE photometry with sigmoid functions to better constrain the time spent in transition (see Figure~\ref{fig:sigmoids}). We consider the time of transition to be the time taken to go from within 10\% of average maximum flux to within 10\% of the average minimum flux and vice versa. For the high to low transition, we find a transition timescale of $528\pm122$ days and for the low to high transition, we find a timescale of $493\pm45$ days. These values are strikingly similar -- the CLQ spends almost exactly the same amount of time transitioning to a low state as it does transitioning to the high state. We also find that J1011 spent about 10.6 years in the low state before returning to the high state, taken as the time between the middle of transition in the CSS lightcurve to the middle of the transition in the ZTF lightcurve.


When we fit NEOWISE photometry with a sigmoid function, we found that the duration of the W1 transition is $1380\pm110$ days -- approximately three times longer than the associated optical transition time. The central date for the IR transition lags 166 days behind the central date for the optical transition.

Although the lightcurves do not cover the transitions as nicely as in the case of J1011, we also fit the IR light curves of J2333 in the return to the high state (see Figure \ref{fig:sigmoids}), and note the location where the optical light curve reaches 90\% of its maximum value. As in J1011, the IR photometry appears to lag behind the optical photometry. In this case, there are $\sim800$ days between the return to a bright state in the optical and the return in the IR, taken as the difference between the time where the optical returns to 90\% of its maximum and the time where the IR returns to 90\% of its maximum. Additionally, the optical transition appears to take around 750 days, if we mark the earliest possible date the transition starts as the date where the latest low state spectrum was taken, at MJD 58429, whereas the IR transition duration is almost twice as long at $1350\pm65$~days.

J2336 does not show the same coordinated behavior in either the IR or the optical light curves. The low-state SDSS spectrum appears to have been observed at the minimum brightness that the CSS light curve reaches, but the same sort of analysis cannot be done because none of the light curves appear to capture the totality of the transitions.

\subsection{Emission Line Variations and their Relation to Continuum Variations}
\begin{figure*}
    \centerline{\hbox{
    \includegraphics[width=0.34\textwidth]{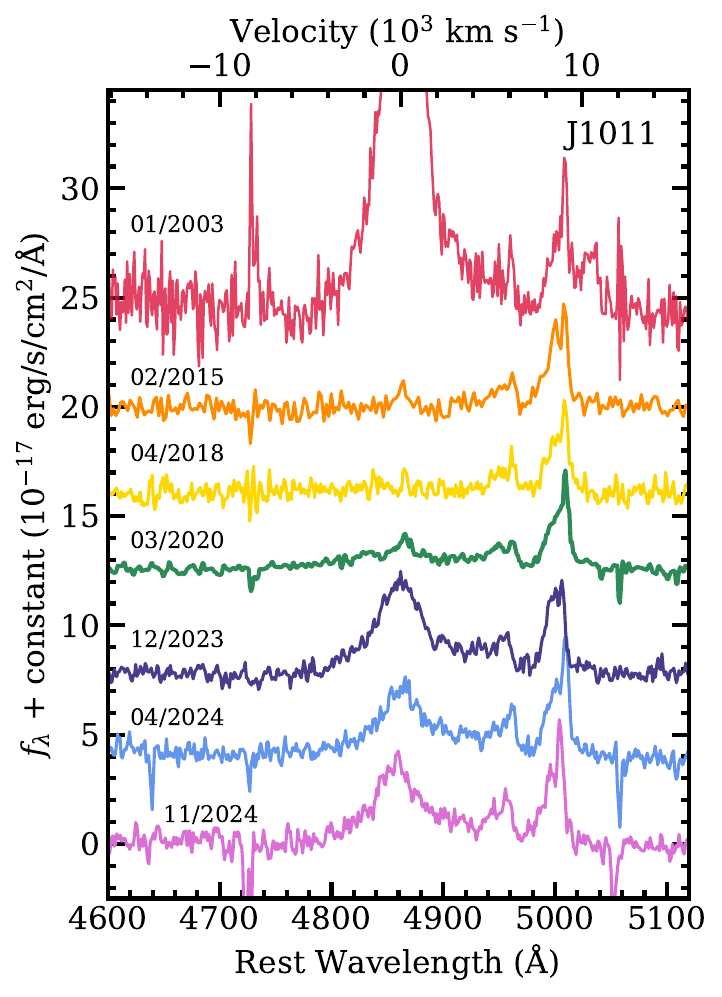}
    \includegraphics[width=0.34\textwidth]{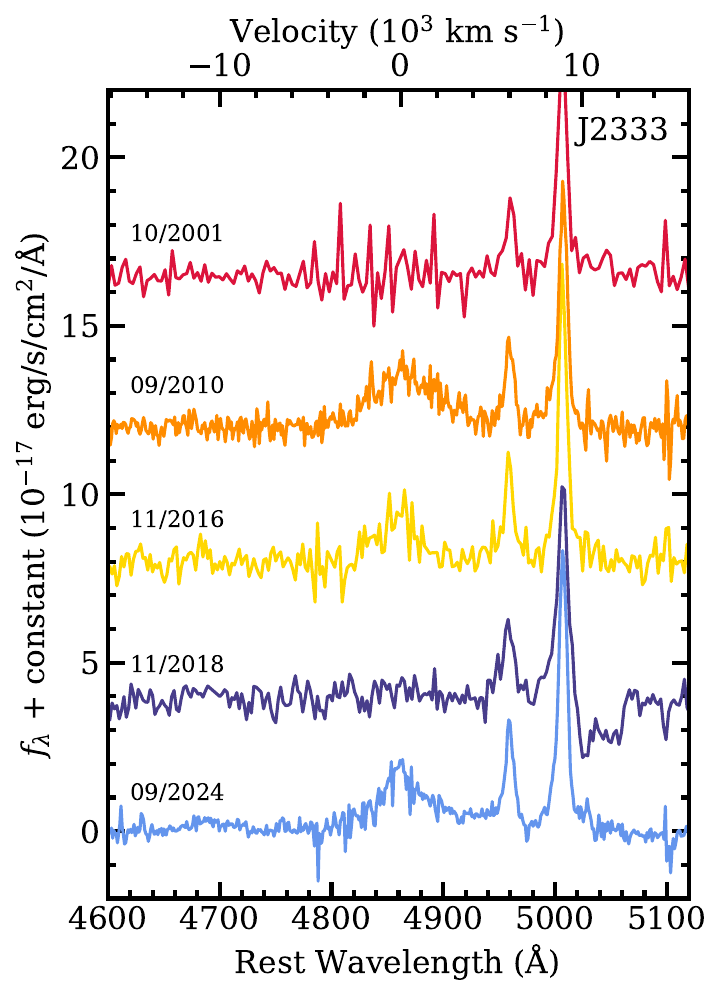}
    \includegraphics[width=0.34\textwidth]{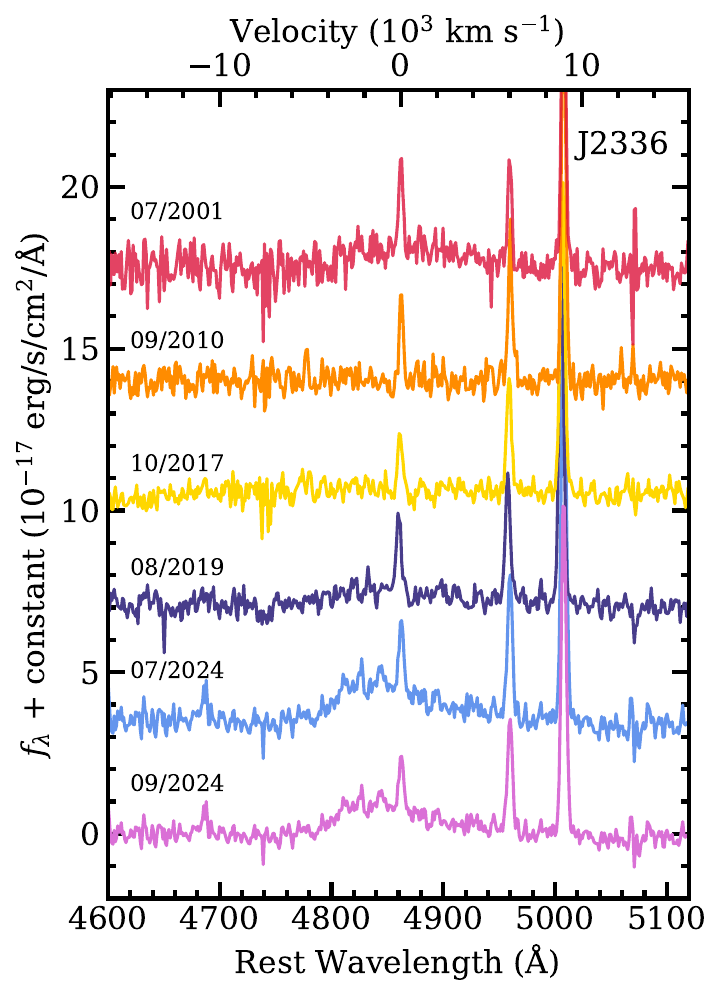}
    }}
    \caption{Spectrally decomposed H$\beta$ profiles of J1011 (left), J2333 (center) and J2336 (right), including the narrow [O~III] doublet. In all three, we show the evolution of the H$\beta$ profile since initial identification, up until the new high state. In both J1011 and J2336, the profiles have evolved from being nearly indiscernible in the initial low state to some higher level in the most recent state. Also in both cases, between the low and new high states, there is a `transition' in the profile of H$\beta$, where it goes from being undetectable to just detectable. For J2333, there was originally no detectable H$\beta$, followed by a rise to a high state, a fall back to a low state, and a new return to a high state.} 
    \label{fig:hbeta}
\end{figure*}

\begin{figure}
    \centering
    \includegraphics[width=0.75\columnwidth]{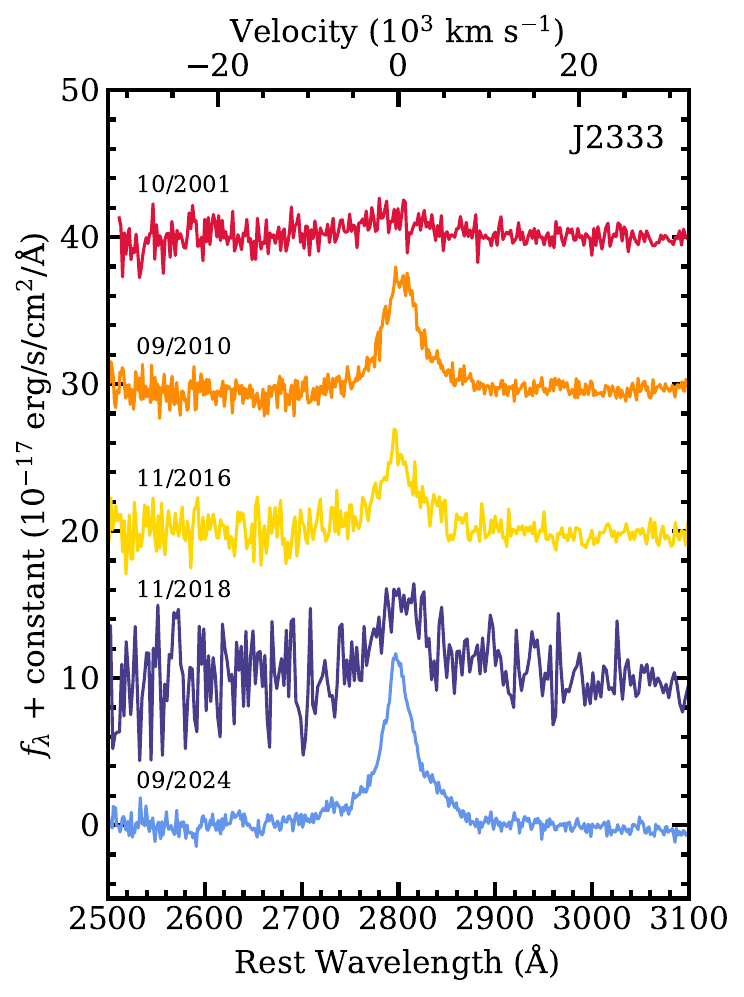}
    \caption{Spectrally decomposed Mg~II profiles of J2333. Even when broad H$\beta$ is undetectable, broad Mg~II is still measurable. The profiles, when both are detectable, are remarkably similar in shape.} 
    \label{fig:mgii}
\end{figure}

\begin{figure*}
    \centering
    \includegraphics[width=0.4\textwidth]{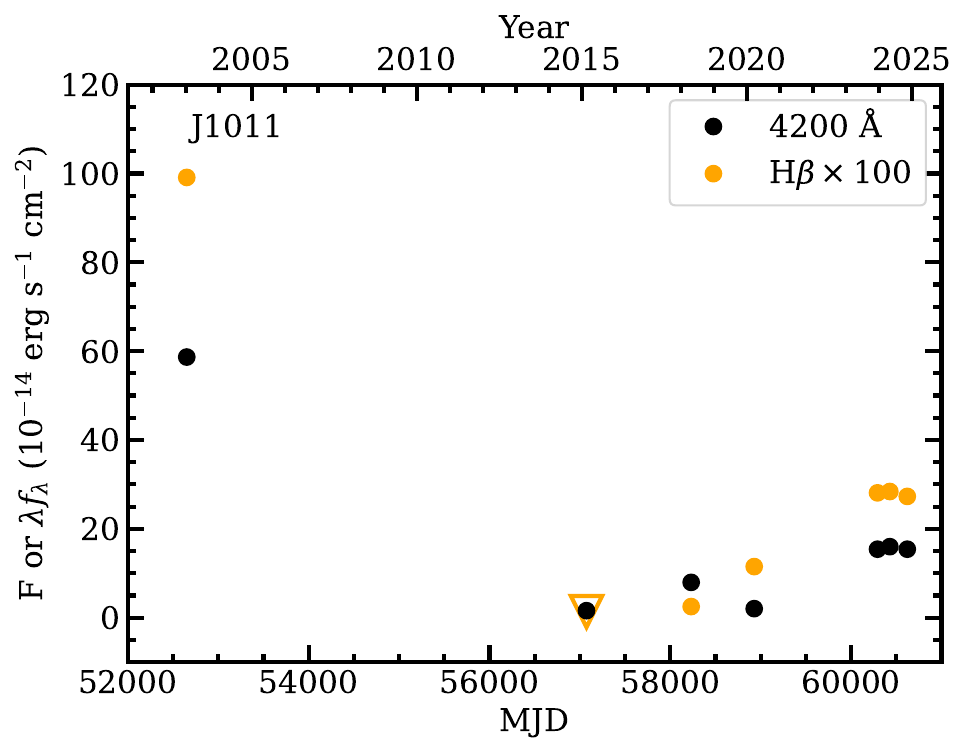}
    \includegraphics[width=0.4\textwidth]{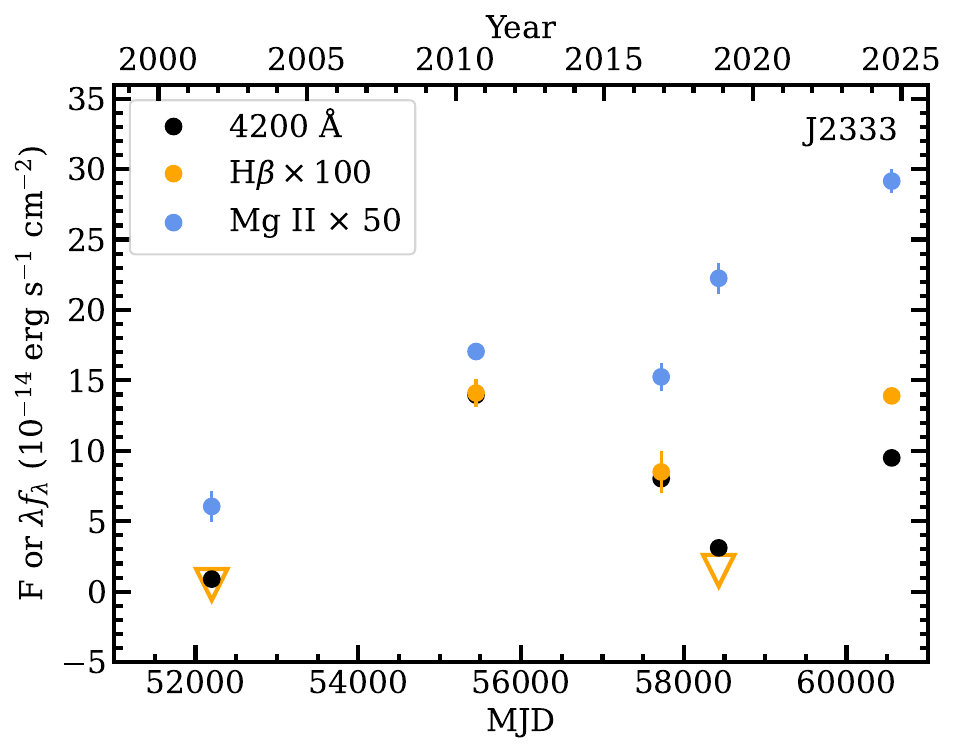}
    \includegraphics[width=0.445\textwidth]{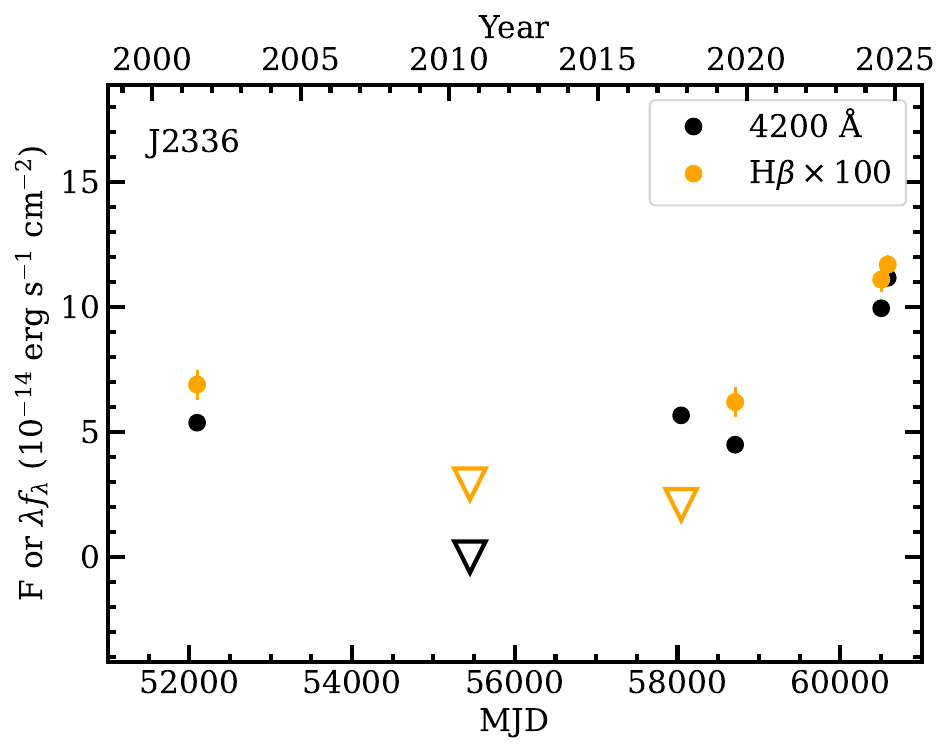}
    \caption{Evolution of broad H$\beta$ (orange points) and Mg~II (blue points) flux and 4200~\AA{} continuum level (black points) over time for J1011 (top left), J2333 (top right) and J2336 (bottom) respectively. Filled points represent measurements, and the hollow inverted triangles represent upper limits on the broad line flux. In all cases, broad H$\beta$ flux has recovered since the initial or most recent low state. In the case of J2336, the recovered high state is, in fact, \textit{brighter} than the initial high state. In all three cases, the continuum strength and broad H$\beta$ flux do not vary in concert. We use these measurements to constrain the maximum time spent in transition.}
    \label{fig:lightcurve_hb}
\end{figure*}

All three quasars have exhibited both continuum and broad emission line changes. We show the progression of the H$\beta$ profiles of all quasars over time in Figure~\ref{fig:hbeta}, and the progression of Mg~II in J2333 in Figure~\ref{fig:mgii}. Interestingly, in J1011 the short-wavelength continuum increases between 2015 and 2018, but decreases again in the 2020 observation (see Figure~\ref{fig:lightcurve_hb}, top and Figure~\ref{fig:low_spec}, top). In the 2020 observation, however, H$\beta$ has begun to return despite the weakness of the blue continuum. Additionally, the profile of H$\beta$ in J1011 is significantly boxier in 2020 than in either the first high state H$\beta$ profile or the 2024 returning profile (see Table \ref{table:emlines}). 

In J2333, the newest observation shows a return of H$\beta$ to the level of the 2010 high state. Finally, in J2336, the spectrum taken in 2019 also shows measurable H$\beta$ flux for the first time since the initial SDSS high-state observations. It is still much weaker than in the 2024 HET spectrum, however.

We compare the flux in the broad H$\beta$ over time, and plot it as a time series in Figure~\ref{fig:lightcurve_hb}. In the same figure, we also plot the evolution of the continuum at 4200~\AA{}. In both J1011 and J2336, broad H$\beta$ recovers substantially since the initial low state observation. In J2336, the broad H$\beta$ flux eventually becomes even higher than the initial high-state observation. It took J2336 between 700 and 2500 days to fully recover from the low state. In this case, where we are unable to rely solely on optical broadband light curves to constrain the transition, we find that the low state had a \textit{minimum} duration of $\sim2500$ days, measured as the time between the first low state spectrum and the final low state spectrum. We also see that in J1011 and J2336, the continuum does not perfectly track the evolution of broad H$\beta$ flux, particularly between the 2018 and 2020 observations for J1011 and between the 2017 and 2019 observations for J2336.

J2333 shows more structure in its light curves. Broad H$\beta$ appears, then drops and vanishes before re-appearing again in the most recent epoch. Here, the continuum and broad H$\beta$ trace each other almost perfectly, but the Mg~II broad flux varies differently than the broad H$\beta$ flux. This difference may be a result of the Mg~II emitting region being more extended than the H$\beta$ region, meaning that the delay between continuum fluctuations and Mg~II fluctuations is longer than for H$\beta$ \citep{guo19, guo20} and that the response is smoother and smaller in amplitude than in H$\beta$ (see Figure~\ref{fig:lightcurve_hb}). Mg~II thus traces a continuum state that is further back in time than the state being traced by H$\beta$ or the continuum itself, although our observations are too sparse to estimate the size of the Mg~II emitting region. Using the broad H$\beta$ light curve that we made here, the maximum duration of J2333's most recent low-state is $\sim3000$ days.

\subsection{$\alpha_{ox}$ and the Eddington Ratio}
For the new 2024 observations, we calculated values of $\alpha_{ox}$ \citep{tananbaum79}, the spectral index of a power law connecting the points in the SED at 2500~\AA{} and 2~keV, using Swift/XRT and UVOT uvw1 data. The uvw1 filter probes the luminosity at rest-frame wavelengths of 2086~\AA{}, 1800~\AA{}, and 2091~\AA{} for J1011, J2333, and J2336 respectively. The new, high-state HET spectra only reliably extend down to rest-frame wavelengths of around 3000~\AA{}, 2500~\AA{}, and 3000~\AA{}, respectively. As a result, we elected to use the uvw1 luminosity in place of the 2500~\AA{} luminosity for J1011 and J2336. This adds a small systematic uncertainty into our calculation, but the difference between the flux at 2500~\AA{} and the flux around 2000~\AA{} is minimal. When we instead calculated $\alpha_{ox}$ using a value for the 2500~\AA{} luminosity inferred from the power law fit to the observed spectra, we found values of $\alpha_{ox}$ that differed from the values calculated using UVOT by less than $0.1$. In J2333, the effect of using the uvw1 flux instead of the flux at 2500~\AA{} from the optical spectra has a similarly small effect. Using the flux at shorter wavelength means that we may overestimate the 2500~\AA{} flux, which in turn would reduce the value of $\alpha_{ox}$. This does not affect our conclusions, and thus we use the values calculated using UVOT measurements for J1011 and J2336.

Two of our three CLQs were not detected in the X-ray band during our Swift/XRT exposures. We thus used the upper limits on the X-ray count rate to obtain an upper limit on the 2 keV luminosity, and used that limit to calculate a lower limit on $\alpha_{ox}$. We report $\alpha_{ox}$ values for the 2024 observations, and compare them with values for previous epochs from the literature in Table~\ref{table:aox}.

We calculated bolometric Eddington ratios using the conversion from \cite{runnoe12} between 5100~\AA{} luminosity and the bolometric luminosity for each state. For each of the three CLQs, we used black hole mass estimates calculated from broad line widths and continuum levels, presented in \cite{runnoe16, jin21} and \cite{ruan16} respectively. Eddington ratios can be found in Table~\ref{table:aox}. In states where contemporaneous X-ray observations are available, we calculated the Eddington ratio using the shape-dependent bolometric corrections from \cite{lusso10}, and report that value instead. Those corrections take into account the observed X-ray-to-UV hardness as parameterized by $\alpha_{ox}$, and thus should provide a more accurate measure of the bolometric luminosity. Values calculated using the two different methods, where both are feasible, agree within error bars \citep[see discussion on difference in bolometric corrections in][]{duffy25}.

\begin{figure}
    \centering
    \includegraphics[width=\columnwidth]{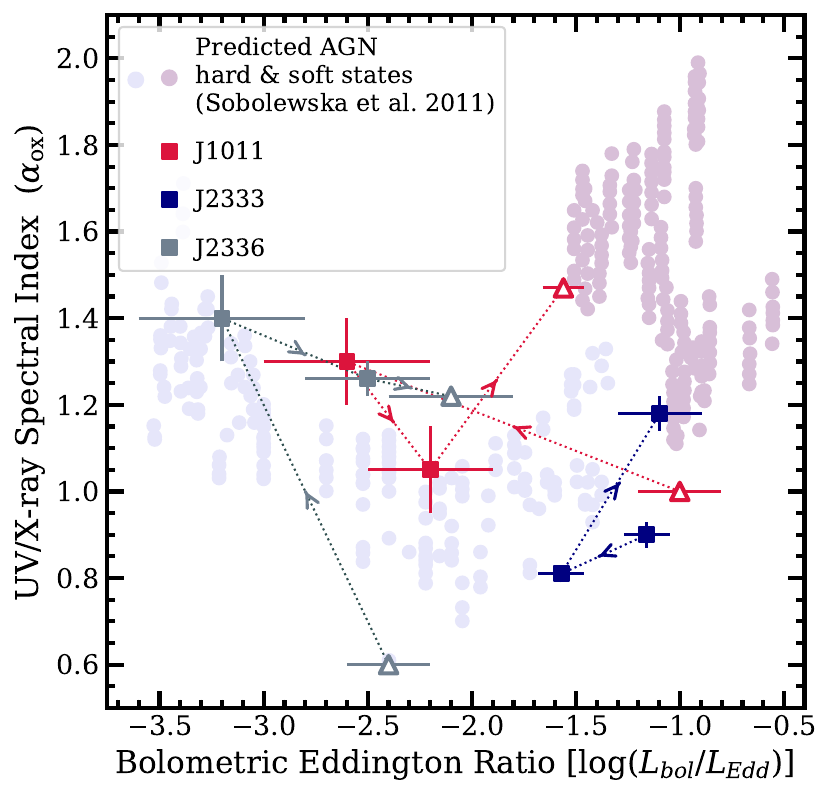}
    \caption{We plot $\alpha_{ox}$ against the bolometric Eddington ratio for J1011 (red squares and hollow triangles), J2333 (navy squares) and J2336 (dark grey squares and hollow triangles). Hollow triangles represent lower limits on $\alpha_{ox}$ in the case of X-ray non-detection. Dashed lines with arrows represent the time evolution of the three CLQs. The light purple and mauve points in the background are points from \cite{sobolewska11} that predict AGN accretion state transitions from a low, hard state (light purple) to a high, soft state (mauve) based on galactic XRBs. All three CLQs evolve along predicted tracks. In particular, J1011 traces the V-shape as it transitions back up to its new high state. } 
    \label{fig:aox}
\end{figure}

We plot the Eddington ratio against $\alpha_{ox}$ in Figure~\ref{fig:aox} for each state where there exist near contemporaneous X-ray and UV/optical observations. We connect states with a dotted line and arrows to show temporal evolution in the diagram. We compared our calculated values to predicted values for a low Eddington ratio, X-ray hard accretion state and a high Eddington ratio, X-ray soft accretion state from \cite{sobolewska11}. Those predictions were made by scaling the SED behavior of an XRB outburst to the level of an AGN -- individual objects are expected to move along the two branches on the plot as they change in Eddington ratio. Qualitatively, all three CLQs follow the expected tracks in $\alpha_{ox}$ vs. $L_{\rm bol}/L_{\rm Edd}$. J1011 in particular evolves from one side of the predicted V to the other -- going from a high/soft state to a low/hard state and back again to a high/soft state in the most recent observation. J2333 and J2336 also evolve along expected paths in this diagram.

\begin{deluxetable}{cccc}
\tablecaption{Calculated properties of CLQs}
\tablewidth{0pt}
\label{table:aox}
\setlength{\tabcolsep}{8pt}
\tablehead{
{}& {}& {} & {}\\
{Object} &{MJD}&{$\alpha_{ox}$}  & {$\log(L_{bol}/L_{Edd})$}\\
{(1)} & (2) & (3) & (4)}
\startdata
{J1011} & {52652$^a$} & {$>1.0$} & {$-1.0 \pm 0.2$}\\
{} & {58231$^a$} & {1.3$\pm$0.1} & {$-2.6 \pm 0.4$}\\
{} & {58928$^b$} & {1.1$\pm$0.1} & {$-2.2 \pm 0.3$}\\
{} & {60428} & {$>1.5$} & {$-1.5 \pm 0.1$}\\
{} & {} & {} & {}\\
{J2333} & {55447$^c$} & {0.90$\pm$0.03} & {$-1.2 \pm 0.1$}\\
{} & {58429$^c$} & {0.81$\pm$0.02} & {$-1.6 \pm 0.1$}\\
{} & {60564} & {1.18$\pm$0.04} & {$-1.1 \pm 0.2$}\\
{} & {} & {} & {}\\
{J2336} & {52096$^a$} & {$>0.6$} & {$-2.4 \pm 0.2$}\\
{} & {54465$^a$} & {1.4$\pm$0.1} & {$-3.2 \pm 0.4$}\\
{} & {52096$^b$} & {1.26$\pm$0.03} & {$-2.5 \pm 0.3$}\\
{} & {60503} & {$>1.2$} & {$-2.1 \pm 0.3$}\\
\enddata
\tablecomments{Column 1: Truncated SDSS object name, Column 2: MJD of associated optical observation, Column 3: Calculated $\alpha_{ox}$ values or lower limits, Column 4: Log of the Eddington ratio calculated using the bolometric corrections from \cite{runnoe12} (where contemporaneous X-ray detections were not available) or \cite{lusso10} (where contemporaneous X-ray luminosities were available).
$^a$Measurements for $\alpha_{ox}$ at this epoch are reported in \cite{ruan19}.\\
$^b$Measurements for $\alpha_{ox}$ at this epoch are reported in Gilbert et al. in prep.\\
$^c$Measurements for all parameters at this epoch are reported in \cite{jin21}.} 
\end{deluxetable}

\section{Discussion}\label{sec:disc}
\subsection{Tidal Disruption Events \& Dust Extinction}
We considered several hypotheses related to the observed behavior. Tidal disruption events are one potential explanation for the changes observed in CLQs (see summary in Section \ref{sec:intro}). The lightcurves for all three CLQs do not follow the predicted decay that most TDEs show. Here, also, the fact that all three quasars are re-brightening after a quiescent period disfavors the TDE explanation -- usual rates of TDEs are expected to be $\sim10^{-4}-10^{-5}$~yr$^{-1}$ in a galaxy \cite[e.g.][]{gezari09, vv18}, and the observed behavior would require a TDE rate three orders of magnitude higher than expected. Although some partial TDEs have been observed to re-brighten periodically \citep[e.g.,][]{wevers23, sun25}, predictions for their light curve decay \citep[e.g.,][]{guillochon13} do not match the behavior we find here.

Dust extinction alone can be ruled out through the abridged version of the extinction test presented in \cite{duffy25}, which confirmed that variable extinction alone was insufficient to explain the variability seen here. In addition, light curves of J1011 and J2333 both show coordinated, dramatic changes in both the optical and IR that are followed by changes in the spectral type, which we would not expect if variable extinction were the culprit.


\subsection{Transition Characteristics}
One way to constrain the mechanism responsible is through characterization of both the duration of the low state and the time it took for the CLQ to transition between a high and low state and vice versa, as well as the behavior of the emission lines over this time. J1011's broad emission lines show some interesting behavior as they return. Broad H$\beta$ in the transition state is boxier than either the initial high state or the new high state (see Table \ref{table:emlines}). Additionally, \cite{wangj24} present a spectrum of J1011 taken in November of 2022 that has a returning broad H$\beta$ profile that appears visually boxier than more recent high states. This adds evidence for a BLR that is evolving over the course of the transition (see further discussion in Section~\ref{sec:blr}).  The changes can be traced both through the evolution of the kurtosis over the course of the observations and in the FWHM, which is significantly larger in transition states than it is in the initial or renewed high states (see Table \ref{table:emlines} for values). The Pearson skewness coefficient of the profile changes over the 20 years of observations, and the centroid wavelength of the profile also shifts by $\sim 1000$ km s$^{-1}$. Interestingly, the new high state of J1011 is significantly dimmer than the previous high state -- the broad H$\beta$ flux is nearly four times lower than the previous high; this new high state has persisted for at least the period between December 2023 and November 2024.

Both J1011 and J2333 took nearly identical amounts of time to `turn-off' and `turn-on', as traced by CSS and ZTF optical light curves, and J1011 and J2336 both remained in a low state for around 10 years. While J1011 and J2336 have made distinct transitions between a high and low state and back again, J2333 has bounced between high and low states intermittently over the past ten years. We also find in J1011 and J2333, where there \textit{is} coordinated behavior between the optical and IR light curves, the IR transition both lags behind the optical transition and takes longer than the observed optical transition. This pattern is consistent with a picture in which the UV continuum is reprocessed into IR emission by a dusty torus, at distances hundreds of light days from the continuum source.

\subsection{Variable Accretion Rates and Transition Time Scales}
We find that the cause of the re-brightening in J1011, J2333 and J2336 is likely a variable accretion rate causing a transition in the accretion flow structure. All three CLQs follow expected tracks in the $\alpha_{ox}$ vs. $L_{\rm bol}/L_{\rm Edd}$ diagram (see Figure~\ref{fig:aox}), moving along the arms of the V-shape predicted by \cite{sobolewska11}. In fact, J1011 transitions from the high/soft state side of the V to the low/hard side when it goes from being quasar-dominated to host-dominated, then evolves further on the low/hard side as its continuum brightens, before moving back to the high/soft side of the diagram when its broad emission lines reappear. This suggests that the accretion flows of these three CLQs are indeed undergoing changes that are associated with the appearance and disappearance of their broad emission lines, such as a transition between a geometrically thin, optically thick accretion disk and an ADAF within a truncated thin disk, as is seen in XRBs \citep[see detailed discussion of the analogies in][especially their Section 5 and Figs. 4, 5]{ruan19} and some LLAGNs \citep[e.g.][]{ho99, maoz07, eracleous10, nemmen14}. J1011 and J2336 in particular span a wide range of Eddington ratios before, during, and after transition.

Despite evidence pointing towards changes in the accretion rate, the timescale for variation that we observe is still challenging to that model. We observe transitions that occur in $\sim500$ days, while basic considerations in the context of the $\alpha$-disk model suggest that large changes in the accretion rate should occur on the viscous time, which is of the order $10^4-10^7$~years \citep[see][]{frank02}. Theoretical models of quasar accretion flows, however, have found that the timescale of CLQ transitions may be achieved with magnetic-pressure dominated disks and outflows \citep[e.g.][]{feng21, wugu23}. But, the \cite{feng21} model (which relies on magnetically driven outflows) suggests significantly different timescales for the change from a high state to a low state than from a low state to a high state, which is not supported by the observations presented here. 

Other models for AGN disks suggest that, similarly to cataclysmic variables (CVs), disks could undergo the same instabilities analogous to dwarf novae \citep{hameury09}. The timescales associated with these instabilities are of order thousands to millions of years and are much longer than the observed transitions, which are of order years. \cite{ross18} invoke similar instabilities, suggesting that the cause of the changing-look phenomenon is the propagation of cooling fronts launched from the innermost stable circular orbit of the accretion disk. Those changes are predicted to occur on the thermal timescale, which is short and qualitatively matches the time scales of CLQ changes. That model, however, predicts that the timescale taken to fall to a low state should be shorter than the rise time to a high state, which is not supported by the observations presented here.

Observations of CVs have informed models for accretion disk evaporation from a thin disk to a corona that then accretes as an ADAF \citep[e.g.][]{meyer94, meyer00, liu02}. In the context of these models one can identify the transition time scales of CLQs from the high state to the low state with the disk evaporation time \cite[e.g.][]{liu09, mmh11}. To scale the evaporation time from a CV \cite[][see their Sections 4 and 5]{meyer94} to an SMBH, we assumed that the disk surface density, $\Sigma$, follows the $\alpha$-disk prescription \cite[see Chapter 8 of][]{frank02}. Thus, we obtained an evaporation time scale of $\sim 10^3\;$years for an ADAF that extends to 100 gravitational radii ($r_{\rm g}\equiv GM_{\rm BH}/c^2$) from the black hole (assuming an Eddington ratio of 0.1 -- appropriate for the three CLQs studied here, a viscosity parameter of 0.1, and an accretion efficiency of 0.1). Although this time scale is much longer than what we observe, it becomes compatible with the observed time scale if we adopt an ADAF radius of 10--20 gravitational radii. The opposite transition from the low state to the high state would be associated with the cooling of the ADAF and its condensation to a geometrically thin disk. The time scale for this process can be estimated using the results of \citet[][see their Section 3]{liu06} and could be consistent with observations, albeit with large uncertainties.

\cite{liska23} suggest that the cause of changing-look events may be due to warped and torn accretion disk causing periods of increased accretion. While the timescales that they predict for recurrent disk tearing events, which would be required to explain multiple high states, is much shorter than what we observe in J1011 and J2336, the recurrent behavior seen in J2333 may be caused by some mechanism such as the one discussed in that paper.

\subsection{Implications for the BLR}\label{sec:blr}


A promising and popular family of models for the BLR associates the broad lines with a wind emanating from the accretion disk \citep[e.g.,][]{emmering92,murray97,waters16,baskin18,naddaf21}. In the context of these models, changes in accretion rate can cause a change in the structure and geometry of the BLR and lead to significant changes in the profiles of the broad emission lines \citep[e.g.,][]{elitzur09,elitzur14,matthews20,matthews23}. As the accretion rate drops, the density (hence the emission measure) of the wind declines and the wind streamlines change from high to low altitude. As a consequence of the changing kinematics and radiative transfer, the fluxes of the broad emission lines decline and the line profiles change from cuspy to boxy. Qualitatively, the predictions from these models agree with our observations -- in these three returning CLQs, we observe that line profiles in intermediate states, where broad emission lines appear to be returning and the accretion rate appears to be increasing, are significantly boxier than the broad line profiles in the final high state. Moreover, our observations of changing broad line profiles suggest that the decline of the broad emission lines in the low state is not only a result of a lower ionizing continuum but also a changing structure of the BLR.

\subsection{CLQs in the Broader Context of Quasar Variability}
More generally, we note that the amplitude of long-term variability shown in the light curves of all three CLQs is not larger than the observed variability of quasars that have been studied extensively through reverberation mapping campaigns, such as the ones in \cite{dexter19} and \cite{fries23}. The quasar in \cite{fries23} shows a three-fold increase in rest-frame 5100~\AA{} flux over the course of less than two years. This type of increase is very similar to the jump seen in J1011's light curve. The quasar studied in \cite{dexter19} is a hypervariable quasar (HVQ), and shows almost order of magnitude fluctuations in its SDSS g-band photometry over $<1000$ days. Even when compared to variability studies of large samples of quasars,  the magnitude of the optical variability we observe in these CLQs is not aberrant \citep[e.g.,][]{macleod10, macleod12, stone22}. However, Eddington ratios of populations of CLQs are generally lower than more typical quasars \citep[see][]{macleod19, zeltyn24}, meaning that CLQs exhibiting normal quasar variability are more likely to cross a critical Eddington ratio and exhibit a change in their accretion flow structures. Thus, CLQs appear to not, in fact, represent a distinct class of objects, but instead represent the portion of the Eddington ratio distribution just above the critical value where changes in the accretion flow and BLR structure come about
\citep[see][]{rumbaugh18, macleod19, ruan19, dexter19}. 

\subsection{Returning CLQ Identification}
With respect to identifying turn-on CLQs in general, we note that the dramatic light curves of J1011 and J2333 make them easy to identify for follow-up. However, behavior like that of J2336 would be difficult to pick out. While there was a small increase in the optical flux of J2336 over the time it took for broad H$\beta$ to return, there was no coordinated behavior in the IR. That is, even quasars that show only small variation in their optical photometry may still vary in spectral type. J2336 would have been missed by many criteria for automatic follow-up identification, where moderate changes in photometry are required. Monitoring previously-identified CLQs spectroscopically may help identify more turn-on cases, even when light curves do not look promising, especially in lower-redshift cases, as the ZTF g-band is better suited to catch variations in the short wavelength continuum for quasars at redshifts $z>0.2$. Changes in g-band magnitude while r-band magnitude remains constant may also be an early indicator that a CLQ is at the early stages of a transition, That is because at these redshifts the observed g-band continuum includes a relatively small contribution from starlight, allowing us to discern changes in the quasar continuum more easily than in the r-band. Moreover, variation at shorter wavelengths is indicative of changes in the non-stellar continuum that may be reflected later in the longer-wavelength continuum and the broad emission lines, and, in general, the variability amplitude at shorter wavelengths is typically larger than at longer wavelengths.

We note that two of these three returning CLQs were analyzed in their low states in \cite{duffy25}. These two CLQs returned to a high state returned fortuitously after a long quiescent period. Thus, we suspect that many turn-off CLQs may continue to evolve and continued monitoring of previously-identified CLQs may prove fruitful.

\section{Conclusions}\label{sec:conc}
We investigated three CLQs that were previously observed to turn off and that have recently turned back on. One of the three was previously reported to have returned to a high state. We observed all three contemporaneously in the optical, UV and X-ray using the HET and Swift. Through data analysis we find:

\begin{enumerate}
    \item 
    Some previously-identified turn-off CLQs turn back on after a long (10 year or greater) period of quiescence (as noted by other authors), and return to states that are not identical to their previous high states. 
    \item 
    In two of the three CLQs, we see a well defined transition in the optical and IR light curves. In both cases, the turn-on and turn-off phases took similar amounts of time.
    \item 
    In two of the three CLQs, NEOWISE mid-IR observations showed coordinated behavior with optical light curves but lagged behind the optical by hundreds of days. This behavior suggests that the cause of the transition in these particular objects is not variable dust extinction, and that the source of the IR light is hundreds of light days away from the continuum source, which is consistent with emission from a dusty torus.
    \item 
    We tracked the evolution of the SEDs of the three turn-on CLQs via the relationship between the Eddington ratio and the X-ray-UV spectral slope. We find that all three turn-on CLQs follow predictions from X-ray binary outbursts. This suggests that the transition observed in CLQs may be due to a changing accretion flow structure, between a geometrically thin, optically thick disk and some combination of an ADAF, truncated thin disk, and jet.
    \item 
    We considered models for accretion state transitions and found that some can produce transitions that are as fast as those we observe (e.g., disk tearing, disk evaporation), for a suitable choice of parameters.
    \item 
    The broad emission lines return at approximately but not exactly the same time as the continuum. The profiles of the Balmer lines are more boxy during the return than in the initial high state. Viewed in the context of BLR models, the behavior of the line profiles suggests that the variations of the broad lines are not just a result of varying illumination of the BLR but also changes in its structure. 
\end{enumerate}

Larger samples of previously-identified changing-look quasars should be followed up in the same manner; identifying the duration of the low and high states and, in particular, the time spent in transition between the states is crucial in identifying the cause of the behavior. To further test the ADAF hypothesis, UV spectra of newly returned high-states would also be useful to confirm if the UV SED of high-state CLQs differs significantly from the low-state UV SED. 

\clearpage
\section{Acknowledgements}
We thank the anonymous referee for their thorough reading of the manuscript and thoughtful comments.

This work is supported by the Penn State Science Achievement Graduate Fellowship Program. L.D.\ acknowledges support from the Graduate School Endowed Fellowship at Penn State, as well as the Pennsylvania Space Grant Consortium Graduate Fellowship. J.J.R.\ acknowledges support from the Canada Research Chairs (CRC) program, the NSERC Discovery Grant program, the Canada Foundation for Innovation (CFI), and the Qu\'{e}bec Ministère de l’\'{E}conomie et de l’Innovation. 

The Low Resolution Spectrograph 2 (LRS2) was developed and funded by the University of Texas at Austin McDonald Observatory and Department of Astronomy, and by Pennsylvania State University. We thank the Leibniz-Institut fur Astrophysik Potsdam (AIP) and the Institut fur Astrophysik Goettingen (IAG) for their contributions to the construction of the integral field units. We acknowledge the Texas Advanced Computing Center (TACC) at The University of Texas at Austin for providing high performance computing, visualization, and storage resources that have contributed to the results reported within this paper.

Based on observations obtained with the Apache Point Observatory 3.5-meter telescope, which is owned and operated by the Astrophysical Research Consortium.

We acknowledge the use of the MMT telescope. Observations reported here were obtained at the MMT Observatory, a joint facility of the Smithsonian Institution and the University of Arizona.

This work made use of data from NASA’s Neil Gehrels Swift Observatory, specifically the Swift-XRT and Swift-UVOT instruments. The observations were carried out as part of the Swift target-of-opportunity program. 

We acknowledge use of SDSS-I/II and SDSS-III data. 

Funding for the SDSS and SDSS-II has been provided by the Alfred P. Sloan Foundation, the Participating Institutions, the National Science Foundation, the U.S. Department of Energy, the National Aeronautics and Space Administration, the Japanese Monbukagakusho, the Max Planck Society, and the Higher Education Funding Council for England. The SDSS Web Site is http://www.sdss.org/.

The SDSS is managed by the Astrophysical Research Consortium for the Participating Institutions. The Participating Institutions are the American Museum of Natural History, Astrophysical Institute Potsdam, University of Basel, University of Cambridge, Case Western Reserve University, University of Chicago, Drexel University, Fermilab, the Institute for Advanced Study, the Japan Participation Group, Johns Hopkins University, the Joint Institute for Nuclear Astrophysics, the Kavli Institute for Particle Astrophysics and Cosmology, the Korean Scientist Group, the Chinese Academy of Sciences (LAMOST), Los Alamos National Laboratory, the Max-Planck-Institute for Astronomy (MPIA), the Max-Planck-Institute for Astrophysics (MPA), New Mexico State University, Ohio State University, University of Pittsburgh, University of Portsmouth, Princeton University, the United States Naval Observatory, and the University of Washington.

Funding for SDSS-III has been provided by the Alfred P. Sloan Foundation, the Participating Institutions, the National Science Foundation, and the U.S. Department of Energy Office of Science. The SDSS-III web site is \url{http://www.sdss3.org/}.

SDSS-III is managed by the Astrophysical Research Consortium for the Participating Institutions of the SDSS-III Collaboration including the University of Arizona, the Brazilian Participation Group, Brookhaven National Laboratory, Carnegie Mellon University, University of Florida, the French Participation Group, the German Participation Group, Harvard University, the Instituto de Astrofisica de Canarias, the Michigan State/Notre Dame/JINA Participation Group, Johns Hopkins University, Lawrence Berkeley National Laboratory, Max Planck Institute for Astrophysics, Max Planck Institute for Extraterrestrial Physics, New Mexico State University, New York University, Ohio State University, Pennsylvania State University, University of Portsmouth, Princeton University, the Spanish Participation Group, University of Tokyo, University of Utah, Vanderbilt University, University of Virginia, University of Washington, and Yale University.

This publication makes use of data products from the Near-Earth Object Wide-field Infrared Survey Explorer (NEOWISE), which is a joint project of the Jet Propulsion Laboratory/California Institute of Technology and the University of Arizona. NEOWISE is funded by the National Aeronautics and Space Administration.

This research has made use of the NASA/IPAC Infrared Science Archive, which is funded by the National Aeronautics and Space Administration and operated by the California Institute of Technology.

We acknowledge the use of ZTF data. Based on observations obtained with the Samuel Oschin Telescope 48 inch and the 60 inch Telescope at the Palomar Observatory as part of the Zwicky Transient Facility project. ZTF is supported by the National Science Foundation under grant No. AST-2034437 and a collaboration including Caltech, IPAC, the Weizmann Institute for Science, the Oskar Klein Center at Stockholm University, the University of Maryland, Deutsches ElektronenSynchrotron and Humboldt University, the TANGO Consortium of Taiwan, the University of Wisconsin at Milwaukee, Trinity College Dublin, Lawrence Livermore National Laboratories, and IN2P3, France. Operations are conducted by COO, IPAC, and UW.

The CSS survey is funded by the National Aeronautics and Space
Administration under Grant No. NNG05GF22G issued through the Science
Mission Directorate Near-Earth Objects Observations Program.  The CRTS
survey is supported by the U.S.~National Science Foundation under
grants AST-0909182 and AST-1313422.

This research made use of Photutils, an Astropy package for
detection and photometry of astronomical sources \citep{photutils}.


%

\vspace{5mm}
\facilities{Sloan (BOSS), HET (LRS2), Swift (XRT, UVOT), ARC 3.5m, MMT (Blue Channel spectrograph), IRSA, NEOWISE, CXO}


\software{astropy \citep{2013A&A...558A..33A,2018AJ....156..123A},  
          extinction \citep{extinction},
          matplotlib \citep{matplotlib},
          pandas \citep{pandas},
          photutils \citep{photutils},
          ppxf \citep{ppxf},
          scipy \citep{scipy},
          specutils \citep{specutils}
          }




\bibliography{sample7}{}

\begin{thebibliography}{}
\expandafter\ifx\csname natexlab\endcsname\relax\def\natexlab#1{#1}\fi
\providecommand{\url}[1]{\href{#1}{#1}}
\providecommand{\dodoi}[1]{doi:~\href{http://doi.org/#1}{\nolinkurl{#1}}}
\providecommand{\doeprint}[1]{\href{http://ascl.net/#1}{\nolinkurl{http://ascl.net/#1}}}
\providecommand{\doarXiv}[1]{\href{https://arxiv.org/abs/#1}{\nolinkurl{https://arxiv.org/abs/#1}}}

\bibitem[{ {Astropy Collaboration} {et~al.}(2013){Astropy Collaboration}, {Robitaille}, {Tollerud}, {Greenfield}, {Droettboom}, {Bray}, {Aldcroft}, {Davis}, {Ginsburg}, {Price-Whelan}, {Kerzendorf}, {Conley}, {Crighton}, {Barbary}, {Muna}, {Ferguson}, {Grollier}, {Parikh}, {Nair}, {Unther}, {Deil}, {Woillez}, {Conseil}, {Kramer}, {Turner}, {Singer}, {Fox}, {Weaver}, {Zabalza}, {Edwards}, {Azalee Bostroem}, {Burke}, {Casey}, {Crawford}, {Dencheva}, {Ely}, {Jenness}, {Labrie}, {Lim}, {Pierfederici}, {Pontzen}, {Ptak}, {Refsdal}, {Servillat}, \& {Streicher}}]{2013A&A...558A..33A}
{Astropy Collaboration}, {Robitaille}, T.~P., {Tollerud}, E.~J., {et~al.} 2013, \bibinfo{title}{{Astropy: A community Python package for astronomy},} \aap, 558, A33, \dodoi{10.1051/0004-6361/201322068}

\bibitem[{ {Astropy Collaboration} {et~al.}(2018){Astropy Collaboration}, {Price-Whelan}, {Sip{\H{o}}cz}, {G{\"u}nther}, {Lim}, {Crawford}, {Conseil}, {Shupe}, {Craig}, {Dencheva}, {Ginsburg}, {VanderPlas}, {Bradley}, {P{\'e}rez-Su{\'a}rez}, {de Val-Borro}, {Aldcroft}, {Cruz}, {Robitaille}, {Tollerud}, {Ardelean}, {Babej}, {Bach}, {Bachetti}, {Bakanov}, {Bamford}, {Barentsen}, {Barmby}, {Baumbach}, {Berry}, {Biscani}, {Boquien}, {Bostroem}, {Bouma}, {Brammer}, {Bray}, {Breytenbach}, {Buddelmeijer}, {Burke}, {Calderone}, {Cano Rodr{\'\i}guez}, {Cara}, {Cardoso}, {Cheedella}, {Copin}, {Corrales}, {Crichton}, {D'Avella}, {Deil}, {Depagne}, {Dietrich}, {Donath}, {Droettboom}, {Earl}, {Erben}, {Fabbro}, {Ferreira}, {Finethy}, {Fox}, {Garrison}, {Gibbons}, {Goldstein}, {Gommers}, {Greco}, {Greenfield}, {Groener}, {Grollier}, {Hagen}, {Hirst}, {Homeier}, {Horton}, {Hosseinzadeh}, {Hu}, {Hunkeler}, {Ivezi{\'c}}, {Jain}, {Jenness}, {Kanarek}, {Kendrew}, {Kern}, {Kerzendorf}, {Khvalko}, {King}, {Kirkby}, {Kulkarni},
  {Kumar}, {Lee}, {Lenz}, {Littlefair}, {Ma}, {Macleod}, {Mastropietro}, {McCully}, {Montagnac}, {Morris}, {Mueller}, {Mumford}, {Muna}, {Murphy}, {Nelson}, {Nguyen}, {Ninan}, {N{\"o}the}, {Ogaz}, {Oh}, {Parejko}, {Parley}, {Pascual}, {Patil}, {Patil}, {Plunkett}, {Prochaska}, {Rastogi}, {Reddy Janga}, {Sabater}, {Sakurikar}, {Seifert}, {Sherbert}, {Sherwood-Taylor}, {Shih}, {Sick}, {Silbiger}, {Singanamalla}, {Singer}, {Sladen}, {Sooley}, {Sornarajah}, {Streicher}, {Teuben}, {Thomas}, {Tremblay}, {Turner}, {Terr{\'o}n}, {van Kerkwijk}, {de la Vega}, {Watkins}, {Weaver}, {Whitmore}, {Woillez}, {Zabalza}, \& {Astropy Contributors}}]{2018AJ....156..123A}
{Astropy Collaboration}, {Price-Whelan}, A.~M., {Sip{\H{o}}cz}, B.~M., {et~al.} 2018, \bibinfo{title}{{The Astropy Project: Building an Open-science Project and Status of the v2.0 Core Package},} \aj, 156, 123, \dodoi{10.3847/1538-3881/aabc4f}

\bibitem[{K. Barbary(2017)Barbary}]{extinction}
Barbary, K. 2017, \bibinfo{title}{extinction v0.3.0,} Zenodo, \dodoi{10.5281/zenodo.804967}

\bibitem[{A. {Baskin} \& A. {Laor}(2018){Baskin} \& {Laor}}]{baskin18}
{Baskin}, A., \& {Laor}, A. 2018, \bibinfo{title}{{Dust inflated accretion disc as the origin of the broad line region in active galactic nuclei},} \mnras, 474, 1970, \dodoi{10.1093/mnras/stx2850}

\bibitem[{E.~C. {Bellm} {et~al.}(2019){Bellm}, {Kulkarni}, {Graham}, {Dekany}, {Smith}, {Riddle}, {Masci}, {Helou}, {Prince}, {Adams}, {Barbarino}, {Barlow}, {Bauer}, {Beck}, {Belicki}, {Biswas}, {Blagorodnova}, {Bodewits}, {Bolin}, {Brinnel}, {Brooke}, {Bue}, {Bulla}, {Burruss}, {Cenko}, {Chang}, {Connolly}, {Coughlin}, {Cromer}, {Cunningham}, {De}, {Delacroix}, {Desai}, {Duev}, {Eadie}, {Farnham}, {Feeney}, {Feindt}, {Flynn}, {Franckowiak}, {Frederick}, {Fremling}, {Gal-Yam}, {Gezari}, {Giomi}, {Goldstein}, {Golkhou}, {Goobar}, {Groom}, {Hacopians}, {Hale}, {Henning}, {Ho}, {Hover}, {Howell}, {Hung}, {Huppenkothen}, {Imel}, {Ip}, {Ivezi{\'c}}, {Jackson}, {Jones}, {Juric}, {Kasliwal}, {Kaspi}, {Kaye}, {Kelley}, {Kowalski}, {Kramer}, {Kupfer}, {Landry}, {Laher}, {Lee}, {Lin}, {Lin}, {Lunnan}, {Giomi}, {Mahabal}, {Mao}, {Miller}, {Monkewitz}, {Murphy}, {Ngeow}, {Nordin}, {Nugent}, {Ofek}, {Patterson}, {Penprase}, {Porter}, {Rauch}, {Rebbapragada}, {Reiley}, {Rigault}, {Rodriguez}, {van Roestel}, {Rusholme},
  {van Santen}, {Schulze}, {Shupe}, {Singer}, {Soumagnac}, {Stein}, {Surace}, {Sollerman}, {Szkody}, {Taddia}, {Terek}, {Van Sistine}, {van Velzen}, {Vestrand}, {Walters}, {Ward}, {Ye}, {Yu}, {Yan}, \& {Zolkower}}]{bellm19}
{Bellm}, E.~C., {Kulkarni}, S.~R., {Graham}, M.~J., {et~al.} 2019, \bibinfo{title}{{The Zwicky Transient Facility: System Overview, Performance, and First Results},} \pasp, 131, 018002, \dodoi{10.1088/1538-3873/aaecbe}

\bibitem[{N. {Bennert} {et~al.}(2002){Bennert}, {Falcke}, {Schulz}, {Wilson}, \& {Wills}}]{bennert02}
{Bennert}, N., {Falcke}, H., {Schulz}, H., {Wilson}, A.~S., \& {Wills}, B.~J. 2002, \bibinfo{title}{{Size and Structure of the Narrow-Line Region of Quasars},} \apjl, 574, L105, \dodoi{10.1086/342420}

\bibitem[{P.~K. {Blanchard} {et~al.}(2017){Blanchard}, {Nicholl}, {Berger}, {Guillochon}, {Margutti}, {Chornock}, {Alexander}, {Leja}, \& {Drout}}]{blanchard17}
{Blanchard}, P.~K., {Nicholl}, M., {Berger}, E., {et~al.} 2017, \bibinfo{title}{{PS16dtm: A Tidal Disruption Event in a Narrow-line Seyfert 1 Galaxy},} \apj, 843, 106, \dodoi{10.3847/1538-4357/aa77f7}

\bibitem[{L. Bradley {et~al.}(2024)Bradley, Sip{\H o}cz, Robitaille, Tollerud, Vin{\'{\i}}cius, Deil, Barbary, Wilson, Busko, Donath, G{\"u}nther, Cara, Lim, Me{\ss}linger, Burnett, Conseil, Droettboom, Bostroem, Bray, Bratholm, Jamieson, Ginsburg, Barentsen, Craig, Pascual, Rathi, Perrin, Morris, \& Perren}]{photutils}
Bradley, L., Sip{\H o}cz, B., Robitaille, T., {et~al.} 2024, \bibinfo{title}{astropy/photutils: 1.12.0,}, 1.12.0 Zenodo, \dodoi{10.5281/zenodo.10967176}

\bibitem[{A.~A. {Breeveld} {et~al.}(2010){Breeveld}, {Curran}, {Hoversten}, {Koch}, {Landsman}, {Marshall}, {Page}, {Poole}, {Roming}, {Smith}, {Still}, {Yershov}, {Blustin}, {Brown}, {Gronwall}, {Holland}, {Kuin}, {McGowan}, {Rosen}, {Boyd}, {Broos}, {Carter}, {Chester}, {Hancock}, {Huckle}, {Immler}, {Ivanushkina}, {Kennedy}, {Mason}, {Morgan}, {Oates}, {de Pasquale}, {Schady}, {Siegel}, \& {vanden Berk}}]{breeveld10}
{Breeveld}, A.~A., {Curran}, P.~A., {Hoversten}, E.~A., {et~al.} 2010, \bibinfo{title}{{Further calibration of the Swift ultraviolet/optical telescope},} \mnras, 406, 1687, \dodoi{10.1111/j.1365-2966.2010.16832.x}

\bibitem[{D.~N. {Burrows} {et~al.}(2005){Burrows}, {Hill}, {Nousek}, {Kennea}, {Wells}, {Osborne}, {Abbey}, {Beardmore}, {Mukerjee}, {Short}, {Chincarini}, {Campana}, {Citterio}, {Moretti}, {Pagani}, {Tagliaferri}, {Giommi}, {Capalbi}, {Tamburelli}, {Angelini}, {Cusumano}, {Br{\"a}uninger}, {Burkert}, \& {Hartner}}]{burrows05}
{Burrows}, D.~N., {Hill}, J.~E., {Nousek}, J.~A., {et~al.} 2005, \bibinfo{title}{{The Swift X-Ray Telescope},} \ssr, 120, 165, \dodoi{10.1007/s11214-005-5097-2}

\bibitem[{M. {Cappellari}(2023{\natexlab{a}}){Cappellari}}]{cappellari23}
{Cappellari}, M. 2023{\natexlab{a}}, \bibinfo{title}{{Full spectrum fitting with photometry in PPXF: stellar population versus dynamical masses, non-parametric star formation history and metallicity for 3200 LEGA-C galaxies at redshift $z\approx0.8$},} MNRAS, 526, 3273, \dodoi{10.1093/mnras/stad2597}

\bibitem[{M. {Cappellari}(2023{\natexlab{b}}){Cappellari}}]{ppxf}
{Cappellari}, M. 2023{\natexlab{b}}, \bibinfo{title}{{Full spectrum fitting with photometry in PPXF: stellar population versus dynamical masses, non-parametric star formation history and metallicity for 3200 LEGA-C galaxies at redshift $z\approx0.8$},} MNRAS, 526, 3273, \dodoi{10.1093/mnras/stad2597}

\bibitem[{J.~A. {Cardelli} {et~al.}(1989){Cardelli}, {Clayton}, \& {Mathis}}]{cardelli89}
{Cardelli}, J.~A., {Clayton}, G.~C., \& {Mathis}, J.~S. 1989, \bibinfo{title}{{The Relationship between Infrared, Optical, and Ultraviolet Extinction},} \apj, 345, 245, \dodoi{10.1086/167900}

\bibitem[{T.~S. {Chonis} {et~al.}(2016){Chonis}, {Hill}, {Lee}, {Tuttle}, {Vattiat}, {Drory}, {Indahl}, {Peterson}, \& {Ramsey}}]{chonis16}
{Chonis}, T.~S., {Hill}, G.~J., {Lee}, H., {et~al.} 2016, in Society of Photo-Optical Instrumentation Engineers (SPIE) Conference Series, Vol. 9908, Ground-based and Airborne Instrumentation for Astronomy VI, ed. C.~J. {Evans}, L.~{Simard}, \& H.~{Takami}, 99084C, \dodoi{10.1117/12.2232209}

\bibitem[{R.~D. {Cohen} {et~al.}(1986){Cohen}, {Rudy}, {Puetter}, {Ake}, \& {Foltz}}]{cohen86}
{Cohen}, R.~D., {Rudy}, R.~J., {Puetter}, R.~C., {Ake}, T.~B., \& {Foltz}, C.~B. 1986, \bibinfo{title}{{Variability of Markarian 1018: Seyfert 1.9 to Seyfert 1},} \apj, 311, 135, \dodoi{10.1086/164758}

\bibitem[{A. {Constantin} {et~al.}(2009){Constantin}, {Green}, {Aldcroft}, {Kim}, {Haggard}, {Barkhouse}, \& {Anderson}}]{constantin09}
{Constantin}, A., {Green}, P., {Aldcroft}, T., {et~al.} 2009, \bibinfo{title}{{Probing the Balance of AGN and Star-forming Activity in the Local Universe with ChaMP},} \apj, 705, 1336, \dodoi{10.1088/0004-637X/705/2/1336}

\bibitem[{K.~S. {Dawson} {et~al.}(2013){Dawson}, {Schlegel}, {Ahn}, {Anderson}, {Aubourg}, {Bailey}, {Barkhouser}, {Bautista}, {Beifiori}, {Berlind}, {Bhardwaj}, {Bizyaev}, {Blake}, {Blanton}, {Blomqvist}, {Bolton}, {Borde}, {Bovy}, {Brandt}, {Brewington}, {Brinkmann}, {Brown}, {Brownstein}, {Bundy}, {Busca}, {Carithers}, {Carnero}, {Carr}, {Chen}, {Comparat}, {Connolly}, {Cope}, {Croft}, {Cuesta}, {da Costa}, {Davenport}, {Delubac}, {de Putter}, {Dhital}, {Ealet}, {Ebelke}, {Eisenstein}, {Escoffier}, {Fan}, {Filiz Ak}, {Finley}, {Font-Ribera}, {G{\'e}nova-Santos}, {Gunn}, {Guo}, {Haggard}, {Hall}, {Hamilton}, {Harris}, {Harris}, {Ho}, {Hogg}, {Holder}, {Honscheid}, {Huehnerhoff}, {Jordan}, {Jordan}, {Kauffmann}, {Kazin}, {Kirkby}, {Klaene}, {Kneib}, {Le Goff}, {Lee}, {Long}, {Loomis}, {Lundgren}, {Lupton}, {Maia}, {Makler}, {Malanushenko}, {Malanushenko}, {Mandelbaum}, {Manera}, {Maraston}, {Margala}, {Masters}, {McBride}, {McDonald}, {McGreer}, {McMahon}, {Mena}, {Miralda-Escud{\'e}}, {Montero-Dorta},
  {Montesano}, {Muna}, {Myers}, {Naugle}, {Nichol}, {Noterdaeme}, {Nuza}, {Olmstead}, {Oravetz}, {Oravetz}, {Owen}, {Padmanabhan}, {Palanque-Delabrouille}, {Pan}, {Parejko}, {P{\^a}ris}, {Percival}, {P{\'e}rez-Fournon}, {P{\'e}rez-R{\`a}fols}, {Petitjean}, {Pfaffenberger}, {Pforr}, {Pieri}, {Prada}, {Price-Whelan}, {Raddick}, {Rebolo}, {Rich}, {Richards}, {Rockosi}, {Roe}, {Ross}, {Ross}, {Rossi}, {Rubi{\~n}o-Martin}, {Samushia}, {S{\'a}nchez}, {Sayres}, {Schmidt}, {Schneider}, {Sc{\'o}ccola}, {Seo}, {Shelden}, {Sheldon}, {Shen}, {Shu}, {Slosar}, {Smee}, {Snedden}, {Stauffer}, {Steele}, {Strauss}, {Streblyanska}, {Suzuki}, {Swanson}, {Tal}, {Tanaka}, {Thomas}, {Tinker}, {Tojeiro}, {Tremonti}, {Vargas Maga{\~n}a}, {Verde}, {Viel}, {Wake}, {Watson}, {Weaver}, {Weinberg}, {Weiner}, {West}, {White}, {Wood-Vasey}, {Yeche}, {Zehavi}, {Zhao}, \& {Zheng}}]{dawson13}
{Dawson}, K.~S., {Schlegel}, D.~J., {Ahn}, C.~P., {et~al.} 2013, \bibinfo{title}{{The Baryon Oscillation Spectroscopic Survey of SDSS-III},} \aj, 145, 10, \dodoi{10.1088/0004-6256/145/1/10}

\bibitem[{K.~S. {Dawson} {et~al.}(2016){Dawson}, {Kneib}, {Percival}, {Alam}, {Albareti}, {Anderson}, {Armengaud}, {Aubourg}, {Bailey}, {Bautista}, {Berlind}, {Bershady}, {Beutler}, {Bizyaev}, {Blanton}, {Blomqvist}, {Bolton}, {Bovy}, {Brandt}, {Brinkmann}, {Brownstein}, {Burtin}, {Busca}, {Cai}, {Chuang}, {Clerc}, {Comparat}, {Cope}, {Croft}, {Cruz-Gonzalez}, {da Costa}, {Cousinou}, {Darling}, {de la Macorra}, {de la Torre}, {Delubac}, {du Mas des Bourboux}, {Dwelly}, {Ealet}, {Eisenstein}, {Eracleous}, {Escoffier}, {Fan}, {Finoguenov}, {Font-Ribera}, {Frinchaboy}, {Gaulme}, {Georgakakis}, {Green}, {Guo}, {Guy}, {Ho}, {Holder}, {Huehnerhoff}, {Hutchinson}, {Jing}, {Jullo}, {Kamble}, {Kinemuchi}, {Kirkby}, {Kitaura}, {Klaene}, {Laher}, {Lang}, {Laurent}, {Le Goff}, {Li}, {Liang}, {Lima}, {Lin}, {Lin}, {Lin}, {Long}, {Lundgren}, {MacDonald}, {Geimba Maia}, {Malanushenko}, {Malanushenko}, {Mariappan}, {McBride}, {McGreer}, {M{\'e}nard}, {Merloni}, {Meza}, {Montero-Dorta}, {Muna}, {Myers}, {Nandra}, {Naugle},
  {Newman}, {Noterdaeme}, {Nugent}, {Ogando}, {Olmstead}, {Oravetz}, {Oravetz}, {Padmanabhan}, {Palanque-Delabrouille}, {Pan}, {Parejko}, {P{\^a}ris}, {Peacock}, {Petitjean}, {Pieri}, {Pisani}, {Prada}, {Prakash}, {Raichoor}, {Reid}, {Rich}, {Ridl}, {Rodriguez-Torres}, {Carnero Rosell}, {Ross}, {Rossi}, {Ruan}, {Salvato}, {Sayres}, {Schneider}, {Schlegel}, {Seljak}, {Seo}, {Sesar}, {Shandera}, {Shu}, {Slosar}, {Sobreira}, {Streblyanska}, {Suzuki}, {Taylor}, {Tao}, {Tinker}, {Tojeiro}, {Vargas-Maga{\~n}a}, {Wang}, {Weaver}, {Weinberg}, {White}, {Wood-Vasey}, {Yeche}, {Zhai}, {Zhao}, {Zhao}, {Zheng}, {Ben Zhu}, \& {Zou}}]{dawson16}
{Dawson}, K.~S., {Kneib}, J.-P., {Percival}, W.~J., {et~al.} 2016, \bibinfo{title}{{The SDSS-IV Extended Baryon Oscillation Spectroscopic Survey: Overview and Early Data},} \aj, 151, 44, \dodoi{10.3847/0004-6256/151/2/44}

\bibitem[{K.~D. {Denney} {et~al.}(2014){Denney}, {De Rosa}, {Croxall}, {Gupta}, {Bentz}, {Fausnaugh}, {Grier}, {Martini}, {Mathur}, {Peterson}, {Pogge}, \& {Shappee}}]{denney14}
{Denney}, K.~D., {De Rosa}, G., {Croxall}, K., {et~al.} 2014, \bibinfo{title}{{The Typecasting of Active Galactic Nuclei: Mrk 590 no Longer Fits the Role},} \apj, 796, 134, \dodoi{10.1088/0004-637X/796/2/134}

\bibitem[{J. {Dexter} \& M.~C. {Begelman}(2019){Dexter} \& {Begelman}}]{DexterBegelman19}
{Dexter}, J., \& {Begelman}, M.~C. 2019, \bibinfo{title}{{Extreme AGN variability: evidence of magnetically elevated accretion?},} \mnras, 483, L17, \dodoi{10.1093/mnrasl/sly213}

\bibitem[{J. {Dexter} {et~al.}(2019){Dexter}, {Xin}, {Shen}, {Grier}, {Liu}, {Gezari}, {McGreer}, {Brandt}, {Hall}, {Horne}, {Simm}, {Merloni}, {Green}, {Vivek}, {Trump}, {Homayouni}, {Peterson}, {Schneider}, {Kinemuchi}, {Pan}, \& {Bizyaev}}]{dexter19}
{Dexter}, J., {Xin}, S., {Shen}, Y., {et~al.} 2019, \bibinfo{title}{{The Sloan Digital Sky Survey Reverberation Mapping Project: Accretion and Broad Emission Line Physics from a Hypervariable Quasar},} \apj, 885, 44, \dodoi{10.3847/1538-4357/ab4354}

\bibitem[{C. {Done} {et~al.}(2007){Done}, {Gierli{\'n}ski}, \& {Kubota}}]{done07}
{Done}, C., {Gierli{\'n}ski}, M., \& {Kubota}, A. 2007, \bibinfo{title}{{Modelling the behaviour of accretion flows in X-ray binaries. Everything you always wanted to know about accretion but were afraid to ask},} \aapr, 15, 1, \dodoi{10.1007/s00159-007-0006-1}

\bibitem[{A.~J. {Drake} {et~al.}(2009){Drake}, {Djorgovski}, {Mahabal}, {Beshore}, {Larson}, {Graham}, {Williams}, {Christensen}, {Catelan}, {Boattini}, {Gibbs}, {Hill}, \& {Kowalski}}]{css}
{Drake}, A.~J., {Djorgovski}, S.~G., {Mahabal}, A., {et~al.} 2009, \bibinfo{title}{{First Results from the Catalina Real-Time Transient Survey},} \apj, 696, 870, \dodoi{10.1088/0004-637X/696/1/870}

\bibitem[{L. {Duffy} {et~al.}(2025){Duffy}, {Eracleous}, {Runnoe}, {Ruan}, {Anderson}, {Dimassimo}, {Green}, \& {LaMassa}}]{duffy25}
{Duffy}, L., {Eracleous}, M., {Runnoe}, J.~C., {et~al.} 2025, \bibinfo{title}{{A Detailed Look at a Trio of Changing-look Quasars: Spectral Energy Distributions and the Dust Extinction Test},} \apj, 981, 127, \dodoi{10.3847/1538-4357/adae0b}

\bibitem[{N. Earl {et~al.}(2020)Earl, Tollerud, Jones, Kerzendorf, shaileshahuja, D'Avella, Robitaille, Ginsburg, Sipőcz, Busko, Ogaz, Günther, rosteen, Barbary, Foster, Torres, Droettboom, Bray, Davies, Casey, Ferguson, Crawford, Teuben, Homeier, Cruz, Pickering, Dencheva, Ninan, gmduvvuri, \& Deil}]{specutils}
Earl, N., Tollerud, E., Jones, C., {et~al.} 2020, \bibinfo{title}{astropy/specutils: v1.0,}, v1.0 Zenodo, \dodoi{10.5281/zenodo.3718589}

\bibitem[{M. {Elitzur} \& L.~C. {Ho}(2009){Elitzur} \& {Ho}}]{elitzur09}
{Elitzur}, M., \& {Ho}, L.~C. 2009, \bibinfo{title}{{On the Disappearance of the Broad-Line Region in Low-Luminosity Active Galactic Nuclei},} \apjl, 701, L91, \dodoi{10.1088/0004-637X/701/2/L91}

\bibitem[{M. {Elitzur} {et~al.}(2014){Elitzur}, {Ho}, \& {Trump}}]{elitzur14}
{Elitzur}, M., {Ho}, L.~C., \& {Trump}, J.~R. 2014, \bibinfo{title}{{Evolution of broad-line emission from active galactic nuclei},} \mnras, 438, 3340, \dodoi{10.1093/mnras/stt2445}

\bibitem[{R.~T. {Emmering} {et~al.}(1992){Emmering}, {Blandford}, \& {Shlosman}}]{emmering92}
{Emmering}, R.~T., {Blandford}, R.~D., \& {Shlosman}, I. 1992, \bibinfo{title}{{Magnetic Acceleration of Broad Emission-Line Clouds in Active Galactic Nuclei},} \apj, 385, 460, \dodoi{10.1086/170955}

\bibitem[{M. {Eracleous} {et~al.}(2012){Eracleous}, {Boroson}, {Halpern}, \& {Liu}}]{eracleous12}
{Eracleous}, M., {Boroson}, T.~A., {Halpern}, J.~P., \& {Liu}, J. 2012, \bibinfo{title}{{A Large Systematic Search for Close Supermassive Binary and Rapidly Recoiling Black Holes},} \apjs, 201, 23, \dodoi{10.1088/0067-0049/201/2/23}

\bibitem[{M. {Eracleous} {et~al.}(2010){Eracleous}, {Hwang}, \& {Flohic}}]{eracleous10}
{Eracleous}, M., {Hwang}, J.~A., \& {Flohic}, H. M.~L.~G. 2010, \bibinfo{title}{{Spectral Energy Distributions of Weak Active Galactic Nuclei Associated with Low-Ionization Nuclear Emission Regions},} \apjs, 187, 135, \dodoi{10.1088/0067-0049/187/1/135}

\bibitem[{H. {Falcke} {et~al.}(2004){Falcke}, {K{\"o}rding}, \& {Markoff}}]{falcke04}
{Falcke}, H., {K{\"o}rding}, E., \& {Markoff}, S. 2004, \bibinfo{title}{{A scheme to unify low-power accreting black holes. Jet-dominated accretion flows and the radio/X-ray correlation},} \aap, 414, 895, \dodoi{10.1051/0004-6361:20031683}

\bibitem[{J. {Feng} {et~al.}(2021){Feng}, {Cao}, {Li}, \& {Gu}}]{feng21}
{Feng}, J., {Cao}, X., {Li}, J.-w., \& {Gu}, W.-M. 2021, \bibinfo{title}{{A Magnetic Disk-outflow Model for Changing Look Active Galactic Nuclei},} \apj, 916, 61, \dodoi{10.3847/1538-4357/ac07a6}

\bibitem[{E.~L. {Fitzpatrick}(1999){Fitzpatrick}}]{fitzpatrick99}
{Fitzpatrick}, E.~L. 1999, \bibinfo{title}{{Correcting for the Effects of Interstellar Extinction},} \pasp, 111, 63, \dodoi{10.1086/316293}

\bibitem[{J. {Frank} {et~al.}(2002){Frank}, {King}, \& {Raine}}]{frank02}
{Frank}, J., {King}, A., \& {Raine}, D.~J. 2002, {Accretion Power in Astrophysics: Third Edition}

\bibitem[{L.~B. {Fries} {et~al.}(2023){Fries}, {Trump}, {Davis}, {Grier}, {Shen}, {Anderson}, {Dwelly}, {Eracleous}, {Homayouni}, {Horne}, {Krumpe}, {Morrison}, {Runnoe}, {Trakhtenbrot}, {Assef}, {Brandt}, {Brownstein}, {Dabbieri}, {Fix}, {Fonseca Alvarez}, {Frederick}, {Hall}, {Koekemoer}, {Li}, {Liu}, {Mart{\'\i}nez-Aldama}, {Ricci}, {Schneider}, {Sharp}, {Temple}, {Yang}, {Zeltyn}, \& {Bizyaev}}]{fries23}
{Fries}, L.~B., {Trump}, J.~R., {Davis}, M.~C., {et~al.} 2023, \bibinfo{title}{{The SDSS-V Black Hole Mapper Reverberation Mapping Project: Unusual Broad-line Variability in a Luminous Quasar},} \apj, 948, 5, \dodoi{10.3847/1538-4357/acbfb7}

\bibitem[{S. {Gezari} {et~al.}(2009){Gezari}, {Heckman}, {Cenko}, {Eracleous}, {Forster}, {Gon{\c{c}}alves}, {Martin}, {Morrissey}, {Neff}, {Seibert}, {Schiminovich}, \& {Wyder}}]{gezari09}
{Gezari}, S., {Heckman}, T., {Cenko}, S.~B., {et~al.} 2009, \bibinfo{title}{{Luminous Thermal Flares from Quiescent Supermassive Black Holes},} \apj, 698, 1367, \dodoi{10.1088/0004-637X/698/2/1367}

\bibitem[{S. {Gezari} {et~al.}(2017){Gezari}, {Hung}, {Cenko}, {Blagorodnova}, {Yan}, {Kulkarni}, {Mooley}, {Kong}, {Cantwell}, {Yu}, {Cao}, {Fremling}, {Neill}, {Ngeow}, {Nugent}, \& {Wozniak}}]{gezari17}
{Gezari}, S., {Hung}, T., {Cenko}, S.~B., {et~al.} 2017, \bibinfo{title}{{iPTF Discovery of the Rapid {\textquotedblleft}Turn-on{\textquotedblright} of a Luminous Quasar},} \apj, 835, 144, \dodoi{10.3847/1538-4357/835/2/144}

\bibitem[{R.~W. {Goodrich}(1989){Goodrich}}]{goodrich89}
{Goodrich}, R.~W. 1989, \bibinfo{title}{{Spectropolarimetry and Variability of Seyfert 1.8 and 1.9 Galaxies},} \apj, 340, 190, \dodoi{10.1086/167384}

\bibitem[{R.~W. {Goodrich}(1990){Goodrich}}]{goodrich90}
{Goodrich}, R.~W. 1990, \bibinfo{title}{{PA beta Measurements and Reddening in Seyfert 1.8 and 1.9 Galaxies},} \apj, 355, 88, \dodoi{10.1086/168743}

\bibitem[{R.~W. {Goodrich}(1995){Goodrich}}]{goodrich95}
{Goodrich}, R.~W. 1995, \bibinfo{title}{{Dust in the Broad-Line Regions of Seyfert Galaxies},} \apj, 440, 141, \dodoi{10.1086/175256}

\bibitem[{M.~J. {Graham} {et~al.}(2019){Graham}, {Kulkarni}, {Bellm}, {Adams}, {Barbarino}, {Blagorodnova}, {Bodewits}, {Bolin}, {Brady}, {Cenko}, {Chang}, {Coughlin}, {De}, {Eadie}, {Farnham}, {Feindt}, {Franckowiak}, {Fremling}, {Gezari}, {Ghosh}, {Goldstein}, {Golkhou}, {Goobar}, {Ho}, {Huppenkothen}, {Ivezi{\'c}}, {Jones}, {Juric}, {Kaplan}, {Kasliwal}, {Kelley}, {Kupfer}, {Lee}, {Lin}, {Lunnan}, {Mahabal}, {Miller}, {Ngeow}, {Nugent}, {Ofek}, {Prince}, {Rauch}, {van Roestel}, {Schulze}, {Singer}, {Sollerman}, {Taddia}, {Yan}, {Ye}, {Yu}, {Barlow}, {Bauer}, {Beck}, {Belicki}, {Biswas}, {Brinnel}, {Brooke}, {Bue}, {Bulla}, {Burruss}, {Connolly}, {Cromer}, {Cunningham}, {Dekany}, {Delacroix}, {Desai}, {Duev}, {Feeney}, {Flynn}, {Frederick}, {Gal-Yam}, {Giomi}, {Groom}, {Hacopians}, {Hale}, {Helou}, {Henning}, {Hover}, {Hillenbrand}, {Howell}, {Hung}, {Imel}, {Ip}, {Jackson}, {Kaspi}, {Kaye}, {Kowalski}, {Kramer}, {Kuhn}, {Landry}, {Laher}, {Mao}, {Masci}, {Monkewitz}, {Murphy}, {Nordin}, {Patterson},
  {Penprase}, {Porter}, {Rebbapragada}, {Reiley}, {Riddle}, {Rigault}, {Rodriguez}, {Rusholme}, {van Santen}, {Shupe}, {Smith}, {Soumagnac}, {Stein}, {Surace}, {Szkody}, {Terek}, {Van Sistine}, {van Velzen}, {Vestrand}, {Walters}, {Ward}, {Zhang}, \& {Zolkower}}]{graham19}
{Graham}, M.~J., {Kulkarni}, S.~R., {Bellm}, E.~C., {et~al.} 2019, \bibinfo{title}{{The Zwicky Transient Facility: Science Objectives},} \pasp, 131, 078001, \dodoi{10.1088/1538-3873/ab006c}

\bibitem[{P.~J. {Green} {et~al.}(2022){Green}, {Pulgarin-Duque}, {Anderson}, {MacLeod}, {Eracleous}, {Ruan}, {Runnoe}, {Graham}, {Roulston}, {Schneider}, {Ahlf}, {Bizyaev}, {Brownstein}, {del Casal}, {Dodd}, {Hoover}, {Matt}, {Merloni}, {Pan}, {Ramirez}, {Ridder}, \& {Moseley}}]{green22}
{Green}, P.~J., {Pulgarin-Duque}, L., {Anderson}, S.~F., {et~al.} 2022, \bibinfo{title}{{The Time Domain Spectroscopic Survey: Changing-look Quasar Candidates from Multi-epoch Spectroscopy in SDSS-IV},} \apj, 933, 180, \dodoi{10.3847/1538-4357/ac743f}

\bibitem[{M. {Gu} \& X. {Cao}(2009){Gu} \& {Cao}}]{gu09}
{Gu}, M., \& {Cao}, X. 2009, \bibinfo{title}{{The anticorrelation between the hard X-ray photon index and the Eddington ratio in low-luminosity active galactic nuclei},} \mnras, 399, 349, \dodoi{10.1111/j.1365-2966.2009.15277.x}

\bibitem[{J. {Guillochon} \& E. {Ramirez-Ruiz}(2013){Guillochon} \& {Ramirez-Ruiz}}]{guillochon13}
{Guillochon}, J., \& {Ramirez-Ruiz}, E. 2013, \bibinfo{title}{{Hydrodynamical Simulations to Determine the Feeding Rate of Black Holes by the Tidal Disruption of Stars: The Importance of the Impact Parameter and Stellar Structure},} \apj, 767, 25, \dodoi{10.1088/0004-637X/767/1/25}

\bibitem[{K. {G{\"u}ltekin} {et~al.}(2022){G{\"u}ltekin}, {Nyland}, {Gray}, {Fehmer}, {Huang}, {Sparkman}, {Reines}, {Greene}, {Cackett}, \& {Baldassare}}]{gultekin22}
{G{\"u}ltekin}, K., {Nyland}, K., {Gray}, N., {et~al.} 2022, \bibinfo{title}{{Intermediate-mass black holes and the Fundamental Plane of black hole accretion},} \mnras, 516, 6123, \dodoi{10.1093/mnras/stac2608}

\bibitem[{J.~E. {Gunn} {et~al.}(2006){Gunn}, {Siegmund}, {Mannery}, {Owen}, {Hull}, {Leger}, {Carey}, {Knapp}, {York}, {Boroski}, {Kent}, {Lupton}, {Rockosi}, {Evans}, {Waddell}, {Anderson}, {Annis}, {Barentine}, {Bartoszek}, {Bastian}, {Bracker}, {Brewington}, {Briegel}, {Brinkmann}, {Brown}, {Carr}, {Czarapata}, {Drennan}, {Dombeck}, {Federwitz}, {Gillespie}, {Gonzales}, {Hansen}, {Harvanek}, {Hayes}, {Jordan}, {Kinney}, {Klaene}, {Kleinman}, {Kron}, {Kresinski}, {Lee}, {Limmongkol}, {Lindenmeyer}, {Long}, {Loomis}, {McGehee}, {Mantsch}, {Neilsen}, {Neswold}, {Newman}, {Nitta}, {Peoples}, {Pier}, {Prieto}, {Prosapio}, {Rivetta}, {Schneider}, {Snedden}, \& {Wang}}]{gunn06}
{Gunn}, J.~E., {Siegmund}, W.~A., {Mannery}, E.~J., {et~al.} 2006, \bibinfo{title}{{The 2.5 m Telescope of the Sloan Digital Sky Survey},} \aj, 131, 2332, \dodoi{10.1086/500975}

\bibitem[{H. {Guo} {et~al.}(2019){Guo}, {Sun}, {Liu}, {Wang}, {Kong}, {Wang}, {Sheng}, \& {He}}]{guo19}
{Guo}, H., {Sun}, M., {Liu}, X., {et~al.} 2019, \bibinfo{title}{{Discovery of an Mg II Changing-look Active Galactic Nucleus and Its Implications for a Unification Sequence of Changing-look Active Galactic Nuclei},} \apjl, 883, L44, \dodoi{10.3847/2041-8213/ab4138}

\bibitem[{H. {Guo} {et~al.}(2020){Guo}, {Shen}, {He}, {Wang}, {Liu}, {Wang}, {Sun}, {Yang}, {Kong}, \& {Sheng}}]{guo20}
{Guo}, H., {Shen}, Y., {He}, Z., {et~al.} 2020, \bibinfo{title}{{Understanding Broad Mg II Variability in Quasars with Photoionization: Implications for Reverberation Mapping and Changing-look Quasars},} \apj, 888, 58, \dodoi{10.3847/1538-4357/ab5db0}

\bibitem[{W.-J. {Guo} {et~al.}(2024{\natexlab{a}}){Guo}, {Zou}, {Fawcett}, {Canning}, {Juneau}, {Davis}, {Alexander}, {Jiang}, {Aguilar}, {Ahlen}, {Brooks}, {Claybaugh}, {de la Macorra}, {Doel}, {Fanning}, {Forero-Romero}, {Gontcho A Gontcho}, {Honscheid}, {Kisner}, {Kremin}, {Landriau}, {Meisner}, {Miquel}, {Moustakas}, {Nie}, {Pan}, {Poppett}, {Prada}, {Rezaie}, {Rossi}, {Siudek}, {Sanchez}, {Schubnell}, {Seo}, {Sui}, {Tarl{\'e}}, \& {Zhou}}]{guo24}
{Guo}, W.-J., {Zou}, H., {Fawcett}, V.~A., {et~al.} 2024{\natexlab{a}}, \bibinfo{title}{{Changing-look Active Galactic Nuclei from the Dark Energy Spectroscopic Instrument. I. Sample from the Early Data},} \apjs, 270, 26, \dodoi{10.3847/1538-4365/ad118a}

\bibitem[{W.-J. {Guo} {et~al.}(2024{\natexlab{b}}){Guo}, {Zou}, {Greenwell}, {Alexander}, {Fawcett}, {Pan}, {Siudek}, {Aguilar}, {Ahlen}, {Brooks}, {Claybaugh}, {Dawson}, {De La Macorra}, {Doel}, {Font-Ribera}, {Gaztanaga}, {Gontcho}, {Gutierrez}, {Kehoe}, {Kisner}, {Landriau}, {Le Guillou}, {Manera}, {Meisner}, {Mique}, {Moustakas}, {Prada}, {Rossi}, {Sanchez}, {Schubnell}, {Sprayberry}, {Sui}, {Tarle}, {Weaver}, {Xiao}, \& {Zou}}]{guo24_2}
{Guo}, W.-J., {Zou}, H., {Greenwell}, C.~L., {et~al.} 2024{\natexlab{b}}, \bibinfo{title}{{Changing-look Active Galactic Nuclei from the Dark Energy Spectroscopic Instrument. II. Statistical Properties from the First Data Release},} arXiv e-prints, arXiv:2408.00402, \dodoi{10.48550/arXiv.2408.00402}

\bibitem[{J.~M. {Hameury} {et~al.}(2009){Hameury}, {Viallet}, \& {Lasota}}]{hameury09}
{Hameury}, J.~M., {Viallet}, M., \& {Lasota}, J.~P. 2009, \bibinfo{title}{{The thermal-viscous disk instability model in the AGN context},} \aap, 496, 413, \dodoi{10.1051/0004-6361/200810928}

\bibitem[{ {HI4PI Collaboration} {et~al.}(2016){HI4PI Collaboration}, {Ben Bekhti}, {Fl{\"o}er}, {Keller}, {Kerp}, {Lenz}, {Winkel}, {Bailin}, {Calabretta}, {Dedes}, {Ford}, {Gibson}, {Haud}, {Janowiecki}, {Kalberla}, {Lockman}, {McClure-Griffiths}, {Murphy}, {Nakanishi}, {Pisano}, \& {Staveley-Smith}}]{hi4pi}
{HI4PI Collaboration}, {Ben Bekhti}, N., {Fl{\"o}er}, L., {et~al.} 2016, \bibinfo{title}{{HI4PI: A full-sky H I survey based on EBHIS and GASS},} \aap, 594, A116, \dodoi{10.1051/0004-6361/201629178}

\bibitem[{G.~J. {Hill} {et~al.}(2021){Hill}, {Lee}, {MacQueen}, {Kelz}, {Drory}, {Vattiat}, {Good}, {Ramsey}, {Kriel}, {Peterson}, {DePoy}, {Gebhardt}, {Marshall}, {Tuttle}, {Bauer}, {Chonis}, {Fabricius}, {Froning}, {H{\"a}user}, {Indahl}, {Jahn}, {Landriau}, {Leck}, {Montesano}, {Prochaska}, {Snigula}, {Zeimann}, {Bryant}, {Damm}, {Fowler}, {Janowiecki}, {Martin}, {Mrozinski}, {Odewahn}, {Rostopchin}, {Shetrone}, {Spencer}, {Mentuch Cooper}, {Armandroff}, {Bender}, {Dalton}, {Hopp}, {Komatsu}, {Nicklas}, {Ramsey}, {Roth}, {Schneider}, {Sneden}, \& {Steinmetz}}]{hill21}
{Hill}, G.~J., {Lee}, H., {MacQueen}, P.~J., {et~al.} 2021, \bibinfo{title}{{The HETDEX Instrumentation: Hobby-Eberly Telescope Wide-field Upgrade and VIRUS},} \aj, 162, 298, \dodoi{10.3847/1538-3881/ac2c02}

\bibitem[{L.~C. {Ho}(1999){Ho}}]{ho99}
{Ho}, L.~C. 1999, \bibinfo{title}{{The Spectral Energy Distributions of Low-Luminosity Active Galactic Nuclei},} \apj, 516, 672, \dodoi{10.1086/307137}

\bibitem[{J.~D. Hunter(2007)Hunter}]{matplotlib}
Hunter, J.~D. 2007, \bibinfo{title}{Matplotlib: A 2D graphics environment,} Computing in Science \& Engineering, 9, 90, \dodoi{10.1109/MCSE.2007.55}

\bibitem[{X. {Jin} {et~al.}(2021){Jin}, {Ruan}, {Haggard}, {Gingras}, {Hountalas}, {MacLeod}, {Anderson}, {Doan}, {Eracleous}, {Green}, \& {Runnoe}}]{jin21}
{Jin}, X., {Ruan}, J.~J., {Haggard}, D., {et~al.} 2021, \bibinfo{title}{{Probing the Disk-Corona Systems and Broad-line Regions of Changing-look Quasars with X-Ray and Optical Observations},} \apj, 912, 20, \dodoi{10.3847/1538-4357/abeb17}

\bibitem[{R.~P. {Kraft} {et~al.}(1991){Kraft}, {Burrows}, \& {Nousek}}]{kraft91}
{Kraft}, R.~P., {Burrows}, D.~N., \& {Nousek}, J.~A. 1991, \bibinfo{title}{{Determination of Confidence Limits for Experiments with Low Numbers of Counts},} \apj, 374, 344, \dodoi{10.1086/170124}

\bibitem[{S.~M. {LaMassa} {et~al.}(2015){LaMassa}, {Cales}, {Moran}, {Myers}, {Richards}, {Eracleous}, {Heckman}, {Gallo}, \& {Urry}}]{lamassa15}
{LaMassa}, S.~M., {Cales}, S., {Moran}, E.~C., {et~al.} 2015, \bibinfo{title}{{The Discovery of the First {\textquotedblleft}Changing Look{\textquotedblright} Quasar: New Insights Into the Physics and Phenomenology of Active Galactic Nucleus},} \apj, 800, 144, \dodoi{10.1088/0004-637X/800/2/144}

\bibitem[{M.~T.~P. {Liska} {et~al.}(2023){Liska}, {Kaaz}, {Musoke}, {Tchekhovskoy}, \& {Porth}}]{liska23}
{Liska}, M.~T.~P., {Kaaz}, N., {Musoke}, G., {Tchekhovskoy}, A., \& {Porth}, O. 2023, \bibinfo{title}{{Radiation Transport Two-temperature GRMHD Simulations of Warped Accretion Disks},} \apjl, 944, L48, \dodoi{10.3847/2041-8213/acb6f4}

\bibitem[{M.~T.~P. {Liska} {et~al.}(2022){Liska}, {Musoke}, {Tchekhovskoy}, {Porth}, \& {Beloborodov}}]{liska22}
{Liska}, M.~T.~P., {Musoke}, G., {Tchekhovskoy}, A., {Porth}, O., \& {Beloborodov}, A.~M. 2022, \bibinfo{title}{{Formation of Magnetically Truncated Accretion Disks in 3D Radiation-transport Two-temperature GRMHD Simulations},} \apjl, 935, L1, \dodoi{10.3847/2041-8213/ac84db}

\bibitem[{B.~F. {Liu} {et~al.}(2006){Liu}, {Meyer}, \& {Meyer-Hofmeister}}]{liu06}
{Liu}, B.~F., {Meyer}, F., \& {Meyer-Hofmeister}, E. 2006, \bibinfo{title}{{An inner disk below the ADAF: the intermediate spectral state of black hole accretion},} \aap, 454, L9, \dodoi{10.1051/0004-6361:20065430}

\bibitem[{B.~F. {Liu} {et~al.}(2002){Liu}, {Mineshige}, {Meyer}, {Meyer-Hofmeister}, \& {Kawaguchi}}]{liu02}
{Liu}, B.~F., {Mineshige}, S., {Meyer}, F., {Meyer-Hofmeister}, E., \& {Kawaguchi}, T. 2002, \bibinfo{title}{{Two-Temperature Coronal Flow above a Thin Disk},} \apj, 575, 117, \dodoi{10.1086/341138}

\bibitem[{B.~F. {Liu} \& R.~E. {Taam}(2009){Liu} \& {Taam}}]{liu09}
{Liu}, B.~F., \& {Taam}, R.~E. 2009, \bibinfo{title}{{Application of the Disk Evaporation Model to Active Galactic Nuclei},} \apj, 707, 233, \dodoi{10.1088/0004-637X/707/1/233}

\bibitem[{E. {L{\'o}pez-Navas} {et~al.}(2022){L{\'o}pez-Navas}, {Mart{\'\i}nez-Aldama}, {Bernal}, {S{\'a}nchez-S{\'a}ez}, {Ar{\'e}valo}, {Graham}, {Hern{\'a}ndez-Garc{\'\i}a}, {Lira}, \& {Rojas Lobos}}]{ln22}
{L{\'o}pez-Navas}, E., {Mart{\'\i}nez-Aldama}, M.~L., {Bernal}, S., {et~al.} 2022, \bibinfo{title}{{Confirming new changing-look AGNs discovered through optical variability using a random forest-based light-curve classifier},} \mnras, 513, L57, \dodoi{10.1093/mnrasl/slac033}

\bibitem[{E. {Lusso} {et~al.}(2010){Lusso}, {Comastri}, {Vignali}, {Zamorani}, {Brusa}, {Gilli}, {Iwasawa}, {Salvato}, {Civano}, {Elvis}, {Merloni}, {Bongiorno}, {Trump}, {Koekemoer}, {Schinnerer}, {Le Floc'h}, {Cappelluti}, {Jahnke}, {Sargent}, {Silverman}, {Mainieri}, {Fiore}, {Bolzonella}, {Le F{\`e}vre}, {Garilli}, {Iovino}, {Kneib}, {Lamareille}, {Lilly}, {Mignoli}, {Scodeggio}, \& {Vergani}}]{lusso10}
{Lusso}, E., {Comastri}, A., {Vignali}, C., {et~al.} 2010, \bibinfo{title}{{The X-ray to optical-UV luminosity ratio of X-ray selected type 1 AGN in XMM-COSMOS},} \aap, 512, A34, \dodoi{10.1051/0004-6361/200913298}

\bibitem[{B. {Lyu} {et~al.}(2022){Lyu}, {Wu}, {Yan}, {Yu}, \& {Liu}}]{lyu22}
{Lyu}, B., {Wu}, Q., {Yan}, Z., {Yu}, W., \& {Liu}, H. 2022, \bibinfo{title}{{WISE View of Changing-look Active Galactic Nuclei: Evidence for a Transitional Stage of AGNs},} \apj, 927, 227, \dodoi{10.3847/1538-4357/ac5256}

\bibitem[{B. {Lyu} {et~al.}(2025){Lyu}, {Wu}, {Pang}, {Wang}, {Zhu}, {Fu}, {Wu}, {Yan}, {Yu}, {Liu}, {Kang}, {Jin}, {Yang}, \& {Wang}}]{lyu24}
{Lyu}, B., {Wu}, X.-B., {Pang}, Y., {et~al.} 2025, \bibinfo{title}{{The changing-look AGN SDSS J101152.98+544206.4 is returning to a type I state},} \aap, 693, A173, \dodoi{10.1051/0004-6361/202451699}

\bibitem[{C.~L. {MacLeod} {et~al.}(2010){MacLeod}, {Ivezi{\'c}}, {Kochanek}, {Koz{\l}owski}, {Kelly}, {Bullock}, {Kimball}, {Sesar}, {Westman}, {Brooks}, {Gibson}, {Becker}, \& {de Vries}}]{macleod10}
{MacLeod}, C.~L., {Ivezi{\'c}}, {\v{Z}}., {Kochanek}, C.~S., {et~al.} 2010, \bibinfo{title}{{Modeling the Time Variability of SDSS Stripe 82 Quasars as a Damped Random Walk},} \apj, 721, 1014, \dodoi{10.1088/0004-637X/721/2/1014}

\bibitem[{C.~L. {MacLeod} {et~al.}(2012){MacLeod}, {Ivezi{\'c}}, {Sesar}, {de Vries}, {Kochanek}, {Kelly}, {Becker}, {Lupton}, {Hall}, {Richards}, {Anderson}, \& {Schneider}}]{macleod12}
{MacLeod}, C.~L., {Ivezi{\'c}}, {\v{Z}}., {Sesar}, B., {et~al.} 2012, \bibinfo{title}{{A Description of Quasar Variability Measured Using Repeated SDSS and POSS Imaging},} \apj, 753, 106, \dodoi{10.1088/0004-637X/753/2/106}

\bibitem[{C.~L. {MacLeod} {et~al.}(2016){MacLeod}, {Ross}, {Lawrence}, {Goad}, {Horne}, {Burgett}, {Chambers}, {Flewelling}, {Hodapp}, {Kaiser}, {Magnier}, {Wainscoat}, \& {Waters}}]{macleod16}
{MacLeod}, C.~L., {Ross}, N.~P., {Lawrence}, A., {et~al.} 2016, \bibinfo{title}{{A systematic search for changing-look quasars in SDSS},} \mnras, 457, 389, \dodoi{10.1093/mnras/stv2997}

\bibitem[{C.~L. {MacLeod} {et~al.}(2019){MacLeod}, {Green}, {Anderson}, {Bruce}, {Eracleous}, {Graham}, {Homan}, {Lawrence}, {LeBleu}, {Ross}, {Ruan}, {Runnoe}, {Stern}, {Burgett}, {Chambers}, {Kaiser}, {Magnier}, \& {Metcalfe}}]{macleod19}
{MacLeod}, C.~L., {Green}, P.~J., {Anderson}, S.~F., {et~al.} 2019, \bibinfo{title}{{Changing-look Quasar Candidates: First Results from Follow-up Spectroscopy of Highly Optically Variable Quasars},} \apj, 874, 8, \dodoi{10.3847/1538-4357/ab05e2}

\bibitem[{A. {Mainzer} {et~al.}(2014){Mainzer}, {Bauer}, {Cutri}, {Grav}, {Masiero}, {Beck}, {Clarkson}, {Conrow}, {Dailey}, {Eisenhardt}, {Fabinsky}, {Fajardo-Acosta}, {Fowler}, {Gelino}, {Grillmair}, {Heinrichsen}, {Kendall}, {Kirkpatrick}, {Liu}, {Masci}, {McCallon}, {Nugent}, {Papin}, {Rice}, {Royer}, {Ryan}, {Sevilla}, {Sonnett}, {Stevenson}, {Thompson}, {Wheelock}, {Wiemer}, {Wittman}, {Wright}, \& {Yan}}]{neowise}
{Mainzer}, A., {Bauer}, J., {Cutri}, R.~M., {et~al.} 2014, \bibinfo{title}{{Initial Performance of the NEOWISE Reactivation Mission},} \apj, 792, 30, \dodoi{10.1088/0004-637X/792/1/30}

\bibitem[{R. {Maiolino} {et~al.}(2010){Maiolino}, {Risaliti}, {Salvati}, {Pietrini}, {Torricelli-Ciamponi}, {Elvis}, {Fabbiano}, {Braito}, \& {Reeves}}]{maiolino10}
{Maiolino}, R., {Risaliti}, G., {Salvati}, M., {et~al.} 2010, \bibinfo{title}{{``Comets'' orbiting a black hole},} \aap, 517, A47, \dodoi{10.1051/0004-6361/200913985}

\bibitem[{D. {Maoz}(2007){Maoz}}]{maoz07}
{Maoz}, D. 2007, \bibinfo{title}{{Low-luminosity active galactic nuclei: are they UV faint and radio loud?},} \mnras, 377, 1696, \dodoi{10.1111/j.1365-2966.2007.11735.x}

\bibitem[{E. {Marchese} {et~al.}(2012){Marchese}, {Braito}, {Della Ceca}, {Caccianiga}, \& {Severgnini}}]{marchese12}
{Marchese}, E., {Braito}, V., {Della Ceca}, R., {Caccianiga}, A., \& {Severgnini}, P. 2012, \bibinfo{title}{{NGC 454: unveiling a new 'changing look' active galactic nucleus},} \mnras, 421, 1803, \dodoi{10.1111/j.1365-2966.2012.20445.x}

\bibitem[{S. {Markoff} {et~al.}(2001){Markoff}, {Falcke}, \& {Fender}}]{markoff01}
{Markoff}, S., {Falcke}, H., \& {Fender}, R. 2001, \bibinfo{title}{{A jet model for the broadband spectrum of XTE J1118+480. Synchrotron emission from radio to X-rays in the Low/Hard spectral state},} \aap, 372, L25, \dodoi{10.1051/0004-6361:20010420}

\bibitem[{A.~G. Markowitz {et~al.}(2014)Markowitz, Krumpe, \& Nikutta}]{markowitz14}
Markowitz, A.~G., Krumpe, M., \& Nikutta, R. 2014, \bibinfo{title}{{First X-ray-based statistical tests for clumpy-torus models: eclipse events from 230 years of monitoring of Seyfert AGN},} Monthly Notices of the Royal Astronomical Society, 439, 1403, \dodoi{10.1093/mnras/stt2492}

\bibitem[{F.~J. {Masci} {et~al.}(2019){Masci}, {Laher}, {Rusholme}, {Shupe}, {Groom}, {Surace}, {Jackson}, {Monkewitz}, {Beck}, {Flynn}, {Terek}, {Landry}, {Hacopians}, {Desai}, {Howell}, {Brooke}, {Imel}, {Wachter}, {Ye}, {Lin}, {Cenko}, {Cunningham}, {Rebbapragada}, {Bue}, {Miller}, {Mahabal}, {Bellm}, {Patterson}, {Juri{\'c}}, {Golkhou}, {Ofek}, {Walters}, {Graham}, {Kasliwal}, {Dekany}, {Kupfer}, {Burdge}, {Cannella}, {Barlow}, {Van Sistine}, {Giomi}, {Fremling}, {Blagorodnova}, {Levitan}, {Riddle}, {Smith}, {Helou}, {Prince}, \& {Kulkarni}}]{masci19}
{Masci}, F.~J., {Laher}, R.~R., {Rusholme}, B., {et~al.} 2019, \bibinfo{title}{{The Zwicky Transient Facility: Data Processing, Products, and Archive},} \pasp, 131, 018003, \dodoi{10.1088/1538-3873/aae8ac}

\bibitem[{G. {Matt} {et~al.}(2003){Matt}, {Guainazzi}, \& {Maiolino}}]{matt03}
{Matt}, G., {Guainazzi}, M., \& {Maiolino}, R. 2003, \bibinfo{title}{{Changing look: from Compton-thick to Compton-thin, or the rebirth of fossil active galactic nuclei},} \mnras, 342, 422, \dodoi{10.1046/j.1365-8711.2003.06539.x}

\bibitem[{J.~H. {Matthews} {et~al.}(2020){Matthews}, {Knigge}, {Higginbottom}, {Long}, {Sim}, {Mangham}, {Parkinson}, \& {Hewitt}}]{matthews20}
{Matthews}, J.~H., {Knigge}, C., {Higginbottom}, N., {et~al.} 2020, \bibinfo{title}{{Stratified disc wind models for the AGN broad-line region: ultraviolet, optical, and X-ray properties},} \mnras, 492, 5540, \dodoi{10.1093/mnras/staa136}

\bibitem[{J.~H. {Matthews} {et~al.}(2023){Matthews}, {Strong-Wright}, {Knigge}, {Hewett}, {Temple}, {Long}, {Rankine}, {Stepney}, {Banerji}, \& {Richards}}]{matthews23}
{Matthews}, J.~H., {Strong-Wright}, J., {Knigge}, C., {et~al.} 2023, \bibinfo{title}{{A disc wind model for blueshifts in quasar broad emission lines},} \mnras, 526, 3967, \dodoi{10.1093/mnras/stad2895}

\bibitem[{I.~M. {McHardy} {et~al.}(2006){McHardy}, {Koerding}, {Knigge}, {Uttley}, \& {Fender}}]{mchardy06}
{McHardy}, I.~M., {Koerding}, E., {Knigge}, C., {Uttley}, P., \& {Fender}, R.~P. 2006, \bibinfo{title}{{Active galactic nuclei as scaled-up Galactic black holes},} \nat, 444, 730, \dodoi{10.1038/nature05389}

\bibitem[{A. {Merloni} {et~al.}(2003){Merloni}, {Heinz}, \& {di Matteo}}]{merloni03}
{Merloni}, A., {Heinz}, S., \& {di Matteo}, T. 2003, \bibinfo{title}{{A Fundamental Plane of black hole activity},} \mnras, 345, 1057, \dodoi{10.1046/j.1365-2966.2003.07017.x}

\bibitem[{A. {Merloni} {et~al.}(2015){Merloni}, {Dwelly}, {Salvato}, {Georgakakis}, {Greiner}, {Krumpe}, {Nandra}, {Ponti}, \& {Rau}}]{merloni15}
{Merloni}, A., {Dwelly}, T., {Salvato}, M., {et~al.} 2015, \bibinfo{title}{{A tidal disruption flare in a massive galaxy? Implications for the fuelling mechanisms of nuclear black holes},} \mnras, 452, 69, \dodoi{10.1093/mnras/stv1095}

\bibitem[{F. {Meyer} {et~al.}(2000){Meyer}, {Liu}, \& {Meyer-Hofmeister}}]{meyer00}
{Meyer}, F., {Liu}, B.~F., \& {Meyer-Hofmeister}, E. 2000, \bibinfo{title}{{Evaporation: The change from accretion via a thin disk to a coronal flow},} \aap, 361, 175, \dodoi{10.48550/arXiv.astro-ph/0007091}

\bibitem[{F. {Meyer} \& E. {Meyer-Hofmeister}(1994){Meyer} \& {Meyer-Hofmeister}}]{meyer94}
{Meyer}, F., \& {Meyer-Hofmeister}, E. 1994, \bibinfo{title}{{Accretion disk evaporation by a coronal siphon flow.},} \aap, 288, 175

\bibitem[{E. {Meyer-Hofmeister} \& F. {Meyer}(2011){Meyer-Hofmeister} \& {Meyer}}]{mmh11}
{Meyer-Hofmeister}, E., \& {Meyer}, F. 2011, \bibinfo{title}{{Broad iron emission lines in Seyfert galaxies - re-condensation of gas onto an inner disk below the ADAF?},} \aap, 527, A127, \dodoi{10.1051/0004-6361/201015478}

\bibitem[{N. {Murray} \& J. {Chiang}(1997){Murray} \& {Chiang}}]{murray97}
{Murray}, N., \& {Chiang}, J. 1997, \bibinfo{title}{{Disk Winds and Disk Emission Lines},} \apj, 474, 91, \dodoi{10.1086/303443}

\bibitem[{M.-H. {Naddaf} {et~al.}(2021){Naddaf}, {Czerny}, \& {Szczerba}}]{naddaf21}
{Naddaf}, M.-H., {Czerny}, B., \& {Szczerba}, R. 2021, \bibinfo{title}{{The Picture of BLR in 2.5D FRADO: Dynamics and Geometry},} \apj, 920, 30, \dodoi{10.3847/1538-4357/ac139d}

\bibitem[{R. {Narayan} \& I. {Yi}(1994){Narayan} \& {Yi}}]{narayan94}
{Narayan}, R., \& {Yi}, I. 1994, \bibinfo{title}{{Advection-dominated Accretion: A Self-similar Solution},} \apjl, 428, L13, \dodoi{10.1086/187381}

\bibitem[{R.~S. {Nemmen} {et~al.}(2014){Nemmen}, {Storchi-Bergmann}, \& {Eracleous}}]{nemmen14}
{Nemmen}, R.~S., {Storchi-Bergmann}, T., \& {Eracleous}, M. 2014, \bibinfo{title}{{Spectral models for low-luminosity active galactic nuclei in LINERs: the role of advection-dominated accretion and jets},} \mnras, 438, 2804, \dodoi{10.1093/mnras/stt2388}

\bibitem[{H. {Noda} \& C. {Done}(2018){Noda} \& {Done}}]{noda18}
{Noda}, H., \& {Done}, C. 2018, \bibinfo{title}{{Explaining changing-look AGN with state transition triggered by rapid mass accretion rate drop},} \mnras, 480, 3898, \dodoi{10.1093/mnras/sty2032}

\bibitem[{B. {Palit} {et~al.}(2025){Palit}, {{\'S}niegowska}, {Markowitz}, {R{\'o}{\.z}a{\'n}ska}, {Farah}, \& {Howell}}]{palit25}
{Palit}, B., {{\'S}niegowska}, M., {Markowitz}, A., {et~al.} 2025, \bibinfo{title}{{Markarian 590: The AGN Awakens},} \mnras, \dodoi{10.1093/mnrasl/slaf027}

\bibitem[{T. pandas~development team(2024)pandas~development team}]{pandas}
pandas~development team, T. 2024, \bibinfo{title}{pandas-dev/pandas: Pandas,}, v2.2.2 Zenodo, \dodoi{10.5281/zenodo.10957263}

\bibitem[{E. {Piconcelli} {et~al.}(2007){Piconcelli}, {Fiore}, {Nicastro}, {Mathur}, {Brusa}, {Comastri}, \& {Puccetti}}]{piconcelli03}
{Piconcelli}, E., {Fiore}, F., {Nicastro}, F., {et~al.} 2007, \bibinfo{title}{{The XMM-Newton view of IRAS 09104+4109: evidence for a changing-look Type 2 quasar?},} \aap, 473, 85, \dodoi{10.1051/0004-6361:20077630}

\bibitem[{T.~S. {Poole} {et~al.}(2008){Poole}, {Breeveld}, {Page}, {Landsman}, {Holland}, {Roming}, {Kuin}, {Brown}, {Gronwall}, {Hunsberger}, {Koch}, {Mason}, {Schady}, {vanden Berk}, {Blustin}, {Boyd}, {Broos}, {Carter}, {Chester}, {Cucchiara}, {Hancock}, {Huckle}, {Immler}, {Ivanushkina}, {Kennedy}, {Marshall}, {Morgan}, {Pandey}, {de Pasquale}, {Smith}, \& {Still}}]{poole08}
{Poole}, T.~S., {Breeveld}, A.~A., {Page}, M.~J., {et~al.} 2008, \bibinfo{title}{{Photometric calibration of the Swift ultraviolet/optical telescope},} \mnras, 383, 627, \dodoi{10.1111/j.1365-2966.2007.12563.x}

\bibitem[{L.~W. {Ramsey} {et~al.}(1998){Ramsey}, {Adams}, {Barnes}, {Booth}, {Cornell}, {Fowler}, {Gaffney}, {Glaspey}, {Good}, {Hill}, {Kelton}, {Krabbendam}, {Long}, {MacQueen}, {Ray}, {Ricklefs}, {Sage}, {Sebring}, {Spiesman}, \& {Steiner}}]{ramsey98}
{Ramsey}, L.~W., {Adams}, M.~T., {Barnes}, T.~G., {et~al.} 1998, in Society of Photo-Optical Instrumentation Engineers (SPIE) Conference Series, Vol. 3352, Advanced Technology Optical/IR Telescopes VI, ed. L.~M. {Stepp}, 34--42, \dodoi{10.1117/12.319287}

\bibitem[{C. {Ricci} \& B. {Trakhtenbrot}(2023){Ricci} \& {Trakhtenbrot}}]{ricci23}
{Ricci}, C., \& {Trakhtenbrot}, B. 2023, \bibinfo{title}{{Changing-look active galactic nuclei},} Nature Astronomy, 7, 1282, \dodoi{10.1038/s41550-023-02108-4}

\bibitem[{C. {Ricci} {et~al.}(2016){Ricci}, {Bauer}, {Arevalo}, {Boggs}, {Brandt}, {Christensen}, {Craig}, {Gandhi}, {Hailey}, {Harrison}, {Koss}, {Markwardt}, {Stern}, {Treister}, \& {Zhang}}]{ricci16}
{Ricci}, C., {Bauer}, F.~E., {Arevalo}, P., {et~al.} 2016, \bibinfo{title}{{IC 751: A New Changing Look AGN Discovered by NuSTAR},} \apj, 820, 5, \dodoi{10.3847/0004-637X/820/1/5}

\bibitem[{P.~W.~A. {Roming} {et~al.}(2005){Roming}, {Kennedy}, {Mason}, {Nousek}, {Ahr}, {Bingham}, {Broos}, {Carter}, {Hancock}, {Huckle}, {Hunsberger}, {Kawakami}, {Killough}, {Koch}, {McLelland}, {Smith}, {Smith}, {Soto}, {Boyd}, {Breeveld}, {Holland}, {Ivanushkina}, {Pryzby}, {Still}, \& {Stock}}]{roming05}
{Roming}, P. W.~A., {Kennedy}, T.~E., {Mason}, K.~O., {et~al.} 2005, \bibinfo{title}{{The Swift Ultra-Violet/Optical Telescope},} \ssr, 120, 95, \dodoi{10.1007/s11214-005-5095-4}

\bibitem[{N.~P. {Ross} {et~al.}(2018){Ross}, {Ford}, {Graham}, {McKernan}, {Stern}, {Meisner}, {Assef}, {Dey}, {Drake}, {Jun}, \& {Lang}}]{ross18}
{Ross}, N.~P., {Ford}, K.~E.~S., {Graham}, M., {et~al.} 2018, \bibinfo{title}{{A new physical interpretation of optical and infrared variability in quasars},} \mnras, 480, 4468, \dodoi{10.1093/mnras/sty2002}

\bibitem[{J.~J. {Ruan} {et~al.}(2019){Ruan}, {Anderson}, {Eracleous}, {Green}, {Haggard}, {MacLeod}, {Runnoe}, \& {Sobolewska}}]{ruan19}
{Ruan}, J.~J., {Anderson}, S.~F., {Eracleous}, M., {et~al.} 2019, \bibinfo{title}{{The Analogous Structure of Accretion Flows in Supermassive and Stellar Mass Black Holes: New Insights from Faded Changing-look Quasars},} \apj, 883, 76, \dodoi{10.3847/1538-4357/ab3c1a}

\bibitem[{J.~J. {Ruan} {et~al.}(2016){Ruan}, {Anderson}, {Cales}, {Eracleous}, {Green}, {Morganson}, {Runnoe}, {Shen}, {Wilkinson}, {Blanton}, {Dwelly}, {Georgakakis}, {Greene}, {LaMassa}, {Merloni}, \& {Schneider}}]{ruan16}
{Ruan}, J.~J., {Anderson}, S.~F., {Cales}, S.~L., {et~al.} 2016, \bibinfo{title}{{Toward an Understanding of Changing-look Quasars: An Archival Spectroscopic Search in SDSS},} \apj, 826, 188, \dodoi{10.3847/0004-637X/826/2/188}

\bibitem[{N. {Rumbaugh} {et~al.}(2018){Rumbaugh}, {Shen}, {Morganson}, {Liu}, {Banerji}, {McMahon}, {Abdalla}, {Benoit-L{\'e}vy}, {Bertin}, {Brooks}, {Buckley-Geer}, {Capozzi}, {Carnero Rosell}, {Carrasco Kind}, {Carretero}, {Cunha}, {D'Andrea}, {da Costa}, {DePoy}, {Desai}, {Doel}, {Frieman}, {Garc{\'\i}a-Bellido}, {Gruen}, {Gruendl}, {Gschwend}, {Gutierrez}, {Honscheid}, {James}, {Kuehn}, {Kuhlmann}, {Kuropatkin}, {Lima}, {Maia}, {Marshall}, {Martini}, {Menanteau}, {Plazas}, {Reil}, {Roodman}, {Sanchez}, {Scarpine}, {Schindler}, {Schubnell}, {Sheldon}, {Smith}, {Soares-Santos}, {Sobreira}, {Suchyta}, {Swanson}, {Walker}, {Wester}, \& {DES Collaboration}}]{rumbaugh18}
{Rumbaugh}, N., {Shen}, Y., {Morganson}, E., {et~al.} 2018, \bibinfo{title}{{Extreme Variability Quasars from the Sloan Digital Sky Survey and the Dark Energy Survey},} \apj, 854, 160, \dodoi{10.3847/1538-4357/aaa9b6}

\bibitem[{J.~C. {Runnoe} {et~al.}(2012){Runnoe}, {Brotherton}, \& {Shang}}]{runnoe12}
{Runnoe}, J.~C., {Brotherton}, M.~S., \& {Shang}, Z. 2012, \bibinfo{title}{{Updating quasar bolometric luminosity corrections},} \mnras, 422, 478, \dodoi{10.1111/j.1365-2966.2012.20620.x}

\bibitem[{J.~C. {Runnoe} {et~al.}(2015){Runnoe}, {Eracleous}, {Mathes}, {Pennell}, {Boroson}, {Sigur{\dh}sson}, {Bogdanovi{\'c}}, {Halpern}, \& {Liu}}]{runnoe15}
{Runnoe}, J.~C., {Eracleous}, M., {Mathes}, G., {et~al.} 2015, \bibinfo{title}{{A Large Systematic Search for Close Supermassive Binary and Rapidly Recoiling Black Holes. II. Continued Spectroscopic Monitoring and Optical Flux Variability},} \apjs, 221, 7, \dodoi{10.1088/0067-0049/221/1/7}

\bibitem[{J.~C. {Runnoe} {et~al.}(2016){Runnoe}, {Cales}, {Ruan}, {Eracleous}, {Anderson}, {Shen}, {Green}, {Morganson}, {LaMassa}, {Greene}, {Dwelly}, {Schneider}, {Merloni}, {Georgakakis}, \& {Roman-Lopes}}]{runnoe16}
{Runnoe}, J.~C., {Cales}, S., {Ruan}, J.~J., {et~al.} 2016, \bibinfo{title}{{Now you see it, now you don't: the disappearing central engine of the quasar J1011+5442},} \mnras, 455, 1691, \dodoi{10.1093/mnras/stv2385}

\bibitem[{S. {Salim} \& D. {Narayanan}(2020){Salim} \& {Narayanan}}]{salim20}
{Salim}, S., \& {Narayanan}, D. 2020, \bibinfo{title}{{The Dust Attenuation Law in Galaxies},} \araa, 58, 529, \dodoi{10.1146/annurev-astro-032620-021933}

\bibitem[{E.~F. {Schlafly} \& D.~P. {Finkbeiner}(2011){Schlafly} \& {Finkbeiner}}]{schlafly11}
{Schlafly}, E.~F., \& {Finkbeiner}, D.~P. 2011, \bibinfo{title}{{Measuring Reddening with Sloan Digital Sky Survey Stellar Spectra and Recalibrating SFD},} \apj, 737, 103, \dodoi{10.1088/0004-637X/737/2/103}

\bibitem[{N.~I. {Shakura} \& R.~A. {Sunyaev}(1973){Shakura} \& {Sunyaev}}]{shakura73}
{Shakura}, N.~I., \& {Sunyaev}, R.~A. 1973, \bibinfo{title}{{Black holes in binary systems. Observational appearance.},} \aap, 24, 337

\bibitem[{S.~L. {Shapiro} {et~al.}(1976){Shapiro}, {Lightman}, \& {Eardley}}]{shapiro76}
{Shapiro}, S.~L., {Lightman}, A.~P., \& {Eardley}, D.~M. 1976, \bibinfo{title}{{A two-temperature accretion disk model for Cygnus X-1: structure and spectrum.},} \apj, 204, 187, \dodoi{10.1086/154162}

\bibitem[{Z. {Sheng} {et~al.}(2020){Sheng}, {Wang}, {Jiang}, {Ding}, {Cai}, {Guo}, {Sun}, {Dou}, \& {Yang}}]{sheng20}
{Sheng}, Z., {Wang}, T., {Jiang}, N., {et~al.} 2020, \bibinfo{title}{{Initial Results from a Systematic Search for Changing-look Active Galactic Nuclei Selected via Mid-infrared Variability},} \apj, 889, 46, \dodoi{10.3847/1538-4357/ab5af9}

\bibitem[{S.~A. {Smee} {et~al.}(2013){Smee}, {Gunn}, {Uomoto}, {Roe}, {Schlegel}, {Rockosi}, {Carr}, {Leger}, {Dawson}, {Olmstead}, {Brinkmann}, {Owen}, {Barkhouser}, {Honscheid}, {Harding}, {Long}, {Lupton}, {Loomis}, {Anderson}, {Annis}, {Bernardi}, {Bhardwaj}, {Bizyaev}, {Bolton}, {Brewington}, {Briggs}, {Burles}, {Burns}, {Castander}, {Connolly}, {Davenport}, {Ebelke}, {Epps}, {Feldman}, {Friedman}, {Frieman}, {Heckman}, {Hull}, {Knapp}, {Lawrence}, {Loveday}, {Mannery}, {Malanushenko}, {Malanushenko}, {Merrelli}, {Muna}, {Newman}, {Nichol}, {Oravetz}, {Pan}, {Pope}, {Ricketts}, {Shelden}, {Sandford}, {Siegmund}, {Simmons}, {Smith}, {Snedden}, {Schneider}, {SubbaRao}, {Tremonti}, {Waddell}, \& {York}}]{smee13}
{Smee}, S.~A., {Gunn}, J.~E., {Uomoto}, A., {et~al.} 2013, \bibinfo{title}{{The Multi-object, Fiber-fed Spectrographs for the Sloan Digital Sky Survey and the Baryon Oscillation Spectroscopic Survey},} \aj, 146, 32, \dodoi{10.1088/0004-6256/146/2/32}

\bibitem[{M. {Sniegowska} {et~al.}(2020){Sniegowska}, {Czerny}, {Bon}, \& {Bon}}]{sniegowska20}
{Sniegowska}, M., {Czerny}, B., {Bon}, E., \& {Bon}, N. 2020, \bibinfo{title}{{Possible mechanism for multiple changing-look phenomena in active galactic nuclei},} \aap, 641, A167, \dodoi{10.1051/0004-6361/202038575}

\bibitem[{M.~A. {Sobolewska} {et~al.}(2011){Sobolewska}, {Siemiginowska}, \& {Gierli{\'n}ski}}]{sobolewska11}
{Sobolewska}, M.~A., {Siemiginowska}, A., \& {Gierli{\'n}ski}, M. 2011, \bibinfo{title}{{Simulated spectral states of active galactic nuclei and observational predictions},} \mnras, 413, 2259, \dodoi{10.1111/j.1365-2966.2011.18302.x}

\bibitem[{J.~J. {Somalwar} {et~al.}(2023){Somalwar}, {Ravi}, {Yao}, {Guolo}, {Graham}, {Hammerstein}, {Lu}, {Nicholl}, {Sharma}, {Stein}, {van Velzen}, {Bellm}, {Coughlin}, {Groom}, {Masci}, \& {Riddle}}]{somalwar23}
{Somalwar}, J.~J., {Ravi}, V., {Yao}, Y., {et~al.} 2023, \bibinfo{title}{{The first systematically identified repeating partial tidal disruption event},} arXiv e-prints, arXiv:2310.03782, \dodoi{10.48550/arXiv.2310.03782}

\bibitem[{D. {Stern} {et~al.}(2018){Stern}, {McKernan}, {Graham}, {Ford}, {Ross}, {Meisner}, {Assef}, {Balokovi{\'c}}, {Brightman}, {Dey}, {Drake}, {Djorgovski}, {Eisenhardt}, \& {Jun}}]{stern18}
{Stern}, D., {McKernan}, B., {Graham}, M.~J., {et~al.} 2018, \bibinfo{title}{{A Mid-IR Selected Changing-look Quasar and Physical Scenarios for Abrupt AGN Fading},} \apj, 864, 27, \dodoi{10.3847/1538-4357/aac726}

\bibitem[{Z. {Stone} {et~al.}(2022){Stone}, {Shen}, {Burke}, {Chen}, {Yang}, {Liu}, {Gruendl}, {Adam{\'o}w}, {Andrade-Oliveira}, {Annis}, {Bacon}, {Bertin}, {Bocquet}, {Brooks}, {Burke}, {Carnero Rosell}, {Carrasco Kind}, {Carretero}, {da Costa}, {Pereira}, {De Vicente}, {Desai}, {Diehl}, {Doel}, {Ferrero}, {Friedel}, {Frieman}, {Garc{\'\i}a-Bellido}, {Gaztanaga}, {Gruen}, {Gutierrez}, {Hinton}, {Hollowood}, {Honscheid}, {James}, {Kuehn}, {Kuropatkin}, {Lidman}, {Maia}, {Menanteau}, {Miquel}, {Morgan}, {Paz-Chinch{\'o}n}, {Pieres}, {Plazas Malag{\'o}n}, {Rodriguez-Monroy}, {Sanchez}, {Scarpine}, {Serrano}, {Sevilla-Noarbe}, {Smith}, {Suchyta}, {Swanson}, {Tarl{\'e}}, {To}, \& {DES Collaboration}}]{stone22}
{Stone}, Z., {Shen}, Y., {Burke}, C.~J., {et~al.} 2022, \bibinfo{title}{{Optical variability of quasars with 20-yr photometric light curves},} \mnras, 514, 164, \dodoi{10.1093/mnras/stac1259}

\bibitem[{J. {Sun} {et~al.}(2025){Sun}, {Guo}, {Gu}, {Li}, {Chen}, {Gonz{\'a}lez-Buitrago}, {Wang}, {Li}, {Feng}, {Xiong}, {Wang}, {Yuan}, {Jin}, {Zhang}, {Deng}, \& {Zhang}}]{sun25}
{Sun}, J., {Guo}, H., {Gu}, M., {et~al.} 2025, \bibinfo{title}{{AT2021aeuk: A Repeating Partial Tidal Disruption Event Candidate in a Narrow-line Seyfert 1 Galaxy},} \apj, 982, 150, \dodoi{10.3847/1538-4357/adb724}

\bibitem[{L. {Sun} {et~al.}(2024){Sun}, {Jiang}, {Dou}, {Shu}, {Zhu}, {Dong}, {Buckley}, {Bradley Cenko}, {Fan}, {Gromadzki}, {Liu}, {Wang}, {Wang}, {Wang}, {Wu}, {Yang}, {Zhang}, {Zhang}, \& {Zhang}}]{sun24}
{Sun}, L., {Jiang}, N., {Dou}, L., {et~al.} 2024, \bibinfo{title}{{Recurring tidal disruption events a decade apart in IRAS F01004-2237},} \aap, 692, A262, \dodoi{10.1051/0004-6361/202452380}

\bibitem[{H. {Tananbaum} {et~al.}(1979){Tananbaum}, {Avni}, {Branduardi}, {Elvis}, {Fabbiano}, {Feigelson}, {Giacconi}, {Henry}, {Pye}, {Soltan}, \& {Zamorani}}]{tananbaum79}
{Tananbaum}, H., {Avni}, Y., {Branduardi}, G., {et~al.} 1979, \bibinfo{title}{{X-ray studies of quasars with the Einstein Observatory.},} \apjl, 234, L9, \dodoi{10.1086/183100}

\bibitem[{B. {Trakhtenbrot} {et~al.}(2019){Trakhtenbrot}, {Arcavi}, {MacLeod}, {Ricci}, {Kara}, {Graham}, {Stern}, {Harrison}, {Burke}, {Hiramatsu}, {Hosseinzadeh}, {Howell}, {Smartt}, {Rest}, {Prieto}, {Shappee}, {Holoien}, {Bersier}, {Filippenko}, {Brink}, {Zheng}, {Li}, {Remillard}, \& {Loewenstein}}]{trakhtenbrot19}
{Trakhtenbrot}, B., {Arcavi}, I., {MacLeod}, C.~L., {et~al.} 2019, \bibinfo{title}{{1ES 1927+654: An AGN Caught Changing Look on a Timescale of Months},} \apj, 883, 94, \dodoi{10.3847/1538-4357/ab39e4}

\bibitem[{S. {van Velzen}(2018){van Velzen}}]{vv18}
{van Velzen}, S. 2018, \bibinfo{title}{{On the Mass and Luminosity Functions of Tidal Disruption Flares: Rate Suppression due to Black Hole Event Horizons},} \apj, 852, 72, \dodoi{10.3847/1538-4357/aa998e}

\bibitem[{D.~E. {Vanden Berk} {et~al.}(2004){Vanden Berk}, {Wilhite}, {Kron}, {Anderson}, {Brunner}, {Hall}, {Ivezi{\'c}}, {Richards}, {Schneider}, {York}, {Brinkmann}, {Lamb}, {Nichol}, \& {Schlegel}}]{vandenberk04}
{Vanden Berk}, D.~E., {Wilhite}, B.~C., {Kron}, R.~G., {et~al.} 2004, \bibinfo{title}{{The Ensemble Photometric Variability of \raisebox{-0.5ex}\textasciitilde25,000 Quasars in the Sloan Digital Sky Survey},} \apj, 601, 692, \dodoi{10.1086/380563}

\bibitem[{P.~M. {Veres} {et~al.}(2024){Veres}, {Franckowiak}, {van Velzen}, {Adebahr}, {Taziaux}, {Necker}, {Stein}, {Kier}, {Mueller}, {Bomans}, {Jordana-Mitjans}, {Kowalski}, {Hammerstein}, {Marci-Boehncke}, {Reusch}, {Garrappa}, {Rose}, \& {Kashyap Das}}]{veres24}
{Veres}, P.~M., {Franckowiak}, A., {van Velzen}, S., {et~al.} 2024, \bibinfo{title}{{Back from the dead: AT2019aalc as a candidate repeating TDE in an AGN},} arXiv e-prints, arXiv:2408.17419, \dodoi{10.48550/arXiv.2408.17419}

\bibitem[{S. {Veronese} {et~al.}(2024){Veronese}, {Vignali}, {Severgnini}, {Matzeu}, \& {Cignoni}}]{veronese24}
{Veronese}, S., {Vignali}, C., {Severgnini}, P., {Matzeu}, G.~A., \& {Cignoni}, M. 2024, \bibinfo{title}{{Interpreting the long-term variability of the changing-look AGN Mrk 1018},} \aap, 683, A131, \dodoi{10.1051/0004-6361/202348098}

\bibitem[{P. Virtanen {et~al.}(2020)Virtanen, Gommers, Oliphant, Haberland, Reddy, Cournapeau, Burovski, Peterson, Weckesser, Bright, {van der Walt}, Brett, Wilson, Millman, Mayorov, Nelson, Jones, Kern, Larson, Carey, Polat, Feng, Moore, {VanderPlas}, Laxalde, Perktold, Cimrman, Henriksen, Quintero, Harris, Archibald, Ribeiro, Pedregosa, {van Mulbregt}, \& {SciPy 1.0 Contributors}}]{scipy}
Virtanen, P., Gommers, R., Oliphant, T.~E., {et~al.} 2020, \bibinfo{title}{{{SciPy} 1.0: Fundamental Algorithms for Scientific Computing in Python},} Nature Methods, 17, 261, \dodoi{10.1038/s41592-019-0686-2}

\bibitem[{J. {Wang} {et~al.}(2024){Wang}, {Xu}, {Cao}, {Gao}, {Xie}, \& {Wei}}]{wangj24}
{Wang}, J., {Xu}, D.~W., {Cao}, X., {et~al.} 2024, \bibinfo{title}{{Instability of Circumnuclear Gas Supply as an Origin of the ``Changing-look'' Phenomenon of Supermassive Black Holes},} \apj, 970, 85, \dodoi{10.3847/1538-4357/ad4d89}

\bibitem[{S. {Wang} {et~al.}(2024{\natexlab{a}}){Wang}, {Woo}, {Gallo}, {Son}, {Yang}, {Jin}, {Guo}, \& {Kong}}]{wang24b}
{Wang}, S., {Woo}, J.-H., {Gallo}, E., {et~al.} 2024{\natexlab{a}}, \bibinfo{title}{{Dormancy and Reawakening Over Years: Eight New Recurrent Changing-Look AGNs},} arXiv e-prints, arXiv:2410.15587, \dodoi{10.48550/arXiv.2410.15587}

\bibitem[{S. {Wang} {et~al.}(2024{\natexlab{b}}){Wang}, {Woo}, {Gallo}, {Guo}, {Son}, {Kong}, {Mandal}, {Cho}, {Kim}, \& {Shin}}]{wang24}
{Wang}, S., {Woo}, J.-H., {Gallo}, E., {et~al.} 2024{\natexlab{b}}, \bibinfo{title}{{Identifying Changing-look AGNs Using Variability Characteristics},} \apj, 966, 128, \dodoi{10.3847/1538-4357/ad3049}

\bibitem[{T. {Waters} {et~al.}(2016){Waters}, {Kashi}, {Proga}, {Eracleous}, {Barth}, \& {Greene}}]{waters16}
{Waters}, T., {Kashi}, A., {Proga}, D., {et~al.} 2016, \bibinfo{title}{{Reverberation Mapping of the Broad Line Region: Application to a Hydrodynamical Line-driven Disk Wind Solution},} \apj, 827, 53, \dodoi{10.3847/0004-637X/827/1/53}

\bibitem[{T. {Wevers} {et~al.}(2023){Wevers}, {Coughlin}, {Pasham}, {Guolo}, {Sun}, {Wen}, {Jonker}, {Zabludoff}, {Malyali}, {Arcodia}, {Liu}, {Merloni}, {Rau}, {Grotova}, {Short}, \& {Cao}}]{wevers23}
{Wevers}, T., {Coughlin}, E.~R., {Pasham}, D.~R., {et~al.} 2023, \bibinfo{title}{{Live to Die Another Day: The Rebrightening of AT 2018fyk as a Repeating Partial Tidal Disruption Event},} \apjl, 942, L33, \dodoi{10.3847/2041-8213/ac9f36}

\bibitem[{E.~L. {Wright} {et~al.}(2010){Wright}, {Eisenhardt}, {Mainzer}, {Ressler}, {Cutri}, {Jarrett}, {Kirkpatrick}, {Padgett}, {McMillan}, {Skrutskie}, {Stanford}, {Cohen}, {Walker}, {Mather}, {Leisawitz}, {Gautier}, {McLean}, {Benford}, {Lonsdale}, {Blain}, {Mendez}, {Irace}, {Duval}, {Liu}, {Royer}, {Heinrichsen}, {Howard}, {Shannon}, {Kendall}, {Walsh}, {Larsen}, {Cardon}, {Schick}, {Schwalm}, {Abid}, {Fabinsky}, {Naes}, \& {Tsai}}]{wright10}
{Wright}, E.~L., {Eisenhardt}, P. R.~M., {Mainzer}, A.~K., {et~al.} 2010, \bibinfo{title}{{The Wide-field Infrared Survey Explorer (WISE): Mission Description and Initial On-orbit Performance},} \aj, 140, 1868, \dodoi{10.1088/0004-6256/140/6/1868}

\bibitem[{W.-B. {Wu} \& W.-M. {Gu}(2023){Wu} \& {Gu}}]{wugu23}
{Wu}, W.-B., \& {Gu}, W.-M. 2023, \bibinfo{title}{{Magnetized Accretion Disks with Outflows for Changing-look AGNs},} \apj, 958, 146, \dodoi{10.3847/1538-4357/acf839}

\bibitem[{Q. {Yang} {et~al.}(2025){Yang}, {Green}, {Wu}, {Eracleous}, {Jiang}, \& {Fu}}]{yang24}
{Yang}, Q., {Green}, P.~J., {Wu}, X.-B., {et~al.} 2025, \bibinfo{title}{{Galaxies Lighting Up: Discovery of Seventy New Turn-on Changing-look Active Galactic Nuclei},} \apj, 980, 91, \dodoi{10.3847/1538-4357/ad94ed}

\bibitem[{Q. {Yang} {et~al.}(2019){Yang}, {Shen}, {Liu}, {Wu}, {Jiang}, {Shangguan}, {Graham}, \& {Yao}}]{yang19}
{Yang}, Q., {Shen}, Y., {Liu}, X., {et~al.} 2019, \bibinfo{title}{{An Unusual Mid-infrared Flare in a Type 2 AGN: An Obscured Turning-on AGN or Tidal Disruption Event?},} \apj, 885, 110, \dodoi{10.3847/1538-4357/ab481a}

\bibitem[{Q. {Yang} {et~al.}(2018){Yang}, {Wu}, {Fan}, {Jiang}, {McGreer}, {Shangguan}, {Yao}, {Wang}, {Joshi}, {Green}, {Wang}, {Feng}, {Fu}, {Yang}, \& {Liu}}]{yang18}
{Yang}, Q., {Wu}, X.-B., {Fan}, X., {et~al.} 2018, \bibinfo{title}{{Discovery of 21 New Changing-look AGNs in the Northern Sky},} \apj, 862, 109, \dodoi{10.3847/1538-4357/aaca3a}

\bibitem[{Q. {Yang} {et~al.}(2023){Yang}, {Green}, {MacLeod}, {Plotkin}, {Anderson}, {Bieryla}, {Civano}, {Eracleous}, {Graham}, {Ruan}, {Runnoe}, \& {Zhao}}]{yang23}
{Yang}, Q., {Green}, P.~J., {MacLeod}, C.~L., {et~al.} 2023, \bibinfo{title}{{Probing the Origin of Changing-look Quasar Transitions with Chandra},} \apj, 953, 61, \dodoi{10.3847/1538-4357/acdedd}

\bibitem[{Q.-X. {Yang} {et~al.}(2015){Yang}, {Xie}, {Yuan}, {Zdziarski}, {Gierli{\'n}ski}, {Ho}, \& {Yu}}]{yang15}
{Yang}, Q.-X., {Xie}, F.-G., {Yuan}, F., {et~al.} 2015, \bibinfo{title}{{Correlation between the photon index and X-ray luminosity of black hole X-ray binaries and active galactic nuclei: observations and interpretation},} \mnras, 447, 1692, \dodoi{10.1093/mnras/stu2571}

\bibitem[{D.~G. {York} {et~al.}(2000){York}, {Adelman}, {Anderson}, {Anderson}, {Annis}, {Bahcall}, {Bakken}, {Barkhouser}, {Bastian}, {Berman}, {Boroski}, {Bracker}, {Briegel}, {Briggs}, {Brinkmann}, {Brunner}, {Burles}, {Carey}, {Carr}, {Castander}, {Chen}, {Colestock}, {Connolly}, {Crocker}, {Csabai}, {Czarapata}, {Davis}, {Doi}, {Dombeck}, {Eisenstein}, {Ellman}, {Elms}, {Evans}, {Fan}, {Federwitz}, {Fiscelli}, {Friedman}, {Frieman}, {Fukugita}, {Gillespie}, {Gunn}, {Gurbani}, {de Haas}, {Haldeman}, {Harris}, {Hayes}, {Heckman}, {Hennessy}, {Hindsley}, {Holm}, {Holmgren}, {Huang}, {Hull}, {Husby}, {Ichikawa}, {Ichikawa}, {Ivezi{\'c}}, {Kent}, {Kim}, {Kinney}, {Klaene}, {Kleinman}, {Kleinman}, {Knapp}, {Korienek}, {Kron}, {Kunszt}, {Lamb}, {Lee}, {Leger}, {Limmongkol}, {Lindenmeyer}, {Long}, {Loomis}, {Loveday}, {Lucinio}, {Lupton}, {MacKinnon}, {Mannery}, {Mantsch}, {Margon}, {McGehee}, {McKay}, {Meiksin}, {Merelli}, {Monet}, {Munn}, {Narayanan}, {Nash}, {Neilsen}, {Neswold}, {Newberg}, {Nichol},
  {Nicinski}, {Nonino}, {Okada}, {Okamura}, {Ostriker}, {Owen}, {Pauls}, {Peoples}, {Peterson}, {Petravick}, {Pier}, {Pope}, {Pordes}, {Prosapio}, {Rechenmacher}, {Quinn}, {Richards}, {Richmond}, {Rivetta}, {Rockosi}, {Ruthmansdorfer}, {Sandford}, {Schlegel}, {Schneider}, {Sekiguchi}, {Sergey}, {Shimasaku}, {Siegmund}, {Smee}, {Smith}, {Snedden}, {Stone}, {Stoughton}, {Strauss}, {Stubbs}, {SubbaRao}, {Szalay}, {Szapudi}, {Szokoly}, {Thakar}, {Tremonti}, {Tucker}, {Uomoto}, {Vanden Berk}, {Vogeley}, {Waddell}, {Wang}, {Watanabe}, {Weinberg}, {Yanny}, {Yasuda}, \& {SDSS Collaboration}}]{york00}
{York}, D.~G., {Adelman}, J., {Anderson}, John~E., J., {et~al.} 2000, \bibinfo{title}{{The Sloan Digital Sky Survey: Technical Summary},} \aj, 120, 1579, \dodoi{10.1086/301513}

\bibitem[{G. {Younes} {et~al.}(2011){Younes}, {Porquet}, {Sabra}, \& {Reeves}}]{younes11}
{Younes}, G., {Porquet}, D., {Sabra}, B., \& {Reeves}, J.~N. 2011, \bibinfo{title}{{Study of LINER sources with broad H{\ensuremath{\alpha}} emission. X-ray properties and comparison to luminous AGN and X-ray binaries},} \aap, 530, A149, \dodoi{10.1051/0004-6361/201116806}

\bibitem[{G. {Zeltyn} {et~al.}(2022){Zeltyn}, {Trakhtenbrot}, {Eracleous}, {Runnoe}, {Trump}, {Stern}, {Shen}, {Hern{\'a}ndez-Garc{\'\i}a}, {Bauer}, {Yang}, {Dwelly}, {Ricci}, {Green}, {Anderson}, {Assef}, {Guolo}, {MacLeod}, {Davis}, {Fries}, {Gezari}, {Grogin}, {Homan}, {Koekemoer}, {Krumpe}, {LaMassa}, {Liu}, {Merloni}, {Mart{\'\i}nez-Aldama}, {Schneider}, {Temple}, {Brownstein}, {Ibarra-Medel}, {Burke}, {Pellegrino}, \& {Kollmeier}}]{zeltyn22}
{Zeltyn}, G., {Trakhtenbrot}, B., {Eracleous}, M., {et~al.} 2022, \bibinfo{title}{{A Transient ``Changing-look'' Active Galactic Nucleus Resolved on Month Timescales from First-year Sloan Digital Sky Survey V Data},} \apjl, 939, L16, \dodoi{10.3847/2041-8213/ac9a47}

\bibitem[{G. {Zeltyn} {et~al.}(2024){Zeltyn}, {Trakhtenbrot}, {Eracleous}, {Yang}, {Green}, {Anderson}, {LaMassa}, {Runnoe}, {Assef}, {Bauer}, {Brandt}, {Davis}, {Frederick}, {Fries}, {Graham}, {Grogin}, {Guolo}, {Hern{\'a}ndez-Garc{\'\i}a}, {Koekemoer}, {Krumpe}, {Liu}, {Mart{\'\i}nez-Aldama}, {Ricci}, {Schneider}, {Shen}, {{\'S}niegowska}, {Temple}, {Trump}, {Xue}, {Brownstein}, {Dwelly}, {Morrison}, {Bizyaev}, {Pan}, \& {Kollmeier}}]{zeltyn24}
{Zeltyn}, G., {Trakhtenbrot}, B., {Eracleous}, M., {et~al.} 2024, \bibinfo{title}{{Exploring Changing-look Active Galactic Nuclei with the Sloan Digital Sky Survey V: First Year Results},} \apj, 966, 85, \dodoi{10.3847/1538-4357/ad2f30}

\end{thebibliography}
\bibliographystyle{aasjournalv7}



\end{document}